\documentclass[traditabstract, twocolumns, longauth,colorlinks=true,linkcolor=blue,citecolor=blue, urlcolor=blue]{aa}  

\usepackage{orcidlink}
\usepackage{txfonts}
\usepackage{graphicx}	
\usepackage{amsmath}	
\usepackage{amssymb}	
\usepackage{xcolor}     
\usepackage{soul}       
\usepackage{array}
\usepackage{makecell}
\usepackage{hyperref}
\usepackage{multirow}

\def\ksmpc{${\, \mathrm{km}\, \mathrm{s}^{-1}\, \mathrm{Mpc}^{-1}}$\xspace}

\newcommand{\Bsixteen}{B1608$+$656\xspace}
\newcommand{\RXJ}{RX\ J1131$-$1231\xspace}
\newcommand{\HEzero}{HE\ 0435$-$1223\xspace}
\newcommand{\Jtwelve}{SDSS\ J1206$+$4332\xspace}
\newcommand{\WFItwenty}{WFI\ 2033$-$4723\xspace}
\newcommand{\PGeleven}{PG\ 1115$+$080\xspace}
\newcommand{\DESzerofour}{DES\ J0408$-$5354\xspace}
\newcommand{\WGDtwenty}{WGD\ 2038$-$4008\xspace}

\newcommand{\lcdm}{$\mathrm{\Lambda CDM}$\xspace}
\newcommand{\Hc}{\ensuremath{H_0}\xspace}

\newcommand{\Ddt}{\ensuremath{D_{\Delta t}}\xspace}
\newcommand{\Om}{\ensuremath{\Omega_{\mathrm m}}\xspace}
\newcommand{\OL}{\ensuremath{\Omega_{\Lambda}}\xspace}

\newcommand{\Ok}{\ensuremath{\Omega_{k}}\xspace}
\newcommand{\kext}{\ensuremath{\kappa_\mathrm{ext}}\xspace}

\newcommand{\thetaE}{\ensuremath{\theta_\mathrm{E}}\xspace}
\newcommand{\nslacsimaging}{34}
\newcommand{\nslacsifu}{14} 
\newcommand{\nslacsifusel}{11}
\newcommand{\nsltwossel}{four}

\newcommand{\sref}[1]{Section~\ref{#1}}

\newcommand{\fref}[1]{Fig.~\ref{#1}}

\newcommand{\tref}[1]{Table~\ref{#1}}

\newcommand{\tdist}{D_{\Delta t}}

\newcommand{\hnotflcdmonea}{71.6^{+3.9}_{-3.3}}

\newcommand{\hnotflcdmoned}{74.3^{+3.1}_{-3.7}}

\begin{document}

\title{
TDCOSMO 2025: Cosmological constraints from strong lensing time delays
}
\author{
TDCOSMO Collaboration:
            Simon~Birrer\orcidlink{0000-0003-3195-5507}\inst{\ref{stonybrook}}\fnmsep\thanks{ Corresponding authors. Emails: \href{mailto:simon.birrer@stonybrook.edu}{simon.birrer@stonybrook.edu}, \href{mailto:mmillon@ethz.ch}{mmillon@ethz.ch}, \href{mailto:ajshajib@uchicago.edu}{ajshajib@uchicago.edu}
            },
            Elizabeth~J.~Buckley-Geer\orcidlink{0000-0002-3304-0733}\inst{\ref{fnal},\ref{Uchicago}},
            Michele~Cappellari\orcidlink{0000-0002-1283-8420}\inst{\ref{oxford}},
            Fr\'ed\'eric~Courbin\orcidlink{0000-0003-0758-6510}\inst{\ref{iccub}, \ref{icrea}},
            Fr\'ed\'eric~Dux\orcidlink{0000-0003-3358-4834}\inst{\ref{esochile},\ref{epfl}},
            Christopher~D.~Fassnacht\orcidlink{0000-0002-4030-5461}\inst{\ref{ucdavis}},
            Joshua~A.~Frieman\orcidlink{0000-0003-4079-3263}\inst{\ref{Uchicago},\ref{Kavlichicago},\ref{slac}},
            Aymeric~Galan\orcidlink{0000-0003-2547-9815}\inst{\ref{mpa}, \ref{tum}},
            Daniel Gilman\orcidlink{0000-0002-5116-7287}\inst{\ref{Uchicago}}\fnmsep\thanks{Brinson Fellow},
            Xiang-Yu~Huang\orcidlink{0000-0001-7113-0599}\inst{\ref{stonybrook}},
            Shawn~Knabel\orcidlink{0000-0001-5110-6241}\inst{\ref{ucla}},
            Danial~Langeroodi\orcidlink{0000-0001-5710-8395}\inst{\ref{dark}},
            Huan~Lin\orcidlink{0000-0002-7825-3206}\inst{\ref{fnal}},
            Martin~Millon\orcidlink{0000-0001-7051-497X}\inst{\ref{ethz}}\fnmsep$^\star$,
            Takahiro~Morishita\orcidlink{0000-0002-8512-1404}\inst{\ref{ipac}},
            Veronica~Motta\orcidlink{0000-0003-4446-7465}\inst{\ref{valparaiso}},
            Pritom~Mozumdar\orcidlink{0000-0002-8593-7243}\inst{\ref{ucla}},
            Eric~Paic\orcidlink{0000-0002-4306-7366}\inst{\ref{utokyo}},
            Anowar~J.~Shajib\orcidlink{0000-0002-5558-888X}\inst{\ref{Uchicago}, \ref{Kavlichicago}, \ref{IndepUniversity}}\fnmsep$^\star$\fnmsep\thanks{NHFP Einstein Fellow},
            William~Sheu\orcidlink{0000-0003-1889-0227}\inst{\ref{ucla}},
            Dominique~Sluse\orcidlink{0000-0001-6116-2095}\inst{\ref{star}},
            Alessandro Sonnenfeld\orcidlink{0000-0002-6061-5977}\inst{\ref{jtus},\ref{sklppc},\ref{klppac}},
            Chiara~Spiniello\orcidlink{0000-0002-3909-6359}\inst{\ref{oxford}},
            Massimo~Stiavelli\orcidlink{0000-0001-9935-6047}\inst{\ref{stsci}},
            Sherry~H.~Suyu\orcidlink{0000-0001-5568-6052}\inst{\ref{tum},\ref{mpa}},
            Chin~Yi~Tan\orcidlink{0000-0003-0478-0473}\inst{\ref{Uchicago}, \ref{Kavlichicago}},
            Tommaso~Treu\orcidlink{0000-0002-8460-0390}\inst{\ref{ucla}},
            Lyne~Van~de~Vyvere\orcidlink{0000-0002-0585-4203}\inst{\ref{star}},
            Han~Wang\orcidlink{0000-0002-1293-5503}\inst{\ref{mpa},\ref{tum}},
            Patrick~Wells\orcidlink{0000-0003-0999-2395}\inst{\ref{argonne}},
            Devon~M.~Williams\orcidlink{0000-0002-8386-0051}\inst{\ref{ucla}}, and
            Kenneth~C.~Wong\orcidlink{0000-0002-8459-7793}\inst{\ref{utokyo}}
}

\institute{
    Department of Physics and Astronomy, Stony Brook University, Stony Brook, NY 11794, USA \label{stonybrook} \goodbreak
    \and
    Fermi National Accelerator Laboratory, PO Box 500, Batavia, IL 60510, USA \label{fnal}
    \goodbreak
    \and
    Department  of  Astronomy  \&  Astrophysics,  University  of Chicago, Chicago, IL 60637, USA \label{Uchicago} \goodbreak
    \and 
    Sub-Department of Astrophysics, Department of Physics, University of Oxford, Denys Wilkinson Building, Keble Road, Oxford, OX1 3RH, UK \label{oxford} \goodbreak
    \and 
    Institut de Ci\`{e}ncies del Cosmos (ICCUB), 
    Universitat de Barcelona (IEEC-UB), 
    Mart\'{i} i Franqu\`{e}s 1, 08028 Barcelona, Spain
    \label{iccub}
    \goodbreak
    \and 
    Instituci\'o Catalana de Recerca i Estudis Avan\c{c}ats (ICREA), 
    Passeig de Llu\'is Companys 23, 08010 Barcelona, Spain
    \label{icrea}
    \goodbreak
    \and    
    European Southern Observatory, Alonso de Córdova 3107, Vitacura, Santiago, Chile \label{esochile}
    \and
    Institute of Physics, Laboratory of Astrophysics, Ecole Polytechnique 
    F\'ed\'erale de Lausanne (EPFL), Observatoire de Sauverny, 1290 Versoix,
    Switzerland \label{epfl} \goodbreak 
    \and  
    Department of Physics and Astronomy, UC Davis, 1 Shields Ave., Davis, CA 95616, USA \label{ucdavis}
    \goodbreak
    \and
    Kavli Institute for Cosmological Physics, University of Chicago, Chicago, IL 60637, USA \label{Kavlichicago} \goodbreak
    \and 
    SLAC National Laboratory, 2575 Sand Hill Rd, Menlo Park, CA 94025 \label{slac}
    \goodbreak
    \and
    Max-Planck-Institut f{\"u}r Astrophysik, Karl-Schwarzschild Stra{\ss}e 1, 85748 Garching, Germany
    \label{mpa}
    \goodbreak
    \and 
    Technical University of Munich, TUM School of Natural Sciences, Physics Department,  James-Franck-Stra{\ss}e 1, 85748 Garching
    \label{tum}
    \goodbreak
    \and
    Department of Physics and Astronomy, University of California, Los Angeles, CA 90095, USA \label{ucla} \goodbreak
    \and
    DARK, Niels Bohr Institute, University of Copenhagen, Jagtvej 155A, 2200 Copenhagen, Denmark
    \label{dark}
    \goodbreak
    \and
    Institute for Particle Physics and Astrophysics, ETH Zurich, Wolfgang-Pauli-Strasse 27, CH-8093 Zurich, Switzerland \label{ethz}
    \and
    IPAC, California Institute of Technology, MC 314-6, 1200 E. California Boulevard, Pasadena, CA 91125, USA \label{ipac}
    \goodbreak
    \and
    Instituto de F\'isica y Astronom\'ia, Universidad de Valpara\'iso, Avda. Gran Breta\~na 1111, Valpara\'iso, Chile.
    \label{valparaiso}
    \goodbreak
    \and 
    Research Center for the Early Universe, Graduate School of Science, The University of Tokyo, 7-3-1 Hongo, Bunkyo-ku, Tokyo 113-0033, Japan
    \label{utokyo}
    \goodbreak
    \and 
    Center for Astronomy, Space Science and Astrophysics, Independent University, Bangladesh, Dhaka 1229, Bangladesh \label{IndepUniversity} \goodbreak
    \and
    STAR Institute, Li\`ege Universit\'e, Quartier Agora - All\'ee du six Ao\^ut, 19c B-4000 Li\`ege, Belgium
    \label{star}
    \goodbreak
    \and
    Department of Astronomy, School of Physics and Astronomy, Shanghai Jiao Tong University, Shanghai 200240, China \label{jtus}
    \goodbreak
    \and
    Shanghai Key Laboratory for Particle Physics and Cosmology, Shanghai Jiao Tong University, Shanghai 200240, China\label{sklppc}
    \goodbreak
    \and 
    Key Laboratory for Particle Physics, Astrophysics and Cosmology, Ministry of Education, Shanghai Jiao Tong University, Shanghai 200240, China \label{klppac}
    \goodbreak
    \and
    Space Telescope Science Institute, 3700 San Martin Dr., Baltimore, MD 21218, USA \label{stsci}
    \goodbreak
    \and
    HEP Division, Argonne National Laboratory, Lemont, IL 60439, USA \label{argonne}
    }

\abstract{
We present cosmological constraints from eight strongly lensed quasars (hereafter, the TDCOSMO-2025 sample). 
Building on previous work, our analysis incorporated new deflector stellar velocity dispersions measured from spectra obtained with the \textit{James Webb} Space Telescope (JWST), the Keck Telescopes, and the Very Large Telescope (VLT), utilizing improved methods.
We used integrated JWST stellar kinematics for five lenses, VLT-MUSE for 2, and resolved kinematics from Keck and JWST for \RXJ. We also considered two samples of non-time-delay lenses: \nslacsifusel\ from the Sloan Lens ACS (SLACS) sample with Keck-KCWI resolved kinematics; and \nsltwossel\ from the Strong Lenses in the Legacy Survey (SL2S) sample.  We improved our analysis of line-of-sight effects, the surface brightness profile of the lens galaxies, and orbital anisotropy, and corrected for projection effects in the dynamics. Our uncertainties are maximally conservative by accounting for the mass-sheet degeneracy in the deflectors' mass density profiles. The analysis was blinded to prevent experimenter bias. 
Our primary result is based on the TDCOSMO-2025 sample, in combination with $\Omega_{\rm m}$ constraints from the  
Pantheon+ Type Ia supernovae (SN) dataset.
In the flat $\Lambda$ cold dark matter (CDM), we find 
$H_0=\hnotflcdmonea$ \ksmpc.
The SLACS and SL2S samples are in excellent agreement with the TDCOSMO-2025 sample, improving the precision on $H_0$ in flat $\Lambda$CDM to 4.6\%. Using the Dark Energy Survey SN Year-5 dataset (DES-SN5YR) or DESI-DR2 baryonic acoustic oscillations (BAO) likelihoods instead of Pantheon+ yields very similar results.
We also present constraints in the open $\Lambda$CDM, $w$CDM, $w_0w_a$CDM, and $w_{\phi}$CDM cosmologies. 
The TDCOSMO $H_0$ inference is robust and consistent across all presented cosmological models, and our cosmological constraints in them agree with those from the BAO and SN.
}

\titlerunning{2025 cosmological constraints}
\authorrunning{TDCOSMO Collaboration}
\maketitle

%
\section{Introduction}
\label{sec:introduction}


Allan Sandage famously described his life's work as the "search for two numbers": the Hubble constant, $H_0$, and the deceleration parameter, $q_0$. Our view of cosmology has changed significantly since then, chiefly with the discovery that the Universe is accelerating \citep{Riess98,Perlmutter99}, the explosion in the number of precise cosmological probes, and the emergence of the concordance $\Lambda$ cold dark matter (\lcdm) model \citep[e.g.,][and references therein]{FTH08}. 

The Hubble constant has become even more central in our quest for understanding the composition and expansion history of the Universe. As the uncertainties in the inferred $H_0$ have shrunk from a factor of two to a percent level, a tension has emerged between direct measurements based on the present-day Universe, and those based on early-Universe probes, extrapolated to the present day under the assumption of a standard flat \lcdm cosmology. For example, whereas analysis of the cosmic microwave background (CMB) data from the \textit{Planck} satellite yields $67.4\pm0.5$\ksmpc in flat $\Lambda$CDM \citep{Planck2020}, the most recent measurement based on the traditional local distance ladder by the SH0ES team, with Cepheids and Type Ia supernova (SN Ia), yields $73.04 \pm 1.04$\ksmpc \citep{Riess22}. Considerable effort has gone into building alternative local distance ladders based on alternatives to Cepheids \citep[e.g.][]{Freedman24,Lee24,Li24a,Li24b}; developing independent methods such as cosmic chronometers \citep{Tomasetti2023}; exploiting early-Universe probes independent of the CMB, including Big Bang nucleosynthesis in combination with baryon acoustic oscillations \citep[BAO, e.g.,][]{DESIDR2}. Overall, this 8\% difference, known as the "Hubble tension", between early-Universe and late-Universe probes, has crossed the traditional gold-standard 5$\sigma$ threshold in terms of statistical significance \citep[see, e.g.,][for a recent review]{cosmoverse25}.

If the Hubble tension is real -- and not due to unknown systematic uncertainties in multiple measurements -- the implications are profound. From a theoretical standpoint, there is no obvious leading contender to reconcile the measurements in tension. Proposed solutions include for example a modification of the early expansion history of the Universe owing to extra relativistic particles beyond the standard model, or an early dark energy phase, 
but this is by no means the only possibility \citep[e.g.,][]{Knox20,Divalentino21,Gelmini21,Abdalla22,Schonberg22,Vagnozzi23,cosmoverse25}.

\section{Overview of current and past time-delay cosmography analyses}
\label{sec:tdcosmo_method}
\subsection{An independent path to \Hc }
Time-delay cosmography \citep{refsdal1964,TreuMarshall2016,Treu_review,TreuShajib_review, Birrer:2024, Suyu2024} provides the opportunity to address the tension in a manner completely independent of all other probes. The delays between the arrival times of multiple images of strongly lensed sources measure absolute distances, and thus yield a measurement of $H_0$, without the need to rely on the local distance ladder for calibration. Furthermore, typical strong lens systems are composed of a deflector at $z\sim0.5$ and a source at $z\sim1-3$. They can therefore be considered a late-time probe, unaffected by pre-recombination physics, and yet measure the Universe on the gigaparsec scale, large enough to be insensitive to local deviations from homogeneity and isotropy \citep{Dalang2023}.

The potential of time-delay cosmography was recognized \citep{refsdal1964} even before the first strong lenses were discovered \citep{Walsh79}. As for many other cosmological probes, however, it took decades of development in methodology and improvement in data quality to reach sufficient precision and accuracy to contribute to the measurement of $H_0$. Breakthroughs were achieved in the measurement of time delays with high-quality light curves \citep{Fassnacht2002, Courbin:2005, Millon:2020b, Dux:2025}, in the modeling of the potential of the main deflector from the quasar host galaxy extended images \citep{Koopmans2003,Suyu2009}, and in the reconstruction of the effects of the inhomogeneity of the mass distribution along the line of sight \citep[LoS, see, e.g.,][for a historical perspective]{Suyu2010,Greene2013,TreuMarshall2016}. A turning point was the publication of the analysis of six quadruply imaged quasars (hereafter, quads) with state-of-the-art quality data by \citet{Wong2019}. By combining the distances obtained from the systems, \citet{Wong2019} measured $H_0=73.3^{+1.7}_{-1.8}$\ksmpc with 2.4\% precision. 
Importantly, apart from the first system \citep{Suyu2010}, the analyses of five out of the six quads were performed blindly to prevent experimenter bias. A seventh quad analyzed in a consistent fashion \citep{Shajib2019} further improved the precision to 2\%, i.e. $H_0=74.2\pm1.6$\ksmpc\citep[][hereafter TDCOSMO-1]{Millon:2020}.  This 2\% measurement was in excellent agreement with the local distance ladder based on Cepheids and SNe Ia from the SH0ES team \citep{Riess22}, thus strengthening the statistical tension \citep{Wong2019,Millon:2020}.

\subsection{TDCOSMO and the challenge of mass profile systematics}
\label{subsec:tdcosmo_collaboration}

After the \citet{Wong2019} result, many scientists working on time-delay cosmography joined their effort in a newly formed collaboration called the Time-Delay COSMOgraphy (TDCOSMO) collaboration, with the goal of uncovering and mitigating unknown systematic effects \citep[e.g.,][]{Millon:2020,TDCOSMO3,TDCOSMO7, TDCOSMO8, Shajib2022}. 

The main potential source of systematic uncertainty was found to be the description of the mass density profile of the main deflector, stemming from the lensing mass-sheet degeneracy \citep[MSD;][]{Falco1985}. Given a mass model that reproduces the multiple images, one can obtain a family of models that reproduce the images equally well via simple mathematical transformations \citep{Schneider2013,Schneider2014}, which modify the inferred source's intrinsic position, size, and luminosity, as well as the inferred Hubble constant. The MSD can be broken with non-lensing data or by invoking theoretical assumptions. For example, in its simple form, the MSD corresponds to the existence of an infinite sheet of mass in the plane of the lens. Therefore, asserting that the mass density profile of the deflector goes to zero at infinity following a power law breaks the MSD. \citet{Wong2019}, \citet{Shajib2019}, and \citet{Millon:2020} broke the MSD by assuming two commonly accepted standard forms for the mass density profile of the deflectors, either a power law \citep{cappellari16} or a composite model with stellar mass following the light and a \citet*[][hereafter NFW]{Navarro1997} profile for the dark matter, and anchoring the mass profiles with stellar velocity dispersion.

Given the high stakes, the TDCOSMO collaboration decided to rebuild the inference of $H_0$ under much weaker and thus conservative assumptions.  \citet[][hereafter, TDCOSMO-4]{TDCOSMOIV} showed that by parametrizing the mass profile in a way that effectively maximizes the MSD and thus the uncertainty on \Hc, the precision drops from 2\% to $\sim$9\% for the same seven lenses and datasets of \cite{Millon:2020}, finding $H_0=74.5_{-6.1}^{+5.6}$\ksmpc. TDCOSMO-4 also proposed that, by combining time-delay lenses with more numerous non-time-delay lenses (hereafter external lens samples), one can improve the precision, assuming that the lens galaxies are drawn from the same population. As an example, they combined the seven time-delay lenses with 33 non-time delay lenses from the Sloan Lens ACS Survey \cite[SLACS;][]{Bolton06,Bolton08} sample. The combination improved the precision to 5\%. While the central value of the posterior shifted to $H_0=67.4_{-3.2}^{+4.1}$\ksmpc, it remained consistent with the TDCOSMO-only result, within the errors. The shift could be due to statistical noise, differences between the time-delay and non-time-delay deflectors, or it could indicate that some of the previous assumptions or measurements were incorrect. The information available at that time was insufficient to distinguish the three hypotheses. Higher quality stellar kinematics,\footnote{To first order, any bias $\delta \sigma$ in the stellar velocity dispersion $\sigma$ results in a bias $\delta H_0 = 2(\delta \sigma/\sigma) \times H_0$ in the Hubble constant \citep{Chen21}.} preferably spatially resolved, is needed to break the mass-anisotropy degeneracy \citep{Courteau14}, and make progress \citep{Treu2002,shajib18,birrer_treu21, yildirim20, yildirim23}. For example, we now know from comparison with Keck spectra with SNR$\sim160$ \citep{Knabel24} that the measurements of stellar velocity dispersion based on relatively low SNR ($\sim9$ \AA$^{-1}$) SDSS spectra of the SLACS lenses used in TDCOSMO-4 suffer from systematic errors at the level of 3.3\% and covariance at the level of 2.3\%, insufficient for precision cosmology. Therefore, the TDCOSMO+SLACS measurement in TDCOSMO-4 should be discarded and replaced with one based on better data. 

\subsection{Advancements in this milestone analysis}
\label{sec:challenges}

In this "milestone" paper, we present a major update to the TDCOSMO collaboration's inference of $H_0$, following the methodology introduced in TDCOSMO-4 with hierarchical Bayesian inference. We combine multiple improvements in data and analysis as follows.

\begin{enumerate}

\item We include the constraints from an eighth time-delay lens, \WGDtwenty, that was recently analyzed by our team \citep{Wong24}, in addition to the seven used in our prior analyses (see Table~\ref{tab:lens_overview}). For convenience, this sample of eight time delay lenses is labeled the TDCOSMO-2025 sample.

\item We use improved methods and template libraries to infer stellar velocity dispersions and estimate residual systematic errors and covariance between galaxies and within spatial bins inside galaxies with spatially resolved kinematics \citep{Knabel25}. We apply these new methods to derive more precise and accurate stellar kinematics for the time-delay lenses and external lens samples, either by reanalyzing old data or using new data when available.

\item We use new spectroscopy of time-delay lenses to improve their kinematic measurements. For the lens \RXJ, we use JWST-NIRSpec integral field unit (IFU) spectroscopy \citep{Shajib25b} and Keck-KCWI ground-based spectroscopy \citep{Shajib23} to derive spatially resolved kinematics. For five systems, we use JWST-NIRSpec spectra to infer integrated stellar velocity dispersions.  For the two remaining systems, we use new spectra obtained with the MUSE instrument on the VLT.

\item We adopt constant anisotropy as our default parametrization instead of the Osipkov--Merritt \citep[][hereafter OM]{Osipkov1979,Merritt1985} form used in TDCOSMO-4. This is motivated by studies of local elliptical galaxies \citep[Section~3.3]{Cappellari2025}, which show that the anisotropy is nearly constant and never fully radial, as in the outer radii of the OM parametrization. 

\item  We use two external lens samples with improved lens models. In addition to a reanalysis of the SLACS sample \citep{Tan24}, we consider the Strong Lenses in the Legacy Survey (SL2S) sample \citep{Ruff11,Gavazzi12,Sonnenfeld13a,Sonnenfeld13b,Sheu25}, which has a larger redshift overlap with the time-delay deflectors.  For the SLACS sample, we use spatially resolved stellar kinematics obtained with KCWI for \nslacsifu\ non-time delay lenses \citep{Knabel24}. Contrary to TDCOSMO-4, we do not use the SDSS-based stellar velocity dispersion, since  \citet{Knabel24} showed that the SDSS spectra of the SLACS galaxies suffer from systematic errors on the stellar velocity dispersion at the level of 3.3\% and covariance at the level of 2.3\%. For the SL2S sample, we use the re-measured stellar velocity dispersion based on the updated methodology and stellar libraries \citep[]{Mozumdar2025}. These authors show that the SLACS, SL2S, and time-delay lenses follow the same mass-fundamental plane \citep{Bolton08,Auger2009}. We are also using new deeper HST observations of the SL2S lenses presented and modeled by \cite{Sheu25}, including a treatment of lens model systematics as a function of the signal-to-noise ratio (SNR) of the observation. We apply rigorous cuts to ensure that the SLACS and SL2S are matched to TDCOSMO-2025 and to enforce high-quality data, ending up utilizing \nslacsifusel\ lenses for SLACS and \nsltwossel\ for SL2S.

\item We account for departures from spherical symmetry in the deflector population and correct for projection effects and the interpretation of the kinematic observables \citep{Huang:2025}.

\item We improve and homogenize our treatment of the LoS effects, extending it to the external lens samples \citep{Wells2024}, including a hierarchical inference of the distribution of external convergence values for our external lenses datasets. 

\item We improve our description of the surface brightness profiles of the deflector galaxies, both in the lensing analysis and in the dynamical models. We have redone the fits and replaced the Hernquist models with multi-gaussian expansion decompositions based on double-S\'ersic fits. For the SL2S lenses, we use Canada--France--Hawaii Telescope Legacy Survey \citep[CFHTLS;][]{Gwyn2012} imaging in combination with HST-based models of the lensed features to derive the deflector light profile \citep{Sheu25}. The CFHTLS images are necessary for cases where the deflector is so red that the HST F475X filter does not capture the deflector light.

\end{enumerate}

As in our previous works, our analysis and cosmographic inference are carried out "blinded" to prevent experimenter bias. This is achieved at the software level by removing the mean of all the cosmographically relevant parameters from posterior probability distribution plots. This was done independently for each dataset, so the collaboration was not able to check for mutual consistency between the TDCOSMO-2025, SLACS, and SL2S datasets prior to unblinding. The posterior was unblinded on May 27, 2025, after the analysis was frozen and approved by internal review, and then the results were published without modification.\footnote{After the release of the first version of this paper on arXiv (v1), we discovered a coding error that incorrectly assigned the external convergence (\kext) distribution of \Bsixteen to all the TDCOSMO lenses, and that was hidden by our blinding procedure. Correcting the error - without any other changes to the analysis - shifted $H_0$ by $-0.5$ \ksmpc w.r.t. the first version of this paper for the TDCOSMO-only result, and by approximately $+2$ \ksmpc when combining with SLACS. Since the product $\lambda_{\rm int}(1-\kext)$ is constrained by the stellar velocity dispersion, correcting \kext\ shifted $\lambda_{\rm int}$ closer to unity. All the numbers and plots have been corrected, with no other substantial edits to the paper.}

The paper is structured as follows: In Section~\ref{sec:tdc}, we provide a brief overview of the time-delay cosmography methodology.  In Section~\ref{sec:lens_sample}, we describe the sample of lenses that we use for the inference. In Section~\ref{sec:newmeas} we describe our new stellar kinematic measurements.  In Section~\ref{sec:sampleselection}, we describe the selection criteria that we apply to the SLACS and SL2S samples to ensure high-quality measurements and that they match the properties of the time-delay lenses. In Section~\ref{sec:hierarch} we describe our joint hierarchical inference of the cosmological parameters and population parameters. In Section~\ref{sec:flat} we describe our results in the standard flat $\Lambda$ cold dark matter (CDM) model. In Section~\ref{sec:alternative_cosmo} we describe our results in alternative cosmological models, including non-flat $\Lambda$CDM, flat $w$CDM, models with dynamical dark energy, such as $w_\phi$CDM and $w_0w_a$CDM, as suggested by recent measurements \citep{DESIDR2}.
In Section~\ref{sec:literature}, we compare our results to our previous work and to other measurements based on time-delay cosmography. In Section~\ref{sec:syst} we discuss residual systematic errors and selection effects. In Section~\ref{sec:summary} we provide a brief summary.
  
The paper was, for the most part, written prior to unblinding.  The work was reviewed by the collaboration prior to unblinding, including four designated reviewers (M.C., J.F., D.S., S.H.S.) and five code reviewers (S.B., A.G., X-Y.H., M.M., A.J.S.). After unblinding, numerical values in plots and tables are automatically updated. Sections~\ref{sec:flat}, \ref{sec:alternative_cosmo}, and the Appendixes were written after unblinding. The abstract, Section~\ref{sec:literature}, Section~\ref{sec:syst}, and Section~\ref{sec:summary} were mostly written prior to unblinding and completed after unblinding. The other Sections were only copy-edited for clarity, uniformity, and style during the final readings of the complete paper. The final version of the paper was further reviewed by the collaboration and approved by the final reader (T.T.). The likelihood, data, and scripts are public on the TDCOSMO public repository\footnote{\url{https://github.com/TDCOSMO/TDCOSMO2025_public}}.

\section{Time-delay cosmography}
\label{sec:tdc}

When the luminosity of a strongly lensed background source, such as an active galactic nucleus (AGN) or a SN, varies over time, the variability pattern manifests in each of the multiple images and is delayed in time due to the different light paths of the different images.
The arrival-time difference $\Delta t_{\rm AB}$ between two images at observed image-plane positions $\boldsymbol{\theta}_{\rm A}$ and $\boldsymbol{\theta}_{\rm B}$ that originated from the source at source-plane position $\boldsymbol{\beta}$, is
\begin{equation}\label{eqn:time_delay}
    \Delta t_{\rm AB} = \frac{D_{\Delta t}}{c} \left[\tau(\boldsymbol{\theta}_{\rm A}, \boldsymbol{\beta}) - \tau(\boldsymbol{\theta}_{\rm B}, \boldsymbol{\beta}) \right],
\end{equation}
where
\begin{equation}\label{eqn:fermat_potential}
    \tau(\boldsymbol{\theta}, \boldsymbol{\beta}) \equiv \left[ \frac{\left(\boldsymbol{\theta} - \boldsymbol{\beta} \right)^2}{2} - \phi(\boldsymbol{\theta})\right]
\end{equation}
is the Fermat potential \citep{Schneider1985, Blandford1986} with $\phi(\boldsymbol{\theta})$ being the lensing potential, and
\begin{equation} \label{eqn:ddt_definition}
    D_{\Delta t} \equiv \left(1 + z_{\rm d}\right) \frac{D_{\rm d}D_{\rm s}}{D_{\rm ds}}
\end{equation}
is the time-delay distance \citep{refsdal1964, Schneider1992, Suyu2010}. In this expression, $z_{\rm d}$ is the redshift of the deflector, while $D_{\rm d}$, $D_{\rm s}$, and $D_{\rm ds}$ are the angular diameter distances from the observer to the deflector, from the observer to the source, and from the deflector to the source, respectively.

Knowledge of the lensing potential (and hence Fermat Potential, Eq.~\ref{eqn:fermat_potential}) paired with a measured time delay between two arriving images can be turned into a measurement of $D_{\Delta t}$.
The Hubble constant is inversely proportional to the absolute distances of objects in the Universe for which redshifts are measured and thus scales with $D_{\Delta t}$ as
\begin{equation} \label{eqn:H0_ddt}
	H_0 \propto D_{\Delta t}^{-1}.
\end{equation}
The inverse proportionality between $H_0$ and the quantity $D_{\Delta t}$ makes $H_0$ the primary cosmological parameter constrained by the time-delay cosmography.

For details on time-delay cosmography, we refer the reader to the reviews, for example, by \cite{TreuMarshall2016, Treu_review, TreuShajib_review, Birrer:2024}.

\subsection{Mass-sheet degeneracy}

A significant challenge in accurately determining the Fermat potential, and thus $D_{\Delta t}$ and $H_0$, is the mass sheet degeneracy (MSD). The MSD arises from a transformation of the lens system that leaves the observed image positions unchanged while rescaling the source position and altering the inferred $D_{\Delta t}$.
Specifically, if a lens model with a (normalized) surface mass density or convergence $\kappa(\boldsymbol{\theta})$ accurately reproduces the observed lensing configuration, then a family of models $\kappa_{\lambda}(\boldsymbol{\theta})$ described by the following transform will also fit the data:
\begin{equation}\label{eqn:mst}
    \kappa_{\lambda}(\boldsymbol{\theta}) = \lambda \kappa(\boldsymbol{\theta}) + \left( 1 - \lambda\right).
\end{equation}
Eq.~\ref{eqn:mst} above is known as the mass-sheet transform (MST) and is a mathematical transformation where the term $(1 - \lambda) \equiv \kappa_{\rm MST}$ is formally equivalent to a uniform surface density sheet of convergence (or mass) that extends to infinite angular scales. Hence, the name mass-sheet transform and the name of the associated degeneracy, the mass-sheet degeneracy. The term $\kappa_{\rm MST}$ can be positive or negative, since it is defined relative to the average positive density of the Universe.
The MST, by means of preserving image positions and being linear, also preserves any higher-order relative differentials of the lens equation. Absolute lensing quantities, such as absolute magnification, source size, and luminosity, however, are not preserved by the MST.
Only observables related to either the unlensed apparent source size ($\boldsymbol{\beta}$ vs. $\lambda\boldsymbol{\beta}$), such as the unlensed apparent brightness, or the lensing potential, are capable of breaking the MSD.

The Fermat potential difference between a pair of images A and B (Eq.~\ref{eqn:fermat_potential}) scales with $\lambda$ as
\begin{equation}\label{eqn:fermat_mst}
    \Delta \tau_{\rm AB , \lambda} =  \lambda \Delta \tau_{\rm AB},
\end{equation}
and so does the relative predicted time delay as
\begin{equation}\label{eqn:time_delay_mst}
    \Delta t_{\rm AB , \lambda} =  \lambda \Delta t_{\rm AB}.
\end{equation}

An MSD effect relative to a specified deflector model might be associated with the mass distribution of the main deflector, referred to as the internal MSD scaled with $\lambda_{\rm int}$, or with inhomogeneities along the LoS of the strong lens system, referred to as the external MSD scaled with the external mass-sheet $\kappa_{\rm ext}$. The quantity $\kappa_{\rm ext}$ can be directly inferred for a given LoS using photometric and spectroscopic observables of the LoS environment \citep[e.g.,][]{Greene2013, Wells2024}.
The effect of all perturbers along a given LoS that are not included explicitly in the lens model itself is quantified with the external convergence$\kappa_{\rm{ext}}$. Conceptually, $\kappa_{\rm{ext}}$ can be thought of as the convergence of a mass sheet, which would produce the same cumulative effect as all LoS perturbers if it were placed coplanar with the lens. For a generic lens, $\kappa_{\rm{ext}}$ is usually of order $10^{-2}$ with positive values corresponding to LoSs that are slightly more dense than the background density of the Universe.

The total MST -- the relevant transform to constrain for an accurate Fermat potential determination and $H_0$ measurement -- is the product of the internal and external MST \citep[e.g.,][]{Schneider2013, Birrer2016, TDCOSMOIV}
\begin{equation}\label{eqn:lambda_combined}
    \lambda = (1-\kappa_{\rm ext}) \times \lambda_{\rm int}.
\end{equation}

To summarize, the prediction of the time delay (Eq.~\ref{eqn:time_delay}) can be generalized to 
\begin{equation}\label{eqn:time_delay_generalized}
    \Delta t_{\rm AB, \lambda} = (1-\kappa_{\rm ext}) \lambda_{\rm int} \frac{D_{\Delta t}}{c} \Delta \tau_{\rm AB},
\end{equation}
where the Fermat potential $\Delta \tau_{\rm AB}$ is the Fermat potential inferred without the inclusion of internal and external mass sheet.

When transforming a lens model with an MST, the inference of the time-delay distance (Eq.~\ref{eqn:ddt_definition}) from a measured time delay and previously inferred Fermat potential transforms as
\begin{equation} \label{eqn:ddt_mst}
    D_{\Delta t , \lambda} = \lambda^{-1}D_{\Delta t}.
\end{equation}
In turn, the Hubble constant, when inferred from the time-delay distance $D_{\Delta t}$, transforms as (from Eq.~\ref{eqn:H0_ddt})
\begin{equation} \label{eqn:h0_mst}
H_{0 , \lambda} =  \lambda H_0.
\end{equation}

\subsection{Combining lensing and kinematics}

Stellar kinematics is the most prominent and commonly used observable to break the total MSD. The collective motion of stars along the LoS is a direct tracer of the 3D gravitational potential and hence provides an independent mass estimate.
Joint lensing+dynamics constraints have been used to provide measurements of galaxy mass profiles \citep[e.g.,][]{Grogin1996, Romanowsky1999, Treu2002, Koopmans2003, Koopmans2004, Barnabe2011, Barnabe2012}.

The prediction of the stellar velocity dispersion projected along the line of sight $\sigma_{\rm los}$ from any model, regardless of the approach, can be decomposed into a cosmology-dependent and cosmology-independent part, as \citep[see e.g.,][]{Birrer2016, Birrer2019}
\begin{equation}\label{eqn:los_sigma_v}
    \sigma_{\rm los}^2 =  \frac{1-\kappa_{\rm s}}{1 - \kappa_{\rm ds}} \frac{D_{\rm s}}{D_{\rm ds}}c^2 J(\boldsymbol{\xi}_{\rm lens}, \boldsymbol{\beta}_{\rm ani},\lambda_{\rm int}),
\end{equation}
where $J$ is a dimensionless quantity dependent on the deflector model parameters ($\boldsymbol{\xi}_{\rm lens}$) and the observational conditions (seeing and spectroscopic aperture) of the velocity dispersion measurement \citep[e.g.,][]{Binney1982, TreuKoopmans2004, Suyu2010};  $c$ is the speed of light; $\boldsymbol{\beta}_{\rm ani}$ characterizes the anisotropy profile of the stellar orbits; and $\kappa_{\rm s}$ and $\kappa_{\rm ds}$ are the LoS external convergences from the observer to the source and the deflector to the source, respectively.  

The internal component $\lambda_{\rm int}$ describing a physical mass component that influences the motion of the stars is incorporated into the kinematics modeling term $J$ \citep{Teodori2022}. In the case of a sheet-like perturbation much more extended than the effective radius of the deflector, we can approximate the effect of $\lambda_{\rm int}$ as
\begin{equation}\label{eqn:j_lambda_int}
    J(\boldsymbol{\xi}_{\rm lens}, \boldsymbol{\beta}_{\rm ani},\lambda_{\rm int}) \approx \lambda_{\rm int} J(\boldsymbol{\xi}_{\rm lens}, \boldsymbol{\beta}_{\rm ani}).
\end{equation}

Joint lensing and dynamics constraints are sensitive to the combination of terms present in Eq.~\ref{eqn:los_sigma_v}. 

Studies of the LoS environment of the lenses and spatially resolved kinematics are needed to break degeneracies between terms within $J$. For example, by constraining the angular diameter distance ratio $D_{\rm s}/D_{\rm ds}$ from relative distance probes, and the LoS contributions $\kappa_{\rm s}$ and $\kappa_{\rm ds}$ from number counts compared to simulations, it is possible to infer $\lambda_{\rm int}$. 
In case of an internal component that is less in extent, detailed radially-binned or 2D kinematics maps can further constrain $\lambda_{\rm int}$.

When combining time delays with lensing and dynamics, one needs to use models that match the lensing configuration and constrain them simultaneously with the observed time delays and stellar kinematics, using Eq.~\ref{eqn:time_delay_generalized} and Eq.~\ref{eqn:los_sigma_v}. These two independent equations can be arbitrarily combined algebraically in 2D angular diameter distance constraints \citep{Birrer2016, Birrer2019, yildirim20}. A convenient transformation of those constraints is the following:
\begin{equation}\label{eqn:Ddt_cosmography}
    \tdist = \frac{1}{(1-\kappa_{\rm ext}) \lambda_{\rm int}}\frac{c \Delta t_{\rm AB}}{\Delta \tau_{\rm AB}}
\end{equation}
and 
\begin{equation}\label{eqn:Dd_cosmography}
    D_{\rm d} = \frac{1}{1 + z_{\rm d}}
    \frac{c \Delta t_{\rm AB}}{\Delta \tau_{\rm AB}}  \frac{c^2 J(\boldsymbol{\xi}_{\rm lens}, \boldsymbol{\beta}_{\rm ani}, \lambda_{\rm int})}{\lambda_{\rm int}\sigma^2_{\rm los}}.
\end{equation}

When mapped into the $D_{\Delta t}$--$D_{\rm d}$ plane as outlined above, the projection on constraints in $D_{\rm d}$ is invariant under any pure external MSD parameter $\kext$
\citep{Jee2015, Birrer2019, yildirim23}.\footnote{$D_{\rm d}$ is still dependent on the LoS between observer and lens, $\kappa_{\rm d}$, \citep[see e.g.,][]{Birrer:2024}.}
If the approximation of Eq.~\ref{eqn:j_lambda_int} holds, $D_{\rm d}$ becomes independent of $\lambda_{\rm int}$.

\section{Lens samples}
\label{sec:lens_sample}

\begin{figure*}
    \centering
    \includegraphics[width=\textwidth]{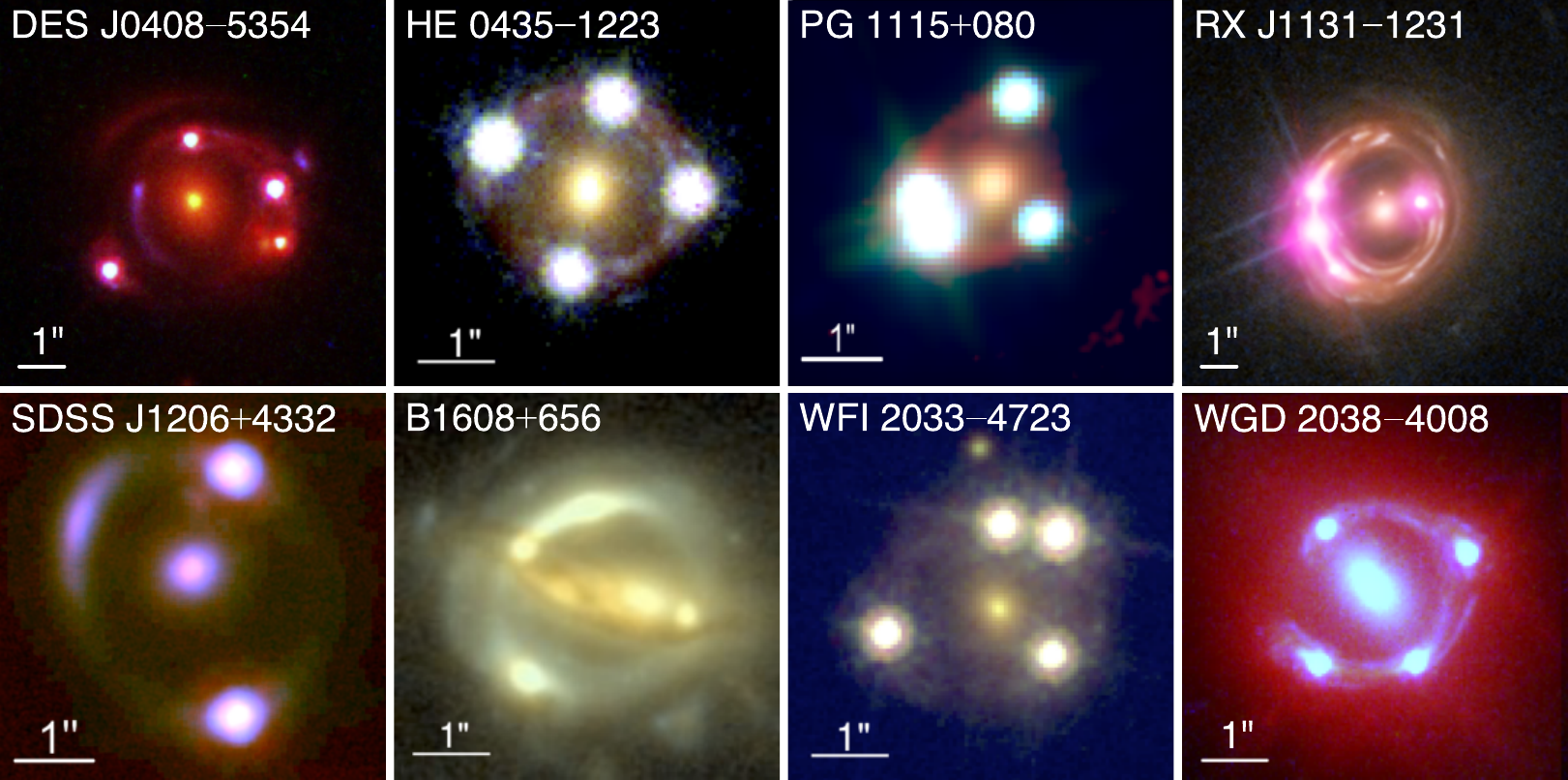}
    \caption{Montage of the eight lensed quasar systems in the TDCOSMO-2025 sample. In each panel, the white bar illustrates the 1\arcsec\ scale. The false-color images are made from two or three bands from the HST, Keck-NIRCam, and \textit{Chandra} X-ray imaging, given their availability. The image for \DESzerofour is adapted from \citet{Shajib2019}, \HEzero, \Bsixteen, and \WFItwenty from \citet{Suyu2017}, \PGeleven and \Jtwelve from \citet{Wong2019}, \RXJ from \citet[][\textit{credits: NASA/ESA/{\rm Chandra}}]{Shajib2024revie}, and \WGDtwenty from \citet{Shajib2022}.}
    \label{fig:montage}
\end{figure*}

Our sample consists of eight time-delay lenses (hereafter, the TDCOSMO-2025 sample; see Fig.~\ref{fig:montage} for a gallery montage). Those have measured time delays and cosmography-grade lens models from previous work. In addition, we present new high-quality spatially unresolved stellar velocity dispersions (except \RXJ, which has spatially resolved kinematics). To enhance our constraining power on the internal structure of the deflectors, following \citet{TDCOSMOIV} and \citet{birrer_treu21}, we added two external samples of lenses that have kinematics but no time delays. The lens models and kinematic data of non-time delay lenses constrain $\lambda_{\rm int}$, and thus indirectly cosmography when combined with time-delay lenses.  These two external samples are the SLACS lenses \citep{Bolton08} and the SL2S lenses \citep{Gavazzi12}. We applied cuts to the two samples of non-time-delay lenses to ensure sufficiently high-quality lensing and kinematic data for cosmography and matched them in properties to the TDCOSMO-2025 sample. After the cuts, the SLACS sample consists of \nslacsifusel\ lenses with spatially resolved kinematics from Keck-KCWI \citep{Knabel24} and updated lens models \citep{Tan24}. After the cuts, the SL2S sample consists of \nsltwossel\ lenses with updated kinematic measurements \citep{Mozumdar2025} and lens models \citep{Sheu25}.
The samples are described in this section, and the new spectroscopic measurements are described next in Section \ref{sec:newmeas}. Then, \sref{sec:sampleselection} describes the selection cuts.

\subsection{Sample of time-delay lenses}

Our TDCOSMO-2025 sample comprises the six original H0LiCOW\footnote{$H_0$ Lenses in COSMOGRAIL's Wellspring \citep{Suyu2017}} lensed quasars -- \HEzero, \PGeleven, \RXJ, \Jtwelve, \Bsixteen, and \WFItwenty\ -- as presented by \citet{Wong2019} and the references listed in Table~\ref{tab:lens_overview}. In addition, we included \DESzerofour, presented by \citet{Shajib2019}, and \WGDtwenty, presented by \citet{Wong24}. The first seven systems are shared with the sample previously published by TDCOSMO-1 and TDCOSMO-4.  These eight lenses in the TDCOSMO-2025 sample provide cosmological information through their time-delay distance measurement, \Ddt. 

In addition to the new data presented in this paper, the analysis presented here relies on data products published in the literature for the \Ddt measurements. This includes time-delay estimations, lens modeling, lens environment characterization, stellar kinematics, and redshift measurements. Individual papers reporting those measurements for each lens are listed in \tref{tab:lens_overview}. 

\begin{table*}[ht]
    \renewcommand{\arraystretch}{1.7}
    \caption{\label{tab:lens_overview}
    Overview of the TDCOSMO-2025 sample of time-delay lenses (strongly lensed quasars).
    }
    \centering
    \resizebox{\textwidth}{!}{%
    \begin{tabular}{llllll}
        \hline \hline
        Lens & Redshifts and Discovery & Kinematics & Time Delays & Lens Modeling & Environment \\
        \hline
        \DESzerofour  & \cite{Lin2017} & This work (MUSE, aperture) & \cite{Courbin2018} & \cite{Shajib2019} & \cite{Buckley-Geer2020} \\
        \hline
        \HEzero       & \makecell{\cite{Wisotzki2002, Morgan2005};\\\cite{Eigenbrod2006, Sluse2012}} & This work (NIRSpec, aperture) & \cite{Bonvin2017} & \makecell[l]{\cite{Wong2017};\\\cite{Chen2019}} & \makecell[l]{\cite{Sluse2017};\\\cite{Rusu2017}}\\
        \hline
        \PGeleven     & \makecell[l]{\cite{Weymann1980}; \\ \cite{Tonry1998}} & This work (NIRSpec, aperture) & \cite{Bonvin2018} & \cite{Chen2019} & \cite{Chen2019} \\
        \hline
        \RXJ          & \makecell[l]{\cite{SLuse2003};\\\cite{Sluse2007}} & \makecell[l]{This work (NIRSpec, resolved),\\ \citet[][KCWI, resolved]{Shajib23}} & \cite{Tewes2013} & \makecell[l]{\cite{Suyu2014};\\\cite{Chen2019}} & \cite{Suyu2013}\\
        \hline
        \Jtwelve      & \makecell[l]{\cite{Oguri2005}; \\ \cite{Agnello2016}}& This work (NIRSpec, aperture) & \makecell[l]{\cite{Eulaers2013};\\\cite{Birrer2019}} & \cite{Birrer2019} & \cite{Birrer2019} \\
        \hline
        \Bsixteen     & \makecell[l]{\cite{Myers1995};\\\cite{Fassnacht1996}} & This work (NIRSpec, aperture) & \cite{Fassnacht2002} & \makecell[l]{\cite{Suyu2009};\\\cite{Jee2019}} & \cite{Suyu2010}\\
        \hline
        \WFItwenty    & \makecell[l]{\cite{Morgan2004}; \\ \cite{Sluse2019, Sluse2012}}& This work (NIRSpec, aperture) & \cite{Bonvin2019} & \cite{Rusu2020} & \cite{Sluse2019} \\
        \hline
        \WGDtwenty    & \makecell[l]{\cite{Agnello2018}; \\ \cite{Buckley-Geer2020}} & This work (MUSE, aperture) & \citet{Wong24} & \cite{Shajib2022} & \cite{Buckley-Geer2020} \\
        \hline
    \end{tabular}
    }
    \tablefoot{The table lists the references where the different data products can be found.}
\end{table*}

\subsection{Samples of non-time-delay lenses}

In this section, we describe the external galaxy-galaxy lens samples with imaging and kinematics data from which we extract information about the mass profiles of the deflector galaxies. We describe the lenses in the SLACS sample in Section~\ref{sec:slacs_lensmodels} and those in the SL2S sample in Section~\ref{sec:sl2s_lensmodels}.

\subsubsection{SLACS} \label{sec:slacs_lensmodels}

The original SLACS sample was selected based on SDSS spectra with multiple redshifts, consistent with a background emission line galaxy being lensed by a foreground massive galaxy, and confirmed by imaging with the Hubble Space Telescope \citep[HST;][]{Bolton06,Bolton08}. For our analysis, we selected lenses from the subset of \nslacsimaging\ lenses that were uniformly modeled by \citet{Tan24}. These lens models adopted the same power-law plus external shear mass model similar to the TDCOSMO-2025 sample, thus putting the two samples on equal footing in terms of the lens models. The uniform modeling of the SLACS lenses was performed with the automated pipeline \textsc{dolphin} \citep{Shajib25}, which uses \textsc{lenstronomy} as the modeling engine \citep{Birrer2016, Birrer2018}.

\citet{Tan24} first selected 50 systems out of the full SLACS sample of 85. This subset was chosen by manually examining the lens images and eliminating systems where: (i) there are nearby satellites or galaxies along the LoS, (ii) the source morphology is extremely complex that would require extensive computational resources to fully capture in the model description, (iii) the deflector is a highly flattened disky galaxy, or (iv) the system lacks archival HST imaging data in the visible band (F555W or F606W). Of these initial 50 systems, the homogeneous and automated modeling performed with \textsc{dolphin} produced successful models for the \nslacsimaging\ systems that passed this quality assurance procedure. The effective or half-light radius for each of these systems was measured from large cutouts that covered the full extent of the galaxy light above the background level after subtracting the best-fit model-predicted arcs from it.
Further selection cuts to match the TDCOSMO-2025 sample and to ensure cosmography grade quality will be discussed in Section~\ref{sec:sampleselection}.

\subsubsection{SL2S} \label{sec:sl2s_lensmodels}

\newcommand{\nsltstotal}{34}
\newcommand{\nsltsdinosi}{13}
\newcommand{\nsltsdinosii}{21}

The SL2S sample was selected on the basis of ground-based images showing a red galaxy surrounded by a blue ring \citep{Gavazzi12}, and confirmed as lenses via space-based HST imaging and ground-based spectroscopy \citep{Ruff11,Sonnenfeld13a,Sonnenfeld13b}. 

The original papers did not fit power-law mass density profiles to the SL2S data, which is needed in our analysis. Therefore, we adopted a set of total \nsltstotal\ lenses for which power-law models are available, including \nsltsdinosi\ from \citet{Tan24} and \nsltsdinosii\ from \citet{Sheu25}. \citet{Tan24} initially modeled 31 systems from the SL2S sample based on the availability of HST imaging, single-aperture velocity dispersion measurement, and redshifts for both the deflector and the source. Out of these 31 attempted systems, the automated modeling with \textsc{dolphin} described in Section~\ref{sec:slacs_lensmodels} produced models that met the quality criteria for 24 systems. However, some of these HST images have low SNR as a result of being short-exposure images obtained from a Snapshot program. To improve the quality of the lens models, we obtained longer-exposure and higher resolution HST images for \nsltsdinosii\ systems (HST-GO-17130, PI: T.~Treu) in the F475X filter, which \citet{Sheu25} modeled using \textsc{lenstronomy}.  Of these \nsltsdinosii\ newly modeled systems, 11 of these systems overlap with the \citet{Tan24} set of 24 lenses.  Therefore, we utilized \nsltsdinosi\ systems modeled by \citet{Tan24}, and \nsltsdinosii\ systems modeled by \citet{Sheu25}.
Further selection cuts to match the TDCOSMO-2025 sample and to ensure cosmography grade quality will be discussed in Section~\ref{sec:sampleselection}.

\section{New stellar kinematics measurements}
\label{sec:newmeas}

In this section, we briefly describe the kinematic measurements that are included in this work, focusing on those that have changed with respect to TDCOSMO-4. Some of the measurements are presented here for the first time, while the rest were previously introduced in previous studies. For those that are discussed elsewhere, we provide only a brief summary and refer the reader to the original papers.

To achieve precise and accurate stellar velocity dispersions for our sample, we undertook a significant effort to refine our measurement techniques and obtain higher-quality data. From the point of view of measurement techniques, \citet{Knabel25} show that for data of sufficient SNR and spectral resolution, sub-percent precision and accuracy can be achieved using appropriately clean libraries of stellar templates and state-of-the-art methods. All of the kinematic measurements used in this analysis are based on this new refined methodology. One important outcome of the \citet{Knabel25} methodology is the estimation of systematic errors and covariance between individual objects and within spatial bins of kinematic maps, arising from stellar template libraries. We included those in our analysis for each dataset. We neglected the covariance between spectroscopic observations obtained with independent instrumental setups, as those are shown to be negligible for our purposes \citep{Mozumdar2025}.

We stress that the stellar velocity dispersions measurements presented in this section are vastly superior to previous published values in terms of data quality and analysis methods. The previously published values should be considered superseded, and we recommend that they not be used for cosmography. They are fine for applications that do not require the same level of precision and accuracy, such as galaxy formation and evolution studies \citep{Shajib21,Tan24,Sheu25}.

\subsection{Stellar kinematics of TDCOSMO-2025 systems}

In this section, we describe the kinematic measurements of the TDCOSMO-2025 systems. All of these measurements were performed using the \textsc{squirrel}\footnote{\url{https://github.com/ajshajib/squirrel}} pipeline \citep{Shajib25b} built on the penalized pixel-fitting (\textsc{pPXF}) software program \citep{Cappellari04, cappellari17, cappellari22_ppxf}. The \textsc{squirrel} pipeline streamlines the process of performing kinematic measurements with multiple stellar template libraries and other model setting combinations to extract the associated systematic uncertainties and covariances, following the methodology of \citet{Knabel25}.

\subsubsection{JWST NIRSpec} 
\label{sec:JWST-Nirspec}

JWST NIRSpec \citep{NIRSpec22} data are available for six out of the eight TDCOSMO-2025 systems. The data were obtained as part of three programs.  \RXJ was observed in Cycle 1 through program JWST-GO-1794 (PI: S.~H.~Suyu; Co-PIs: A.~Y{\i}ld{\i}r{\i}m, T.~Treu). \PGeleven, \HEzero, and \WFItwenty were observed through program JWST-GTO-1198 (PI: M.~Stiavelli). \Bsixteen and \Jtwelve were observed in Cycle 2 through program JWST-GO-2974 (PI: T.~Treu; Co-PI: A.~J.~Shajib). The IFU on NIRSpec was used in all cases centered on the deflector galaxy. The grating/filter setup G140M/F100LP was chosen to include the near-infrared Calcium triplet (hereafter, CaT) at the redshift of the deflector.

Achieving our target accuracy and precision required substantial effort to improve data reduction and calibration with respect to the standard pipeline's output. We used the custom data reduction pipeline \textsc{RegalJumper}\footnote{\url{https://github.com/ajshajib/regaljumper}} developed by \citet{Shajib25b} to reduce all the JWST data. \textsc{RegalJumper} extends the official JWST data reduction pipeline by including additional cleaning steps for artifacts, cosmic rays \citep[using \textsc{la-cosmic};][]{vanDokkum01}, and $1/f$ noise \citep[using \textsc{NSClean};][]{Rauscher24}. The version of the official JWST pipeline that we used is 1.17.1, with \textsc{stcal} version 1.11.1 and JWST Calibration References Data System context 1322.  Further steps were performed on the reduced datacube of \RXJ\ to correct for the oscillatory pattern (also referred to as "wiggles") introduced by undersampling \citep[e.g.,][]{Law23} using the software package \textsc{raccoon}\footnote{\url{https://github.com/ajshajib/raccoon}} \citep{Shajib25raccoon}. The wavelength-dependent line spread function (LSF) was obtained from robust fits to the emission lines of a planetary nebula \citep{Shajib25lsf}. The effective PSF for the datacube's spatial directions at the position of the CaT wavelength was directly estimated by \citet{Shajib25b} through lens modeling of a 2D image obtained by summing the datacube across a narrow wavelength range encompassing the CaT, while simultaneously reconstructing the PSF. We took the full width at half maximum (FWHM) of a circular Gaussian profile fitted to this reconstructed pixelated PSF, finding a FWHM of 0\farcs148. All these steps are described in detail by \citet{Shajib25b} in the case of \RXJ. 

\begin{figure*}
    \centering
    \includegraphics[width=\textwidth]{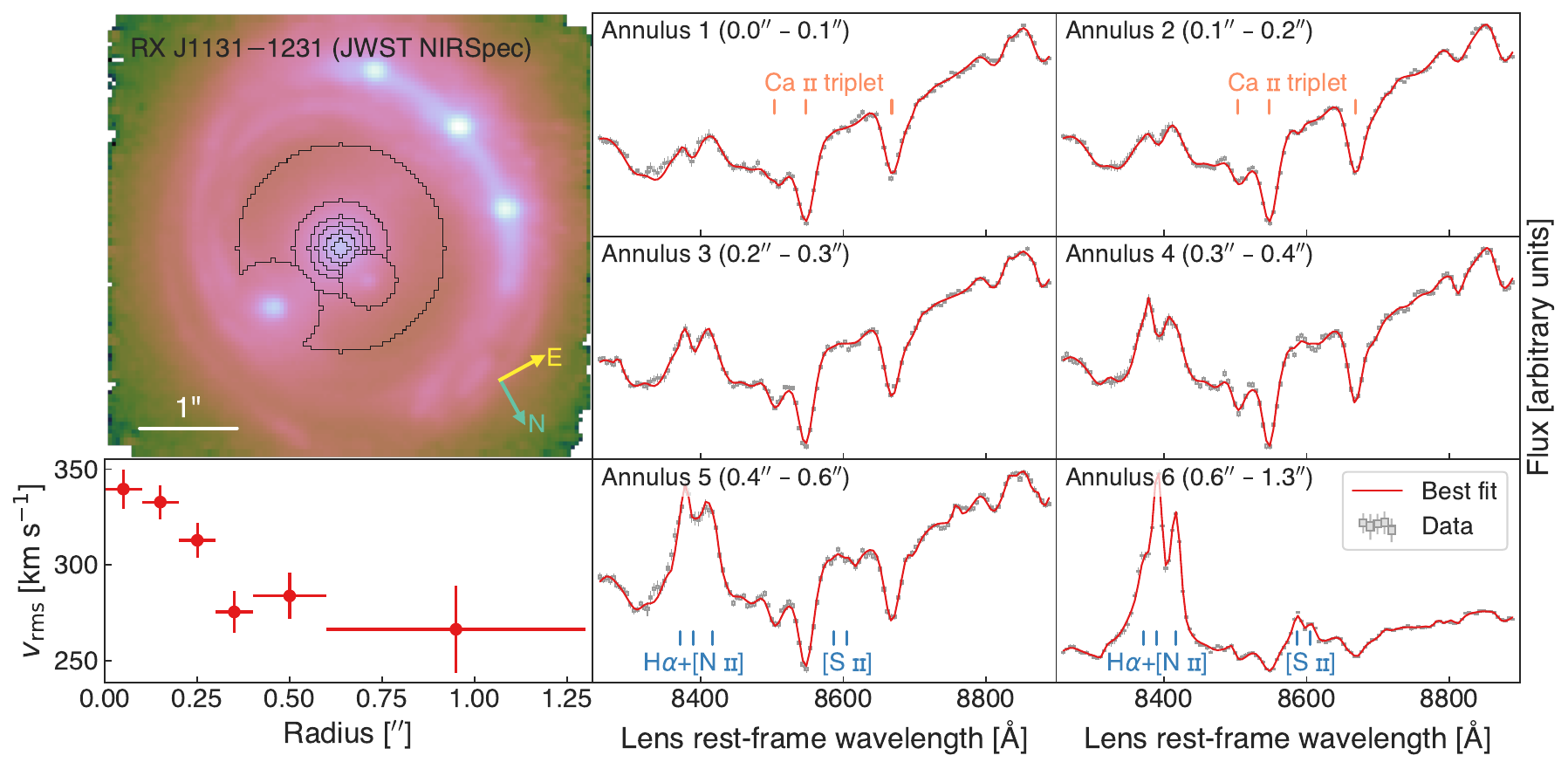}
    \caption{JWST-NIRSpec spectra and kinematic fits for \RXJ. \textit{Top left:} The six annuli (black contours), from which summed spectra are extracted, are illustrated on top of the NIRSpec white-light image. The regions around the satellite galaxy and the closest quasar image are excluded. The white bar represents 1$\arcsec$ scale, and the North and East directions are pointed with emerald and yellow arrows, respectively. \textit{Bottom left:} The measured $v_{\mathrm{rms}}$ in the six annuli. The horizontal error bars show the annulus widths, and the vertical error bars show the combined statistical and systematic uncertainty for each measurement. The measured values have 0.66\% covariance on average. \textit{Second and third columns:} The six panels show the integrated spectra in each annulus (gray boxes) and the kinematic fit (red line). The height of the gray box represents the nominal uncertainty levels estimated by the JWST pipeline, and the size of the vertical error bars represents the total boosted uncertainty levels to achieve $\chi^2_{\rm red} = 1$ for each fit. The vertical orange lines mark the wavelengths of the Calcium triplet lines from the lens galaxy, and the vertical blue lines mark the wavelengths of the H$\alpha$, [N \textsc{ii}], and [S \textsc{ii}] lines from the quasar host galaxy.}
    \label{fig:RXJspectra}
\end{figure*}

To obtain the 1D resolved kinematics of \RXJ\ used in this analysis, we took six annular bins centered on the lens galaxy with the outermost annulus extending to 1\farcs3 (Fig.~\ref{fig:RXJspectra}). As noticeable for the outer annuli, the H$\alpha$, [N \textsc{ii}], and [S \textsc{ii}] emission lines from the quasar host galaxy coincidentally fall close to the CaT lines of the lens galaxy. For that reason, following \citet{Shajib25b}, we simultaneously modeled these emission lines from the quasar host galaxy in our kinematic fitting with \textsc{pPXF}. We followed \citet{Shajib25b} in modeling spurious spike-like artifacts, which could have astrophysical origins, either from faint sources with undetected continuum at other redshifts along the LoS or the zodiac, or they may be artifacts in the data. However, our tests show that not directly modeling these spurious spikes would change the average velocity dispersion only within a sub-percent level. After an initial fit was performed for the spectra of a given annulus with the model settings described above, we boosted the noise levels to achieve $\chi^2_{\rm red} = 1$ before fitting once again, leading to the uncertainty on the measured kinematics being inflated accordingly. The median SNR of the noise-boosted spectra in the annuli is $\gtrsim$90 \AA$^{-1}$. We produced the covariance matrix of the measured values that includes modeling systematics by marginalizing over the stellar template library choice between Indo-US \citep{Valdes2004} and X-shooter Spectral Library \citep[XSL, DR3;][]{Verro2022a} following the methodology presented by \citet{Knabel25},\footnote{We did not use the MILES library for the JWST spectra since the original MILES library does not cover wavelengths up to the CaT, and the CaT library from \citet{Cenarro01}, which contains a subset of the MILES stars, was not "cleaned" by \citet{Knabel25} to incorporate it within their methodology.} detection thresholds for spurious spikes, combination of additive and multiplicative polynomials, and the fitted wavelength range. We used the "cleaned" versions of the libraries provided by \citet{Knabel25}. The marginalization across model choice combinations was done using Bayesian Information Criterion (BIC) weighting and Bessel correction. The average magnitude of the off-diagonal terms in the covariance matrix is 0.66\%. For the spherical Jeans modeling performed in this analysis, we fit to the $v_{\rm rms}$ profile shown in  
Fig.~\ref{fig:RXJspectra} (lower left panel), where $v_{\rm rms}^2 \equiv v_{\rm los}^2+\sigma_{\rm los}^2$ with $v_{\rm los}$ being the LoS velocity after subtracting off the systemic velocity.

It is important to match the light profile used for the luminosity weighting done in the Jeans modeling to that of the kinematic tracer. For the light profile of the kinematic tracer in \RXJ's NIRSpec data, we adopted a "stitched" surface brightness profile. Within $R \leq 1\farcs3$ in this stitched profile, we took the surface brightness distribution of the lens galaxy from the double S\'ersic fit that was obtained from the lens modeling by \citet{Shajib25b} that was performed with a 2D image obtained from the NIRSpec datacube by summing within 8700--8800 $\rm \AA$ in the lens rest-frame. Outside of this radius (i.e., at $R > 1\farcs3$), we adopted the light profile obtained from the double S\'ersic fit performed by \citet{Shajib23} on a large cutout of the HST F814W imaging (with the pivot wavelength corresponding to 6208 $\rm \AA$ in the lens rest-frame) after subtracting off the quasar images and the lens-model-predicted arcs. We perform the stitching as the limited field of view of the NIRSpec datacube (roughly $5\arcsec\times5\arcsec$) does not provide a robust handle on the light profile that is much further out, which is still relevant in the integration of the 3D light profile performed along the LoS for the kinematic modeling. We choose $R = 1\farcs3$ as the stitching radius, as that is the outermost extent of the measured kinematics from the NIRSpec data.

For the remainder of the lenses with JWST spectroscopy, we present, for the first time, spatially integrated kinematics in this paper. Obtaining spatially resolved maps requires additional work to correct for resampling noise or wiggles \citep{Shajib25b}, which is beyond the scope of this paper. The maps will be presented in future work and will be used in the next TDCOSMO milestone paper. 

\begin{figure}
    \centering
    \includegraphics[width=\linewidth]{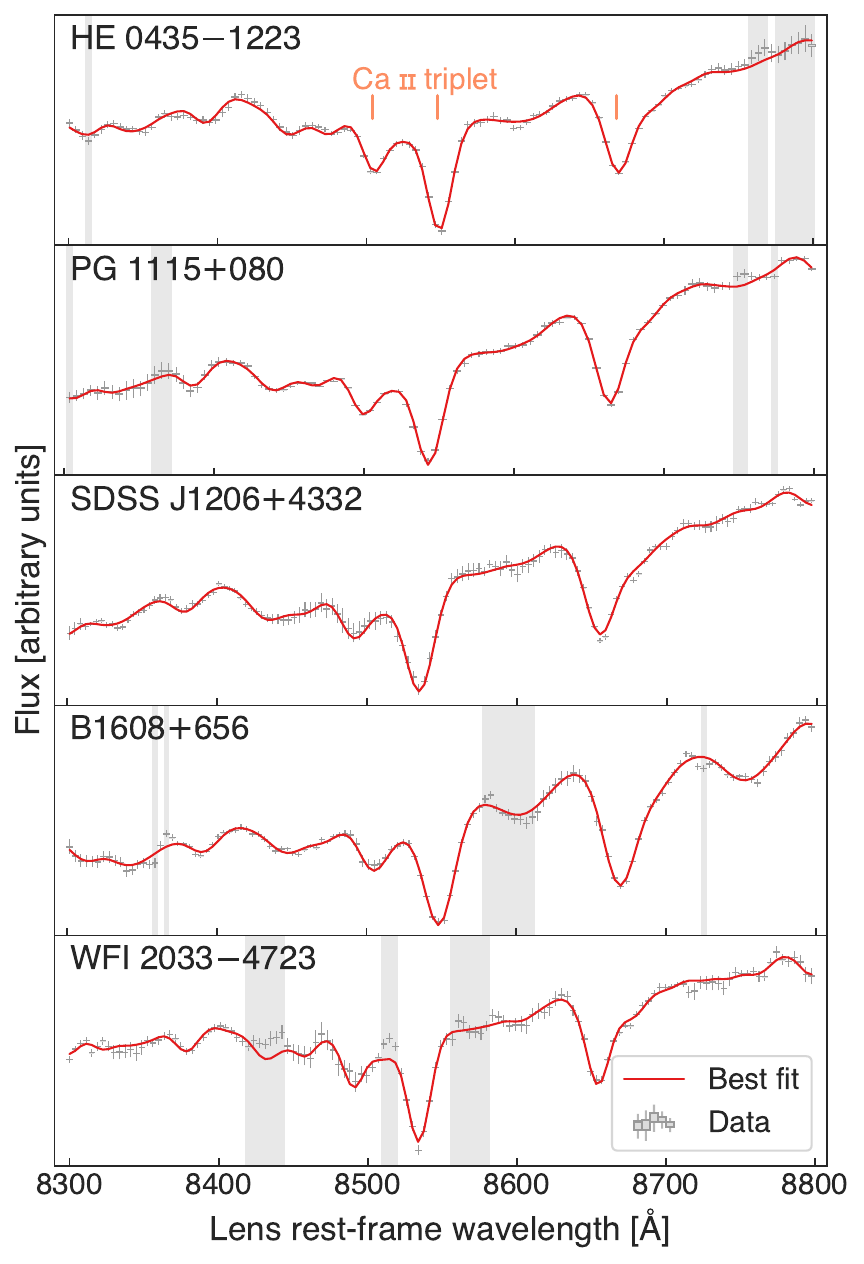}
    \caption{JWST-NIRSpec integrated spectra (gray bars) and kinematic fits (red lines) for five TDCOSMO-2025 lenses. The height of the gray boxes represents the nominal uncertainty levels estimated by the JWST pipeline, the size of the vertical error bars represents the total boosted uncertainty levels to achieve $\chi^2_{\rm red} = 1$, and the width of the boxes represents the wavelength-pixel size. The $x$-axis shows the lens rest-frame wavelength. The gray-shaded vertical bands were masked during the fits due to contamination by the lensed quasar features.}
    \label{fig:JWSTspectra}
\end{figure}

We obtained the PSF FWHM corresponding to the CaT's observed wavelength by extrapolating from the estimated FWHM value for \RXJ\ using a wavelength-dependent scaling extracted from simulated PSFs with {\sc stpsf} \citep{Perrin2014_stpsf}. Similar to the case of \RXJ, the LSF's FWHM was obtained at the observed wavelength of the CaT using the values provided by \citet{Shajib25lsf}. The aperture-integrated spectra and fits are shown in Fig.~\ref{fig:JWSTspectra}. Table~\ref{tab:kinematics} lists these integrated measurements, along with the statistical and systematic uncertainties after marginalizing over choices of stellar template libraries and polynomial degrees following the methodology of \citet{Knabel25}, with BIC weighting and Bessel correction applied. To estimate the noise for each integrated spectrum, we take the nominal uncertainty levels estimated by the JWST pipeline and perform an initial fit for each of 30 combinations of stellar template libraries and polynomial choices. For each of these fits, we take a noise boosting factor required to achieve $\chi^2_{\rm red}=1$. We average over these factors for each object's spectrum and rescale the noise according to that value. The median SNR for the noise-boosted spectra of each of the five lens galaxies ranges within $\sim$34--59 \AA$^{-1}$. This inflated noise is shown in Fig.~\ref{fig:JWSTspectra} and used for all subsequent spectral fittings. All spectra were fitted with \textsc{pPXF} using "cleaned" Indo-US and XSL stellar template libraries in the wavelength ranges 8300--8800 $\rm \AA$. Systematic uncertainties due to stellar template library selection are on average 0.77\% when isolated from the effect of the continuum polynomial degrees and wavelength range, as shown by \citet{Knabel25}. We tested additive (multiplicative) continuum polynomial degrees of 2, 3, 4, 5, and 6 (0, 1, and 2), and for each individual object, we selected the best polynomial degrees as the combinations that minimize the difference between the stellar libraries. Only one object, \HEzero, prefers a multiplicative polynomial degree greater than zero, which is clearly needed by the fit to account for a continuum slope issue likely due to imperfect flux calibration. The effect of varying the additive polynomial degree for a fixed template library, wavelength range, and multiplicative polynomial degree is on average 0.55\% for the sample. We also tested the choice of fitted wavelength range by shifting it by 25 $\rm \AA$ in either direction from the baseline range (8300--8800 $\rm \AA$), resulting in an average 0.38\% systematic effect. Since the change in wavelength range involves a change in the noise realization, and the effect is well below the statistical uncertainties, we did not include the wavelength range as an additional systematic term. To estimate the systematic uncertainties associated with the choice of template library and polynomial degrees following \citet{Knabel25}, we adopted the options of using BIC weights and Bessel correction. The combined systematic uncertainty from the template libraries and polynomial degrees for the sample as a whole is therefore at a level of 0.95\% after adding them in quadrature. The total systematic uncertainty was then added in quadrature from these contributions, which is the value reported in Table~\ref{tab:kinematics} for each lens. The covariance of stellar velocity dispersion between galaxies is at the level of 0.4\%, which we considered negligible given the levels of statistical and systematic variance.

\subsubsection{Reanalysis of Keck-KCWI data for \RXJ}

\begin{figure*}
    \centering
    \includegraphics[width=\linewidth]{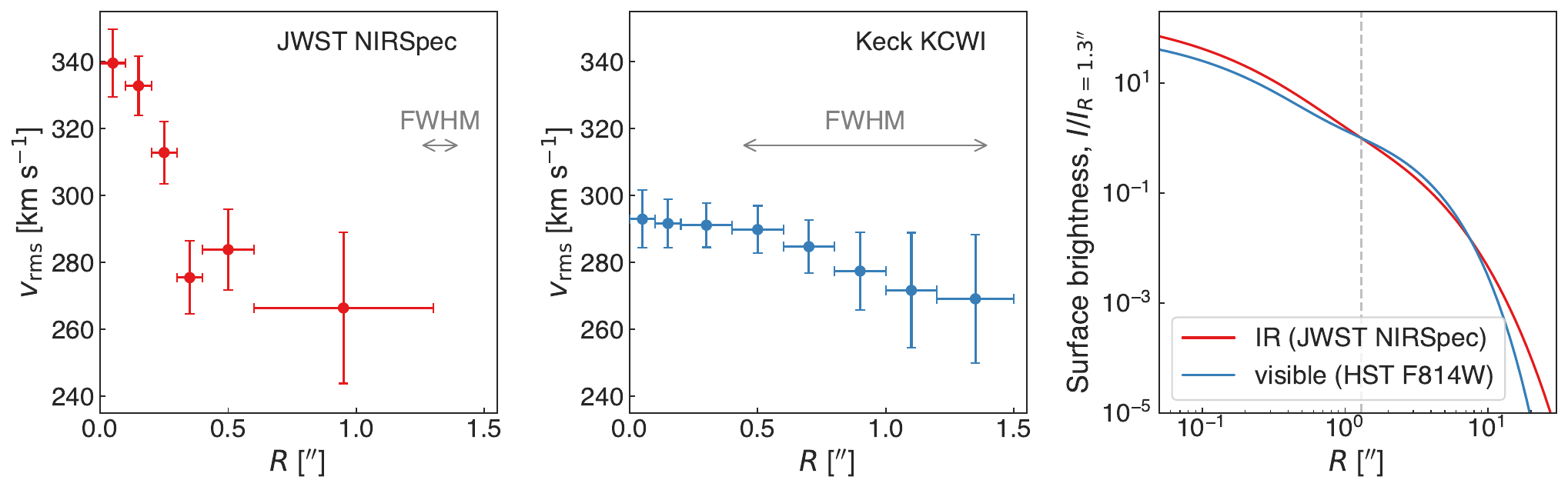}
    \caption{Measured values of $v_{\rm rms}$ for \RXJ in radial annuli from the JWST NIRSpec (left-hand panel, the same ones shown in Fig.~\ref{fig:RXJspectra}) and Keck KCWI (middle panel). Both of these measurements marginalize over the same choices of template libraries, namely the Indo-US and the XSL DR3, in addition to separate choice combinations for polynomial orders and fitted wavelength range. The arrows in these panels illustrate the size of the PSF FWHM for each case ($0\farcs15$ for JWST-NIRSpec and $0\farcs96$ for Keck-KCWI): the higher resolution of the JWST-NIRSpec data enabled the identification of the sharp rise of the velocity dispersion profile in the center. The right-hand panel illustrates the 1D surface brightness profile in optical (blue line) and in the IR (red line). The optical light profile was extracted from double S\'ersic fitting by \citet{Shajib23} from a large cutout that encapsulates the full extent of the galaxy in the HST F814W imaging (which corresponds to a pivot wavelength of 6208 $\rm \AA$ in the lens rest-frame). The IR light profile was extracted from the double S\'ersic light profile fitted as part of the lens modeling done with a 2D image obtained from the NIRSpec datacube by summing within the wavelength range 8700--8800 $\rm \AA$ in the lens rest-frame. The surface brightness profiles are normalized by the corresponding amplitudes at $R=1\farcs3$ (vertical dashed line). For kinematic modeling of the NIRSpec kinematics, we use a stitched light profile that transitions from the IR light profile at $R \leq 1\farcs3$ to the optical light profile shape at $R > 1\farcs3$.}
    \label{fig:KCWIspectra}
\end{figure*}

The Keck-KCWI IFU data for \RXJ were presented by \citet{Shajib23}. Here, we updated the measurement using the new methodology of \citet{Knabel25}. \citet{Shajib23} used only one template library, XSL DR2, in their measurement of kinematics. In this paper, we used XSL DR3 and Indo-US libraries to account for the systematic uncertainty stemming from the template library choice \citep{Knabel25}. We also accounted for systematics from other model setting choices adopted by \citet{Shajib23}, namely the wavelength range fitted and the combination of polynomial orders, to produce a covariance matrix that encapsulates all the above sources of systematics. Furthermore, we identified a small systematic bias in \citet{Shajib23}, where the XSL DR2 template library was read out in a manner that assumed uniform wavelength sampling across the library, whereas in reality, the wavelength sampling varies slightly across the templates. Thus, ignoring the slight variations in the wavelength sampling led to a slight underestimation of the stellar velocity dispersion. Overall, the reanalysis of the KCWI performed here resulted in a 1.3\% increase in velocity dispersion on average (weighted by the inverse variance of the radial annuli; individual changes ranged from 0.74\% to 2.5\% from the inner annulus to the outer). In Fig.~\ref{fig:KCWIspectra}, we illustrate the updated $v_{\rm rms}$ measurements for \RXJ.

\subsubsection{VLT MUSE}

Two TDCOSMO-2025 lenses, namely \DESzerofour and \WGDtwenty, were observed with the Multi Unit Spectroscopic Explorer (MUSE) on the European Southern Observatory’s (ESO) Very Large Telescope (VLT). The observations were carried out as part of two ESO programs: 0102.A-0600(E) (PI: A.~Agnello) and 105.205V.002 (PI: C.~Ducourant). They consist of 7 $\times$ 900s exposures for \WGDtwenty and 20 $\times$ 675s exposures for \DESzerofour. Observations were performed in wide-field mode with adaptive-optics correction, covering the wavelength ranges 4700--5803 $\rm \AA$ and 5966--9350 $\rm \AA$. Details of the data reduction can be found in \citet{Buckley-Geer2020} for \DESzerofour. The reduction of \WGDtwenty follows \cite{Sluse2019}, with improvement on sky subtraction and frame combination as prescribed by \citet{Bacon2023}.

Similarly to \sref{sec:JWST-Nirspec}, we focused on obtaining aperture-integrated spectra of the deflector, while the analysis of resolved kinematic maps is left for future work. We summed the flux within circular apertures of 1\farcs5 and 1\farcs0 diameters, centered on the lensing galaxy, for \WGDtwenty and \DESzerofour, respectively. The median SNR of the integrated spectra is $\sim$38 \AA$^{-1}$ for \WGDtwenty and $\sim$33 \AA$^{-1}$  for \DESzerofour, sufficient to ensure accurate velocity dispersion measurements \citep{Knabel25}.
For each object, we removed the quasar components before measuring the velocity dispersion. For \WGDtwenty, the 2D model of the system was constituted of 4 Moffat components for the quasar images and of a PSF-convolved de Vaucouleurs model for the lensing galaxy \citep[see][for details]{Sluse2019}. For \DESzerofour, we used {\textsc{lenstronomy}} iteratively on each slice of the data cube to model the lensed image. During this process, we froze the mass model to the best model used for the cosmographic analysis in \citet{Shajib2019}, while allowing the light profiles of the lens, source, and point-like images to vary for each wavelength. 

We measured the aperture velocity dispersions with \textsc{pPXF}, following the methodology introduced by \citet{Knabel25} for estimating statistical and systematic uncertainties. We marginalized over the choice of template libraries by combining measurements using Indo-US, XSL, and MILES \citep{Valdes2004,SanchezBlazquez2006,Falcon-Barroso11,Verro2022a} stellar template libraries. The fitting was restricted to the region around the Ca H \& K absorption lines — the spectral range used spans 3820--4200 $\rm \AA$ in the rest frame of the lens galaxy. The LSF of the MUSE instrument over this spectral range was estimated from the empirical relation described by \citet{Bacon2023}. The results of these measurements, including their statistical and systematic uncertainties, are summarized in \tref{tab:kinematics}. Our measurement for \DESzerofour represents a methodological improvement over the previous analysis by \citet{Buckley-Geer2020} and is performed within a slightly different aperture.

\begin{table}[h!]
\centering
\renewcommand{\arraystretch}{1.3}
\caption{Unresolved stellar velocity dispersion measurements based on JWST-NIRSpec and VLT-MUSE spectra.}
\resizebox{\columnwidth}{!}{%
\begin{tabular}{lcccc}
\hline  \hline
Lens & Instrument & Aperture & $\sigma_{\rm los}$ [km s$^{-1}$] & FWHM [\arcsec] \\
\hline
\DESzerofour & MUSE & 1\farcs0 diameter & 242 $\pm$ 10 $\pm$ 7 & 0.52 \\
\HEzero & NIRSpec & 0\farcs55 $\times$ 0\farcs55  & 227 $ \pm$ 6 $\pm$ 2 & 0.15 \\
\PGeleven & NIRSpec & 0\farcs55 $\times$ 0\farcs55  & 236 $ \pm$ 7 $\pm$ 1 & 0.15 \\
\Jtwelve & NIRSpec & 0\farcs55 $\times$ 0\farcs55  & 291 $ \pm$ 9 $\pm$ 4 &  0.15 \\
\Bsixteen & NIRSpec & 0\farcs55 $\times$ 0\farcs55  & 306 $ \pm$ 9 $\pm$ 7 &  0.15 \\
\WFItwenty & NIRSpec & 0\farcs55 $\times$ 0\farcs55  & 211 $ \pm$ 11 $\pm$ 2 &  0.15 \\
\WGDtwenty & MUSE & 1\farcs5 diameter & 255 $\pm$ 16 $\pm$ 2 & 1.26 \\
\hline
\end{tabular}}

\tablefoot{The aperture column lists the sizes of the square extraction aperture for the NIRSpec and the diameters of the circular aperture for the MUSE. The $\sigma_{\rm los}$ column reports the best-fit value, followed by the random and systematic uncertainty. The reported PSF FWHMs correspond to the observed wavelength of the Ca H\&K lines for MUSE and of the CaT lines for NIRSpec.}
\label{tab:kinematics}
\end{table}

\begin{figure*}
    \centering
    \includegraphics[width=\textwidth]{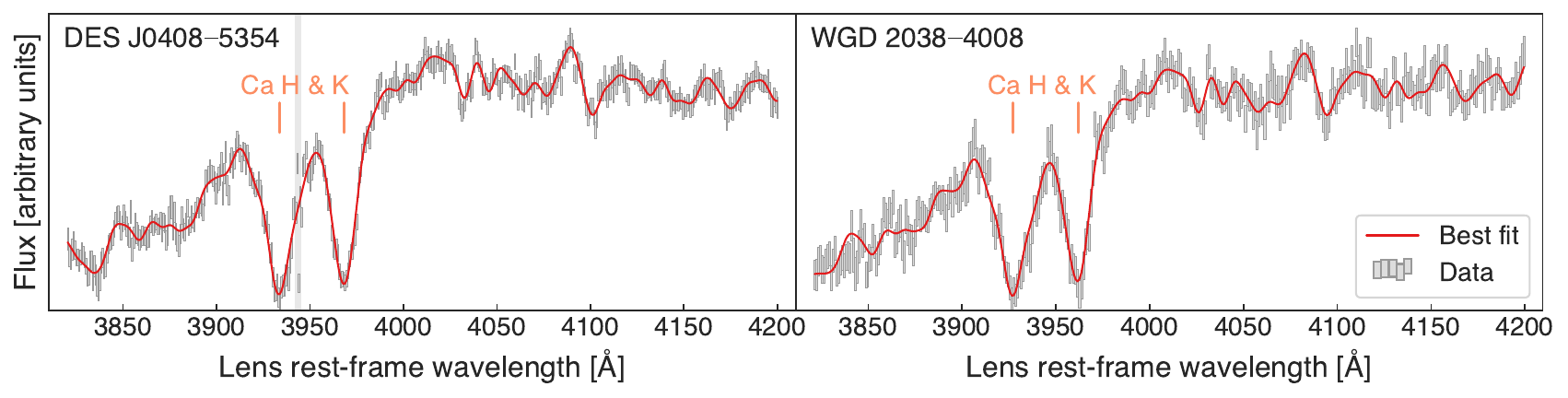}
    \caption{VLT-MUSE spectra and kinematics fits with \textsc{pPXF} for \DESzerofour\ (left-hand panel) and \WGDtwenty\ (right-hand panel). The gray rectangles illustrate the data with the height representing the 1$\sigma$ uncertainty and the width representing the wavelength-pixel size. The red line illustrates the best-fit model. The principal spectral features probing the kinematics are the Ca H \& K absorption lines marked with vertical orange lines. The gray-shaded region on the left-hand panel represents the wavelength range excluded from the fit.}
    \label{fig:muse_spectra}
\end{figure*}

\subsection{Stellar kinematics of non-time-delay lenses}

\subsubsection{SLACS} 

All of the \nslacsimaging\ SLACS systems have single-aperture velocity dispersions measured from the SDSS spectra. The fiber diameter for these measurements was 3$^{\prime\prime}$ and the average seeing was 1\farcs4, which were used in TDCOSMO-4. Recent analysis by \citet{Knabel24} shows that the SDSS spectra of SLACS lenses have SNR ($\sim9/$\AA) insufficient for cosmographic analysis, suffering from systematic errors at the level of 3.3\% \cite[see also][]{Bolton08} and covariance at the level of 2.3\%. Therefore, we did not use the SDSS spectra to obtain the stellar velocity dispersions in this analysis.

New kinematic data were obtained for \nslacsifu\ lenses using optical Keck Cosmic Web Imager (KCWI) integral-field spectroscopy on the W.~M.~Keck Observatory \citep{morrissey12_kcwi,morrissey18_kcwi}. The new data are vastly superior to any previous data obtained for this sample, both in terms of SNR (the integrated SNR of the spectra is 160 \AA$^{-1}$ on average over the sample) and spectral resolution. The data were fully described by \citet{Knabel24}. The effective parent sample for the SLACS external lens sample was thus reduced to these \nslacsifu\ systems. For this analysis, the 2D kinematic maps were rebinned in radial annuli. The radial annuli were determined in terms of fractions of the lens effective radius, as measured from S\'ersic lens light models from \citet{Sheu25}. To avoid mismatched centering of the datacube with respect to the HST photometry, we summed the datacube over the wavelength range in which we fit the kinematics (rest frame 3600--4500 $\rm \AA$) and fit the resultant 2D image with a S\'ersic profile, accounting for the average value of $0\farcs9$ arcsecond seeing (in FWHM) over the observations. The background source arcs are blended with the foreground deflector light due to the limited spatial resolution and PSF blurring. Therefore, for this S\'ersic profile fit, we focused on a 5$\times$5 spaxel grid centered around the brightest one, where the spaxel size is 0\farcs1457.

To rebin the 2D maps from the Voronoi bins \citep{Cappellari03} to the circular annuli, we took a luminosity-weighted average contribution over the Voronoi bins' spaxels that lie within the circular annuli. Given Voronoi bins $j$ that have some portion lying within annulus $k$, the weighted combined $v_{\rm{rms},k}$ is

\begin{equation}
    v_{\rm{rms},\it k}^2 = \frac{\sum_{j} I_{jk} v_{\rm{rms},\it j}^2}{\sum_j I_{jk}}
\end{equation}

\noindent where $I_{jk}$ is the surface brightness of the portion of Voronoi bin $j$ that contributes to annulus $k$. 

In general, each Voronoi bin contributes to multiple annuli. Any annulus that does not contain a luminosity-weighted Voronoi bin center is folded into the adjacent annulus, which in practice only affects the innermost bin with 0\farcs1 radius. Since the Voronoi bins were constructed from an asymmetric subset of spaxels from the datacube (all those spaxels whose SNR per $\AA$ > 1), the outer edges of the kinematic maps are not circular. Therefore, to calculate $v_{\rm{rms}}$ of the outermost annulus, we averaged over all of the spaxels beyond the inner edge of the annulus, as we did with the other annuli. Then, we assigned the outer edge of this annulus as the radius at which the area enclosed is equivalent to the total area covered by pixels that contribute to that annulus. From \citet{Knabel25}, we included systematic uncertainties of 0.67\% from template libraries and 0.60\% from continuum polynomial degrees to be added in quadrature to the statistical uncertainties, as well as a 0.47\% covariance term associated with the template library selection.
The profiles are shown in Fig.~\ref{fig:KCWIfits}.  

\subsubsection{SL2S} 

\citet{Mozumdar2025} reanalyzed all the available Keck and VLT spectra for SL2S galaxies using the methodology introduced by \citet{Knabel25}. They estimate the statistical and systematic errors and the covariance between sample galaxies. We refer the reader to \citet{Mozumdar2025} for a description of the data and measurements.

\section{External lens sample selection}
\label{sec:sampleselection}

Before we could include non-time-delay lenses (SL2S and SLACS) in the hierarchical inference to improve our constraints on the population-level mass profile of deflector galaxies -- and thus on cosmological parameters -- we needed to apply some selection criteria. First, they need to have all the relevant data. Thus, we applied a pre-selection based on the existence of kinematic measurements \citep{Knabel24, Mozumdar2025}, lens models \citep{Tan24, Sheu25}, and LoS measurements \citep{TDCOSMOIV, Wells2024}. We then have 12 lenses left in the SLACS sample and 21 lenses in the SL2S sample.  Second, we needed to ensure that the data quality of the non-time delay lenses is sufficient for cosmological applications. The original SL2S and SLACS data were more than adequate for galaxy formation and evolution studies, and cosmography was not an intended goal, in contrast to the data for the TDCOSMO-2025 lenses that were obtained with the specific goal of cosmography. As described in the previous section, some cosmography-grade data for the non-time-delay lenses were subsequently obtained; however, it is important to apply stringent quality cuts. Third, we needed to match the observable properties of the deflectors in the time-delay and non-time-delay lenses in order to minimize residual differences between the populations. 

\subsection{Lens morphology selection}

We applied a light-ellipticity cut of $q_\mathrm{light} > 0.7$, where $q_\mathrm{light}$ is the apparent axis ratio of the deflector's isophotes. This cut is applied to match the range of the $q_\mathrm{light}$ values of the TDCOSMO-2025 lenses, among which \WFItwenty has a minimum $q_\mathrm{light} = 0.73$. This cut also removes the more flattened galaxies, which have a higher possibility of being spiral galaxies, whose kinematics are not fully modeled by the spherical Jeans equations. This cut removes 1 out of 12 SLACS lenses and 11 out of 21 SL2S lenses. 

\subsection{Lensing information selection}

We also applied a quality cut based on the lensing information quantity $\mathcal{I}$ defined in Eqs.~(8--9) of \citet{Tan24}, which is a weighted SNR of the lensed arcs from the extended source, with the lens light subtracted.  The purpose of this cut was to ensure that the data contained enough information to constrain the power-law slope $\gamma$. There are two free dimensionless parameters $a$ and $b$ in the weights, and \citet{Sheu25} adopted values $a = 2$ and $b = 0.1$ that minimize the correlation between $\log \mathcal{I}$ and $\log \sigma_\gamma$ ($\sigma_\gamma$ is the statistical uncertainty of the power-law slope $\gamma$). In this work, we also used these values for all SLACS and SL2S lenses. We selected lenses with higher lensing information. For lenses with single-aperture spectroscopy, we applied a lower-end cut $\mathcal{I} > 150$. For lenses with KCWI spatially resolved spectroscopy, which directly constrains the mass density profile slope to better precision than lensing, we did not use the lensing-derived slope. Thus, we did not apply the lensing information cut. As a result, this cut removes 5 out of 21 SL2S lenses, with 2 already removed by the lens morphology selection. 

\subsection{Kinematics data quality selection}

To avoid potential biases in the stellar velocity dispersion and mitigate systematics, we applied a quality cut on the stellar kinematics of the SL2S lenses. We kept only those based on spectra with ${\rm SNR}>13\ \rm\AA^{-1}$, for which systematic errors are below 2\% \citep{Mozumdar2025}. This cut removes 5 out of 21 SL2S lenses, with 2 already removed by the lens morphology selection and 1 already removed by the lensing information selection. In practice, the four final selected systems have spectra with SNR greater than 18 \AA$^{-1}$.

\subsection{Velocity dispersion selection}

Velocity dispersion is a quantity tightly correlated with mass and all other observables in elliptical galaxies \citep{Auger2010,cappellari16}. For galaxy-scale lenses, measured stellar velocity dispersion $\sigma_{\rm los}$ and the lensing velocity dispersion $\sigma_{\rm SIS}$ derived from the Einstein radius and cosmological parameter\footnote{For the calculation of $\sigma_\mathrm{SIS}$, we use $\Omega_{\rm m} = 0.3$ for a flat $\Lambda$CDM cosmology.} are the same within 7\% intrinsic scatter \citep{Treu2006,Treu10}. To match the SLACS and SL2S lenses with the TDCOSMO-2025 lenses, we selected subsamples based on  $\sigma_\mathrm{SIS} \in [150, 350]$ km/s, the range spanned by our time-delay lens sample. This cut does not remove any of the SLACS lenses and removes 1 out of 21 SL2S lenses. 

\subsection{Environment selection}
 
We applied selection cuts to the local environment of the deflector galaxies, as we require their lensing effects to be described by the composition of a power-law lens mass model and an external, uncorrelated LoS contribution. Therefore, we specifically excluded lenses in an overdense region or those with a perturber nearby. 

The local over(under)-density is characterized by the relative over-density quantity $\zeta_{1/R}$. For the SLACS sample, \citet{TDCOSMOIV} used the $r$-band data in the DESI Legacy Imaging Surveys \citep[][]{Dey2019} to count objects with $18 < r < 23$ near the lens galaxies within 3\arcsec\ to 120\arcsec\ annuli. The perturbative effect of these nearby objects is quantified by the number $N_{1/R}$ defined as \citep{Greene2013}
\begin{equation}
    N_{1/R} = \sum_{i, R < 2\arcmin} \frac{1}{R_i},
\end{equation}
where $R_i$ is the inverse projected distance from the nearby $i$-th object to the lens galaxy, and the summation was performed within the 3\arcsec\ to 120\arcsec\ annulus. To compare with random fields, $N_{1/R}$ was also calculated for 10,000 random points in the Legacy Imaging Surveys footprint. The relative over-density $\zeta_{1/R}$ is 
\begin{equation}\label{eqn:rel_overdensity_slacs}
    \zeta_{1/R} = \frac{N_{1/R}}{\langle N_{1/R} \rangle _\mathrm{rand}},
\end{equation}
where the denominator is the median of the $N_{1/R}$ of the random LoSs used for calibration. We refer the reader to Section 6.3.3 of \citet{TDCOSMOIV} for more details on the LoS characteristics of the SLACS sample. 

For the SL2S sample, we utilized the LoS measurement of the LoS summary statistics from the CFHTLS, using a magnitude cut of $i < 24$ in the $i$-band \citep{Wells2024}. The inverse distances $1/R$ of nearby objects are calculated within an aperture of 5\arcsec\ to 120\arcsec\ for both the SL2S lenses and the reference fields. The over-density $\zeta_{1/R}$ is calculated as 
\begin{equation}\label{eqn:rel_overdensity_sl2s}
    \zeta_{1/R} = \mathrm{median} \left[ \frac{N_\mathrm{lens}^{{R < 2\arcmin}} \times \mathrm{median} \left[ 1/{R}_\mathrm{lens} \right]} {N_\mathrm{ref}^{{R < 2\arcmin}} \times \mathrm{median} \left[ 1/{R}_\mathrm{ref} \right]} \right],
\end{equation}
where $N_\mathrm{lens}^{R < 2\arcmin}$ and $N_\mathrm{ref}^{R < 2\arcmin}$ denote the total number of objects within the 120\arcsec\ aperture under the magnitude cut for the lens and reference field points, and the median in the numerator and the denominator is taken over the aperture. Note that the definition of $\zeta_{1/R}$ is different for the SLACS and SL2S sample, as by visual inspection of $\zeta_{1/R}$ versus $\kappa_\mathrm{ext}$ we find that using the expression in Eq.(\ref{eqn:rel_overdensity_slacs}) for the SLACS sample leads to positive correlation between both variables as expected, while for the SL2S sample we do not see the correlation except when using the expression for $\zeta_{1/R}$ in Eq.~(\ref{eqn:rel_overdensity_sl2s}).

\begin{figure}
    \centering
    \includegraphics[width=\linewidth]{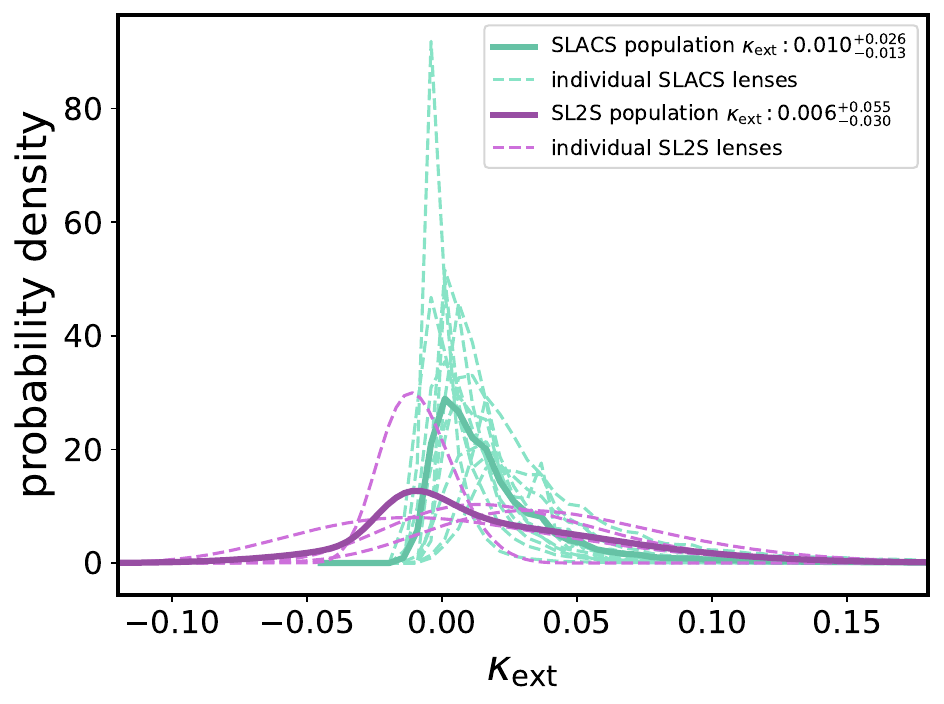}
    \caption{Probability distribution function of external convergence $\kappa_\mathrm{ext}$ for the quality samples of the SLACS and the SL2S lenses. Dashed lines are $\kappa_\mathrm{ext}$ of individual lenses, and solid lines are their combination. The median of $\kappa_\mathrm{ext}$ for both samples is shown in the figure legend.}
    \label{fig:kappa_ext_slacs_sl2s}
\end{figure}

For the SLACS sample, we applied an over-density cut $\zeta_{1/R} < 2.10$, which is the same cut as in \citet{TDCOSMOIV}. However, no lens was rejected by this selection cut. For the SL2S sample, we did not apply selection cut on $\zeta_{1/R}$ for two reasons: (i) all the values of the relative over-density $\zeta_{1/R}$ (defined as in Eq.~(\ref{eqn:rel_overdensity_sl2s})) for our SL2S sample \citep{Wells2024} do not exceed 2.10, and (ii) 
the selected SL2S sample lenses are not in overdense regions, as shown in Fig.~\ref{fig:kappa_ext_slacs_sl2s}. 
Fig.~\ref{fig:kappa_ext_slacs_sl2s} also shows the $\kappa_\mathrm{ext}$ distributions for the selected SLACS sample. The population distribution of $\kappa_\mathrm{ext}$ is consistent between our selected SLACS and SL2S samples, although this is not required since we marginalized over $\kappa_\mathrm{ext}$ for individual lenses in the inference.

\subsection{Selected SLACS and SL2S lenses}
\label{ssec:selected}

We present a summary of the selected SLACS and SL2S lenses used in our joint inference on cosmology and mass density profiles, along with our sample of TDCOSMO-2025 time-delay lenses. After the cuts, we end up with \nslacsifusel\ lenses from the SLACS sample and \nsltwossel\ lenses from the SL2S sample. We reiterate that the SLACS sample is significantly smaller than the one used by \citet{TDCOSMOIV}, but it is vastly superior in terms of quality and information content, having replaced the SDSS stellar velocity dispersion with Keck-KCWI spatially resolved measurements, and having updated the lens models \citep{Tan24}. 
Their properties are listed in Table~\ref{tab:quality_lens_property}. 

\begin{figure*}
    \centering
    \includegraphics[width=\linewidth]{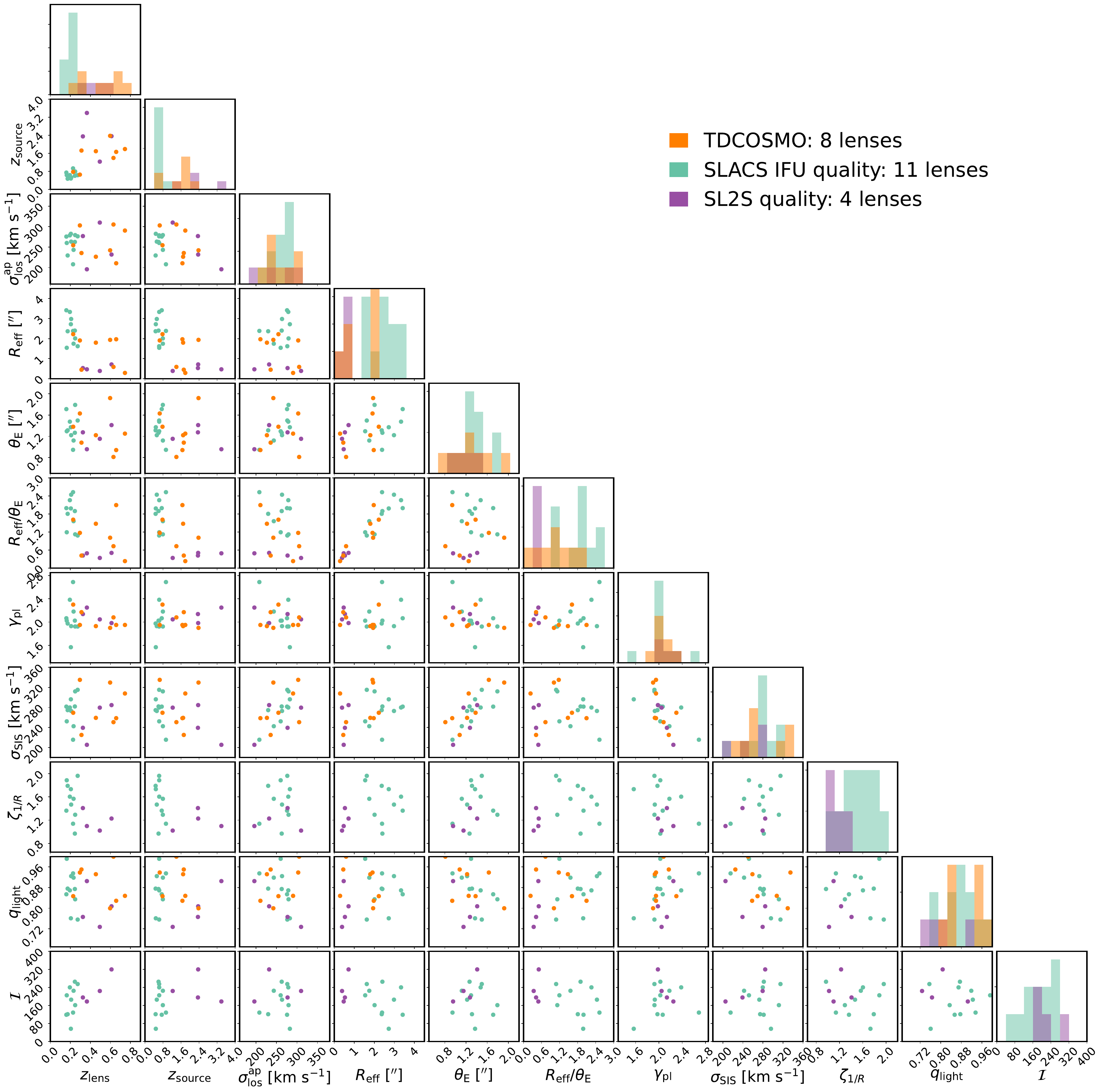}
    \caption{Properties of selected lens samples that enter the cosmology inference. 
    The sample selection is only based on (i) lens and source existing spectroscopic redshift measurements; (ii) lensing analysis based on imaging data; (iii) LoS analysis. From left to right, the quantities are: lens redshift $z_\mathrm{lens}$, background source redshift $z_\mathrm{source}$, stellar velocity dispersion of the deflector $\sigma^\mathrm{ap}_\mathrm{los}$, half-light radius of the deflector $R_\mathrm{eff}$, Einstein radius \thetaE, $R_\mathrm{eff}/\theta_\mathrm{E}$, lensing power-law slope $\gamma_\mathrm{pl}$, lensing stellar velocity dispersion $\sigma_\mathrm{SIS}$, relative over-density $\zeta_{1/R}$, apparent axis ratio of light profile of the deflector $q_\mathrm{light}$, and lensing information $\mathcal{I}$ calculated under the choice of scaling parameters from \citet{Sheu25}. We demonstrate for the selected lenses, there is no significant difference from sample to sample in the parameter spaces of this plot, except for the distribution in redshifts and the difference in effective radius expected from the evolution of the size-mass relation \citep{vd10}, see \sref{ssec:selected} for more information. }
    \label{fig:sample_selection}
\end{figure*}

The key physical quantities of the deflectors together with the source redshifts are also shown in Fig.~\ref{fig:sample_selection}. We compare these quantities of the SLACS and SL2S lenses with those of the TDCOSMO-2025 lenses. Even though the samples are small and shot noise prevents strong statistical conclusions, apparent differences from sample to sample include: (i) the distribution of lens and source redshifts between the SLACS and the SL2S sample, as a result of their selection processes \citep{Bolton06,Gavazzi12} -- the SL2S being better matched to the TDCOSMO-2025; (ii) the half-light radii $R_\mathrm{eff}$ are generally smaller for SL2S and TDCOSMO-2025 than for SLACS, consistent with the evolution of the size--mass relation \citep{vd10}. Except for these differences, the other properties of the SLACS and the SL2S lenses agree well with the TDCOSMO-2025. We note also that the SL2S, SLACS, and TDCOSMO-2025 lenses follow the same lens-mass fundamental plane, as recently shown by \citet{Mozumdar2025}, lending further support to the assumption that they are drawn from the same parent population. 

Next in Section~\ref{sec:hierarch}, we describe how we combined the quality-assured sample of non-time-delay lenses with the TDCOSMO-2025 lenses to obtain better constraints on the mass density slope and ultimately $H_0$.

\section{Hierarchical inference}
\label{sec:hierarch}

\subsection{General concept}
 
For the hierarchical analysis, we follow the method presented by \cite{TDCOSMOIV}. We provide a brief overview of the method here, highlighting the detailed parameterizations, assumptions, and priors that inform our inference. The reader is referred to \citet{TDCOSMOIV} for more details.

In Bayesian language, we want to calculate the probability of the cosmological parameters $\boldsymbol{\pi}$, given the strong lensing data set $p(\boldsymbol{\pi} | \{\mathcal{D}_{i} \}_{N})$, where $\mathcal{D}_i$ is the dataset of an individual lens (including imaging data, time-delay measurements, kinematic observations and the properties of the LoS galaxies) and $N$ is the total number of lenses in the sample.

In addition to $\boldsymbol{\pi}$, we denote all the model parameters -- including both those at the single lens level and those at the population level -- with $\boldsymbol{\xi}$. Using Bayes' rule and considering that the data of each individual lens $\mathcal{D}_{i}$ are independent, we can write:

\begin{multline} \label{eqn:full_inference}
    p(\boldsymbol{\pi} \mid \{\mathcal{D}_{i} \}_{N}) \propto \mathcal{L}(\{\mathcal{D}_{i} \}_{N}| \boldsymbol{\pi}) p(\boldsymbol{\pi})  = \int \mathcal{L}(\{\mathcal{D}_{i} \}_{N} \mid \boldsymbol{\pi}, \boldsymbol{\xi})p(\boldsymbol{\pi}, \boldsymbol{\xi}) d \boldsymbol{\xi} \\
    = \int \prod_i^N \mathcal{L}(\mathcal{D}_{i}| \boldsymbol{\pi}, \boldsymbol{\xi})p(\boldsymbol{\pi}, \boldsymbol{\xi}) d \boldsymbol{\xi}.
\end{multline}

In the following, we divide the nuisance parameter, $\boldsymbol{\xi}$, into a subset of parameters $\boldsymbol{\xi}_i$ that we constrained independently for each lens, and a set of parameters $\boldsymbol{\xi}_{\rm pop}$ that require to be sampled across the lens sample population globally. The parameters of each individual lens $\boldsymbol{\xi}_i$ include the lens model, source, and lens light surface brightness and any other relevant parameter of the model needed to predict the imaging data.
Hence, we can express the hierarchical inference (Eq.~\ref{eqn:full_inference}) as
\begin{multline} \label{eqn:full_inference_extended}
    p(\boldsymbol{\pi} \mid \{\mathcal{D}_{i} \}_{N})
     \propto \int \prod_i \left[ \mathcal{L}(\mathcal{D}_i \mid D_{\rm d, s, ds}(\boldsymbol{\pi}), \boldsymbol{\xi}_i,  \boldsymbol{\xi}_{\rm pop})
     p(\boldsymbol{\xi}_i) \right]
     \\ \times
     \frac{p(\boldsymbol{\pi}, \{\boldsymbol{\xi}_i \}_{N}, \boldsymbol{\xi}_{\rm pop})}{\prod_i p(\boldsymbol{\xi}_i)} d \boldsymbol{\xi}_{\{i\}} d \boldsymbol{\xi}_{\rm pop},
\end{multline}
where $\{\boldsymbol{\xi}_i \}_{N} = \{\boldsymbol{\xi}_1, \boldsymbol{\xi}_2, ..., \boldsymbol{\xi}_N \}$ is the set of parameters describing the individual lenses and $p( \boldsymbol{\xi}_i)$ are the interim priors on the model parameters in the inference of an individual lens. The cosmological parameters $\boldsymbol{\pi}$ are fully encompassed in the set of angular diameter distances $\{D_{\rm d}, D_{\rm s}, D_{\rm ds} \} \equiv D_{\rm d, s, ds}$, and thus, instead of stating $\boldsymbol{\pi}$ in Eq.~\ref{eqn:full_inference_extended}, we now state $D_{\rm d, s, ds}(\boldsymbol{\pi})$.
Up to this point, no approximation was applied to the full hierarchical expression (Eq.~\ref{eqn:full_inference}).

To infer $H_0$ from a set of lenses, some priors describing the population hyperparameters and their correlations need to be imposed.
Our choices and parameterization are presented in \sref{sec:deflector_population}.

\subsection{Deflector population model}
\label{sec:deflector_population}

In the following, we describe the model components and their hierarchical treatment for the deflector population. We describe the deflector mass density profiles in Section~\ref{sec:mass_density_profiles}, the stellar anisotropy and kinematics description in Section~\ref{sec:slacs_ifu_models}, and the LoS convergence in Section~\ref{sec:los_modeling}.

\subsubsection{Mass density profiles}\label{sec:mass_density_profiles}

For the mass density profile assumptions, we followed closely the approach of \citet{TDCOSMOIV}.
The deflectors in the TDCOSMO-2025 lenses are massive elliptical galaxies. These galaxies, observationally, follow a tight relation in a parameter space of luminosity, size, and velocity dispersion \citep[e.g.,][]{FaberJackson:1976, Auger2010, cappellari16, Bernardi:2020, Mozumdar2025}, exhibiting a high degree of self-similarity among the population.

For the deflector mass profile, to maximize the uncertainties allowed by the MSD, we chose not to break the MSD based on imaging data and the assumption of a mass profile as in \citet{Wong2019}. Instead, we used stellar kinematics to break the MSD and constrain the mass profile.
To do so, we chose as a baseline model a power-law elliptical mass distribution \citep[PEMD,][]{Kormann:1994, Barkana:1998}
\begin{equation} \label{eqn:pl_profile}
  \kappa(\theta_1, \theta_2) = \frac{3-\gamma_{\rm pl}}{2} \left[\frac{\theta_{\rm E}}{\sqrt{q_{\rm m}\theta_1^2 + \theta_2^2/q_{\rm m}}} \right]^{\gamma_{\rm pl}-1},
\end{equation}
where $\gamma_{\rm pl}$ is the logarithmic slope of the profile, $q_{\rm m}$ is the axis ratio of the minor and the major axes of the elliptical profile, and $\theta_{\rm E}$ is the Einstein radius. In the models, we added an additional external shear component.
This model is constrained on the lens-by-lens case based on high-resolution imaging data.
The PEMD lens profile inherently breaks the MSD, and the parameters of the PEMD profile can be precisely constrained (within a few per cent) by exquisite imaging data. In this work, we avoided describing the PEMD parameters at the population level, such as redshift, mass, or galaxy environment, and utilized the individual lens inference posterior products derived from the data based on uniform priors.
To allow for the full flexibility in the radial density profiles that can be produced by an MST, we added a global internal MST distribution specified at the population level.
We note that, physically, $\lambda_{\rm int}$ does not imply that galaxies contain an infinite sheet of mass. The MST works as long as the "sheet" tapers off sufficiently far from the images \citep{Blum2020, TDCOSMOIV}. This parameter should be viewed as a parametrization of the deviations from a power-law profile, which is particularly powerful because it is directly degenerate with $H_0$ and therefore allows us to capture the full uncertainty.

We note that the power-law slope $\gamma_{\rm pl}$ of the PEMD profile inferred from imaging data is a local quantity at the Einstein radius of the deflector. The Einstein radius is a geometrical quantity that depends on the mass of the deflector and the redshifts of the lens and the source. Thus, the physical location of the measured $\gamma_{\rm pl}$ from imaging data depends on the redshift configuration of the lens system. In a scenario where the mass profiles of massive elliptical galaxies deviate from an MST of a PEMD, resulting in a gradient in the measured slope $\gamma_{\rm pl}$ as a function of physical projected distance, a global joint MST correction on top of the individually inferred PEMD profiles may lead to inaccuracies.

To address such potential inaccuracies, we allowed for a radial trend in the applied MST relative to the local quantities inferred from imaging. This was achieved by parameterizing the global MST population with a linear relation in $R_{\rm eff}/\theta_{\rm E}$ as
\begin{equation}\label{eqn:lambda_scaling}
 \lambda_{\rm int}(R_{\rm eff}/ \theta_{\rm E}) = \lambda_{\rm int, 0} + \alpha_{\lambda}\left( \frac{R_{\rm eff}}{\theta_{\rm E}}-1 \right),
\end{equation}
where $\lambda_{\rm int, 0}$ is the global MST parameter when the Einstein radius is at the half-light radius of the deflector, i.e., $R_{\rm eff}/\theta_{\rm E} = 1$, and $\alpha_{\lambda}$ is the linear slope in the expected MST parameter as a function of $R_{\rm eff}/\theta_{\rm E}$ \citep{TDCOSMOIV}. In this form, we assumed self-similarity in the lenses with regard to their half-light radii.
In addition to the global MST normalization and trend parameterization, we added a Gaussian distributed intrinsic scatter with standard deviation $\sigma(\lambda_{\rm int})$ at fixed $R_{\rm eff}/\theta_{\rm E}$, to allow for a degree of non-uniformity between the deflectors.

For \RXJ and for the SLACS lenses, for which we have IFU kinematics data, we also sample individually the power-law slope $\gamma_{\rm pl}$ for each of these lenses. The IFU data is constraining enough to supersede the imaging-inferred $\gamma_{\rm pl}$ values for the SLACS lenses. For \RXJ we include the full lensing information and the covariance between inferred $D_{\Delta t}$ and $\gamma_{\rm pl}$. Details about the procedure and analysis of \RXJ can be found in Appendix~\ref{app:rxj}.

\begin{table*}[]
\caption{\label{tab:param_prior}
Summary of the parameters and priors for the baseline hierarchical model.}
\renewcommand{\arraystretch}{1.2}
\centering
\begin{tabular}{l l l}
    \hline \hline
    Parameters & Prior & Description \\
    \hline
    \textbf{Cosmology (flat $\Lambda$CDM)} \\
    $H_0$ [\ksmpc] & $\mathcal{U}\,(0, 150)$  & Hubble constant \\
    $\Omega_{\rm m}$ & $\mathcal{U}\,(0.05, 0.5)$ & Matter density \\
    \hline
    \textbf{Mass profile} \\
    $\lambda_{\rm int, 0}$ & $\mathcal{U}\,(0.5, 1.5)$ & Internal MST population mean \\
    $\alpha_{\lambda}$ & $\mathcal{U}\,(-1, 1)$  & Slope of $\lambda_{\rm int}$ with $R_{\rm eff}/\theta_{\rm E}$ of the deflector \\
    $\sigma(\lambda_{\rm int})$ & $\textrm{Loguniform}\,(0.001, 0.5)$ & 1$\sigma$ Gaussian scatter in the internal MST \\
    $\gamma_{\mathrm{pl}, i}$ & $\mathcal{U}\,(1.5, 2.5)$ & Individual power-law slope of the IFU lenses  \\
    $\gamma_{\mathrm{pl}, {\rm rxj}}$ & $p\,(\gamma_{\mathrm{pl}, {\rm rxj}})$ & Lens model-only prior on the power-law slope for \RXJ \\
    \hline
    \textbf{Stellar kinematics} \\
    $\langle \sigma_{\rm t}/\sigma_{\rm r} \rangle$ & $\mathcal{U}\,(0.87,1.12)$ & Mean anisotropy parameter for a spatially constant anisotropy \\
    $\sigma(\sigma_{\rm t}/\sigma_{\rm r})$ & $\textrm{Loguniform}\,(0.01, 1)$  & 1$\sigma$ Gaussian scatter in $\sigma_{\rm t}/\sigma_{\rm r} $\\
    \hline
    \textbf{LoS} \\
    $\kappa_{\rm ext}$ & $p\,(\kappa_{\rm ext})$ of individual lenses & External convergence of lenses \\
    \hline
\end{tabular}
\end{table*}

\subsubsection{Stellar anisotropy and kinematics} \label{sec:slacs_ifu_models}

Strong lensing achieves more constraining power when combined with dynamical mass measurements from kinematics \citep{Treu2002, Auger2010, shajib18, yildirim20, yildirim23}. However, LoS velocity dispersions from single-slit spectroscopy cannot constrain the anisotropy of stellar orbits; spatially-resolved kinematics are required \citep{cappellari08, barnabe09, shajib18, TDCOSMOIV, yildirim20, birrer_treu21, Shajib23} to break the degeneracy between the dynamical mass profile and the anisotropy, the mass--anisotropy degeneracy (MAD) \citep{Binney1982, Gerhard1993,Courteau14}.

After the early pioneering works \citep[e.g., TIGER,][]{bacon95_tiger}, at the beginning of the century, spatially resolved kinematics studies of elliptical galaxies started being carried out for statistically significant samples of nearby galaxies \citep[e.g., the SAURON survey:][]{bacon01_sauron, dezeeuw02, emsellem04, Emsellem07, cappellari06, cappellari07}. Following the success of these early experiments, integral-field spectroscopy (IFS) became the norm in most major observatories \citep[e.g., MUSE,][]{bacon10_muse}. The SAURON and following IFU surveys (ATLAS$^{\rm 3D}$, \citealt{Cappellari11}; CALIFA, \citealt{Sanchez2012}; SAMI, \citealt{bryant15}; SDSS MaNGA, \citealt{bundy15}) have developed consistent pictures of the kinematic properties of elliptical galaxies in the local Universe \citep{Cappellari2025}. At the same time, methods and systematic uncertainties in the extraction of stellar kinematics at the spaxel level using high-resolution stellar templates have become standardized and well understood \citep{Knabel25}, using, for example, {\sc pPXF}.

Dynamical models have been constructed with the spatially resolved kinematics of the aforementioned IFU surveys by solving 2-integral \citep{emsellem94} and 3-integral Jeans equations under assumptions of spherical or axial symmetry using the Jeans Anisotropic Modeling (JAM) method \citep{cappellari08,Cappellari2020} and more generally with Schwarzschild models of orbital superposition \citep{schwarzschild79, richstone_tremaine88, rix97, vandermarel98, cappellari06, 2004SchwarzchildAXi, 2021SchwarzchildAXi, 2021SchwarzschildTriaxial, 2024SchwarzschildTriOS}. Despite some intrinsic model degeneracies, these studies have shown that one can reasonably well constrain the orbital anisotropy of nearby elliptical galaxies using the 2D projection of stellar LoS velocity distributions (LoSVD) and the projected tracer population, i.e., the surface brightness profile \citep[see][for a full review]{Cappellari2025}. 

The velocity anisotropy at a 3D location in the galaxy can be parameterized by $\beta \equiv 1-\sigma^2_{\rm t}/\sigma^2_{\rm r}$ that describes the ratio of tangential to radial velocity dispersions, where $\beta=0$ corresponds to the isotropic case. Under the assumption of axisymmetry, the velocity ellipsoid (in velocity space) can be assumed to align with the symmetry axis and described in cylindrical coordinates so that $\beta_z = 1-\sigma^2_z/\sigma^2_R$. This was found to describe well the stellar kinematics of regularly rotating elliptical galaxies \citep[e.g.,][Section~3.4]{cappellari16}. The other extreme assumption for the alignment of the velocity ellipsoid is radial alignment from the center of the galaxy. In this case, the anisotropy can be described in spherical coordinates as $\beta_{r} = 1-\sigma^2_{\theta}/\sigma^2_{r}$. In the spherical limit, these models converge to spherical models. This is appropriate for the mildly triaxial family of "pressure-supported" non-regular rotators, which are always close to spherical. Spatially resolved kinematics of nearby elliptical galaxies have been shown to be well-fit by a single spatially-uniform nearly-isotropic anisotropy (slightly radial, i.e., $\beta > 0$) with the assumption of an oblate axisymmetric geometry. These models are generally preferred over models with radially variant anisotropy profiles (e.g., the OM profile). 
A general agreement has been demonstrated between axisymmetric and spherical models \citep{Cappellari2020, zhu23_dynpop1}, indicating the robustness of the methods. 

For our joint lensing and dynamical analysis, we used the posteriors of the lens mass model as priors and obtained the posteriors of the mass model by requiring the kinematic data to be well described by the updated model. As the dynamical model requires a full 3D description, we accounted for the projection effect introduced by the intrinsic shape of the deflector galaxies. Using a Singular Isothermal Ellipsoid (SIE) mass model and a Jaffe \citep{Jaffe1983} stellar light tracer profile, \citet{Huang:2025} show that the aperture integrated stellar velocity dispersion within $R_\mathrm{eff}$ modeled with spherical JAM $\sigma^\mathrm{ap}_\mathrm{sph}$ 
can be biased at the level of around $-4$\% to $+1$\% \citep[Fig.~8 in][]{Huang:2025}, depending on the shape and apparent projected axis ratio of the galaxy.  We corrected for this bias statistically by a probabilistic deprojection based on the observed projected axis ratio and a realistic proposed intrinsic shape distribution. Residual random uncertainties are 1\% for typical ellipticity values \citep{Huang:2025}. 
We adopted the "axisymmetric correction factor" $\sigma^\mathrm{ap}_\mathrm{axi}/\sigma^\mathrm{ap}_\mathrm{sph}$ proposed and computed in a forward-modeling fashion by \citet{Huang:2025}. There, the authors used an axisymmetric intrinsic shape distribution from \citet{Li2018_intrinsic_shape} and assumed random distribution of the inclination angle for the model predictions of $\sigma^\mathrm{ap}_\mathrm{axi}$ within $R_\mathrm{eff}$. By circularizing the projected mass and stellar tracer profile, they calculated the model prediction of $\sigma^\mathrm{ap}_\mathrm{sph}$, and constructed a distribution of $\sigma^\mathrm{ap}_\mathrm{axi}/\sigma^\mathrm{ap}_\mathrm{sph}$ as a function of the projected axis ratio. In this work, we further accounted for the radial dependence of $\sigma^\mathrm{ap}_\mathrm{axi}/\sigma^\mathrm{ap}_\mathrm{sph}$ for both aperture kinematics and IFU-type stellar velocity dispersion, which are integrated with shell bins. The final $\sigma^\mathrm{ap}_\mathrm{axi}/\sigma^\mathrm{ap}_\mathrm{sph}$ distribution we applied to the stellar velocity prediction $\sigma^\mathrm{ap}_\mathrm{sph}$ as calculated from Eq.~(\ref{eqn:los_sigma_v}) is matched to the aperture size or annulus radii for each lens. Furthermore, we included an empirical Gaussian PSF with FWHM = $0\farcs8$ in the prediction of $\sigma^\mathrm{ap}_\mathrm{axi}$ and $\sigma^\mathrm{ap}_\mathrm{sph}$, which reduces numerical stochasticity in the correction factor and better matches the seeing of the data, but does not rescale $\sigma^\mathrm{ap}_\mathrm{axi}/\sigma^\mathrm{ap}_\mathrm{sph}$ significantly enough so as to introduce biases. 

A further improvement can be obtained by modeling them using the state-of-the-art axisymmetric JAM method directly \citep[e.g., ][]{Shajib23, yildirim23, Wang2025} on the 2D velocity dispersion map, gaining more access to the inclination angle and anisotropy, as well as information about the deflector ellipticity. However, applying a full parameter inference with axisymmetric JAM to the complete lensing+kinematics parameter space is extremely computationally demanding due to the difficulty of simultaneously fitting the lensing and dynamical variables in high-dimensional parameter spaces, and it is beyond the scope of this paper. Methods for accelerating the modeling process have been explored \citep{Gomer2023, Wang2025}, and those methods will be employed in future analyses by our collaboration.

\subsubsection{LoS convergence}\label{sec:los_modeling}

The external convergence $\kappa_{\rm{ext}}$ is inferred by using weighted number count statistics to estimate the relative density of a given lens field. We used data from the Millennium Simulation \citep{Springel05} to map these statistics to similar LoSs in Millennium, which has the value of $\kappa$ measured at high angular resolution \citep[see][]{Hilbert2009}. These techniques have been used extensively by the TDCOSMO collaboration and its predecessors. For complete overviews, see \citet{Rusu2017} and \citet{Wells2024}. 

We used the same set of lenses from the SL2S survey, discussed in Section~\ref{sec:sl2s_lensmodels}, to build a hierarchical model of the LoS convergence across the sample. This model is discussed in detail by \citet{Wells2024}. In brief, the population-level distribution of $\kappa_{\rm ext}$ follows a generalized extreme-value distribution with a mean and scatter in their respective parameters. These parameters are then constrained on the population level, which allows the population-level prior to deviate from random LoSs of the Millennium Simulation. The population-level constraints are then applied to the individual LoSs of the lenses, providing a posterior $p(\kappa_{\rm ext})$ that includes as a prior the population-level inference results.
Our model demonstrates with high confidence that the LoSs in our sample are drawn from a population that is overdense compared to the population of all LoSs in the Universe, as expected for massive early-type galaxies \citep{Auger2008,Treu2009}. This demonstrates that correcting for LoS effects on a population level is a necessary step when doing time-delay cosmography with large samples of lenses.
For our hierarchical inference, we used the individual posterior distributions of $\kappa_{\rm ext}$ (see e.g., Fig.~\ref{fig:kappa_ext_slacs_sl2s}).

\begin{table*}[h!]
\renewcommand{\arraystretch}{1.3} 
\caption{\label{tab:cosmo_model_LCDM} Combinations of cosmological probes and strong lensing datasets considered in this work, assuming a flat \lcdm cosmology.} 
\centering
\begin{tabular}{lllll}
\hline \hline
\multicolumn{5}{c}{Baseline flat-\lcdm cosmology} \\ \hline 
Model ID & Cosmological Model & Experiments/datasets & External lenses & Cosmological priors \\ \hline
U$\Lambda$CDM1 & Flat \lcdm & TDCOSMO & -- & \multirow{4}{*}{\begin{tabular}[c]{@{}l@{}}$H_0 \sim \mathcal{U}\, (0,150)$ \ksmpc \\ \Om $\sim \mathcal{U}\, (0.05, 0.5)$\end{tabular}} \\
U$\Lambda$CDM2 & Flat \lcdm & TDCOSMO & SLACS & \\
U$\Lambda$CDM3 & Flat \lcdm & TDCOSMO & SL2S & \\
U$\Lambda$CDM4 & Flat \lcdm & TDCOSMO & SLACS + SL2S & \\ \hline

$\Lambda$CDM1a & Flat \lcdm & TDCOSMO + {Pantheon+} & -- & \multirow{8}{*}{\begin{tabular}[c]{@{}l@{}}$H_0 \sim \mathcal{U}\, (0,150)$ \ksmpc \\ \Om $\sim \mathcal{U}\, (0.05, 0.5)$\end{tabular}} \\
$\Lambda$CDM1b & Flat \lcdm & TDCOSMO + {Pantheon+} & SLACS & \\
$\Lambda$CDM1c & Flat \lcdm & TDCOSMO + {Pantheon+} & SL2S & \\
$\Lambda$CDM1d & Flat \lcdm & TDCOSMO + {Pantheon+} & SLACS + SL2S & \\

$\Lambda$CDM2a & Flat \lcdm & TDCOSMO + {DES-SN5YR} & -- & \\
$\Lambda$CDM2b & Flat \lcdm & TDCOSMO + {DES-SN5YR} & SLACS + SL2S & \\

$\Lambda$CDM3a & Flat \lcdm & TDCOSMO + {DESI BAO} & -- &  \\
$\Lambda$CDM3b & Flat \lcdm & TDCOSMO + {DESI BAO} & SLACS + SL2S & \\
\hline
\end{tabular}
\tablefoot{The group of parameter priors within each row block is used for all cosmological models within that row block.}
\end{table*}

\begin{table*}[h!]
\renewcommand{\arraystretch}{1.3} 
\caption{Cosmological models, probes, and strong lensing datasets considered in extensions of the \lcdm model.} 
\centering
\begin{tabular}{llll}
 \hline \hline
\multicolumn{4}{c}{\lcdm extensions} \\ \hline
Model ID & Cosmological Model & Experiments/datasets & Cosmological priors \\ \hline
 &  &  & \multirow{4}{*}{\begin{tabular}[c]{@{}l@{}}$H_0 \sim \mathcal{U}\, (0,150)$ \ksmpc \\ \Om $\sim \mathcal{U}\, (0.05, 0.5)$ \\ \Ok $\sim \mathcal{U}\, (-0.5, 0.5)$ \\ $\OL = 1 - \Om - \Ok > 0$ \end{tabular}} \\ 
Uo$\Lambda$CDM & Open \lcdm & TDCOSMO &    \\
o$\Lambda$CDM & Open \lcdm & TDCOSMO + \textit{Planck} & \\ 

 &  &  &   \\ \hline 

U$w$CDM & Flat $w$CDM & TDCOSMO & \multirow{4}{*}{\begin{tabular}[c]{@{}l@{}}$H_0 \sim \mathcal{U}\, (0,150)$ \ksmpc \\ \Om $\sim \mathcal{U}\, (0.05, 0.5)$ \\ $w_0$ $\sim \mathcal{U}\, (-1.5, 0.5)$ \end{tabular}} \\

$w$CDM1 & Flat $w$CDM & TDCOSMO + \textit{Planck}  &  \\

$w$CDM2 & Flat $w$CDM & TDCOSMO + {DESI BAO}  &  \\

$w$CDM3 & Flat $w$CDM & TDCOSMO + {Pantheon+} & \\ \hline

U$w_0w_a$CDM & Flat $w_0w_a$CDM & TDCOSMO  & \multirow{4}{*}{\begin{tabular}[c]{@{}l@{}}$H_0 \sim \mathcal{U}\, (0,150)$ \ksmpc \\ \Om $\sim \mathcal{U}\, (0.05, 0.5)$ \\ $w_0$ $\sim \mathcal{U}\, (-1.5, 0.5)$ \\ $w_a \sim \mathcal{U}\, (-10,10)$ \end{tabular}}  \\

$w_0w_a$CDM1 & Flat $w_0w_a$CDM & TDCOSMO + \textit{Planck} & \\

$w_0w_a$CDM2 & Flat $w_0w_a$CDM & TDCOSMO + {DESI BAO} & \\

$w_0w_a$CDM3 & Flat $w_0w_a$CDM & TDCOSMO + {Pantheon+} & \\ 

$w_0w_a$CDM4 & Flat $w_0w_a$CDM & TDCOSMO + {DESI BAO} + {Pantheon+} & \\ \hline 

U$w_\phi$CDM & Flat $w_\phi$CDM & TDCOSMO  & \multirow{4}{*}{\begin{tabular}[c]{@{}l@{}}$H_0 \sim \mathcal{U}\, (0,150)$ \ksmpc \\ \Om $\sim \mathcal{U}\, (0.05, 0.5)$ \\ $w_0$ $\sim \mathcal{U}\, (-1.5, 0.5)$ \\ $\alpha \sim \mathcal{U}\, (1.35,1.55)$ \end{tabular}}  \\

$w_\phi$CDM1 & Flat $w_\phi$CDM & TDCOSMO + {DESI BAO} &  \\

$w_\phi$CDM2 & Flat $w_\phi$CDM & TDCOSMO + {Pantheon+} & \\ 
& & & \\ \hline

\end{tabular}
\tablefoot{All cases in this table include the external lensing datasets from the SLACS and SL2S. The group of parameter priors within each row block is used for all cosmological models within that row block.}
\end{table*}

\section{Cosmological results in flat \lcdm}
\label{sec:flat}

In this section, we adopted the flat \lcdm model as our baseline cosmological model. The cosmological constraints in alternative cosmological models are presented in \sref{sec:alternative_cosmo}. We first considered the cosmological constraints obtained only from the lensing information in \sref{sec:tdcosmo_flcdm} before combining them with other cosmological probes in \sref{subsec:external_datasets}.

\subsection{Lensing-only constraints} 
\label{sec:tdcosmo_flcdm}

For the lensing-only analysis, we took uniform priors on \Hc and $\Omega_{\rm m}$ in the range [0, 150] \ksmpc and [0.05, 0.5], respectively. Other priors on the lens population model parameters are summarized in \tref{tab:param_prior}.

We considered separately the addition of the SLACS and SL2S datasets of galaxy-galaxy lenses to the sample of eight TDCOSMO-2025 time-delay lenses. We list the combinations of external lens samples and external cosmological probes considered in this work in \tref{tab:cosmo_model_LCDM}.

In our baseline \lcdm analysis, we find $\Hc = 73.7^{+4.7}_{-4.4}$ \ksmpc from only the eight TDCOSMO-2025 lenses, at 6.2\% precision. The TDCOSMO-2025, SLACS, and SL2S samples yielded consistent constraints in terms of $\lambda_{\rm int}$ and can thus be combined. Including the SLACS lenses to further constrain the mass profile of our deflector population, we obtain $\Hc = 76.8^{+3.9}_{-4.6}$ \ksmpc, at 5.5\% precision. The inclusion of the SL2S lenses yields $\Hc = 74.2^{+4.6}_{-4.5}$ \ksmpc. Combining all lens samples together (TDCOSMO-2025 + SLACS + SL2S) gives similar constraints as TDCOSMO-2025 + SLACS, $\Hc = 77.8^{+3.7}_{-4.7}$ \ksmpc, at 5.4\% precision. The median, 16\textsuperscript{th} and 84\textsuperscript{th} percentiles of the posterior distributions for the cosmological parameters and deflector population model parameters are reported in \tref{tab:results_LCDM}. 

\subsection{Combination with independent cosmological probes}
\label{subsec:external_datasets}

Time-delay cosmography primarily measures \Hc but offers limited information on \Om. In this section, to mitigate the degeneracy between \Hc and \Om and enhance the precision of our measurements, we included external probes that provide stronger constraints on \Om than time-delay cosmography alone. All resulting constraints are summarized in \tref{tab:results_LCDM}.

\subsubsection{Pantheon+}
\label{subsec:Pantheon}

We used the Pantheon+ SN dataset \citep{Brout:2022, Scolnic2022}, which includes 1701 light curves of 1550 distinct SNe Ia spanning a redshift range from $z$ = 0.001 to 2.26. We incorporated the data and covariance matrix (including systematic uncertainties) provided by \cite{Brout:2022} but excluded the SH0ES absolute SN Ia magnitude calibration. Pantheon+ effectively provided a prior on $\Omega_{\mathrm{m}}$ (i.e., $\Omega_{\mathrm{m}} = 0.334 \pm 0.018$), while the absolute SNe Ia distances were anchored to our sample of time-delay lenses.

\begin{figure*}
    \centering
    \includegraphics[width=\textwidth]{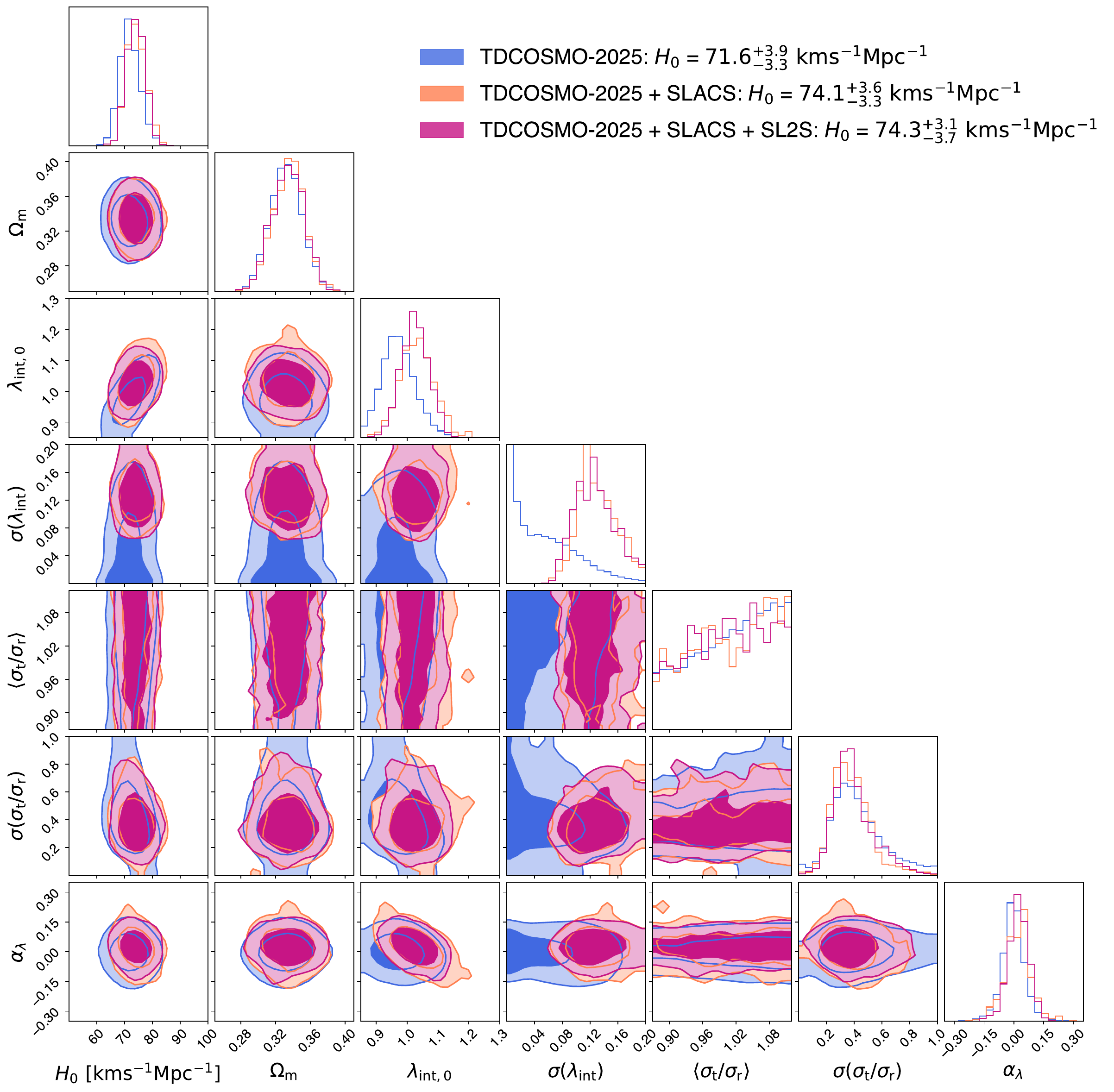}
    \caption{
    Posterior distributions for a subset of the parameters described in  \tref{tab:param_prior}. Blue: constraints from the eight TDCOSMO-2025 lenses, including JWST kinematic data for \RXJ. Orange: constraints from the TDCOSMO-2025 sample + \nslacsifusel\ SLACS lenses with KCWI IFU kinematics data. Pink: constraints from TDCOSMO-2025 + SLACS + SL2S sample. All contours include $\Omega_{\rm m}$ constraint coming from the Pantheon+ sample \citep{Scolnic2022, Brout:2022}. Priors are described in \tref{tab:param_prior}. The posteriors distribution of \Hc and $\lambda_{\rm int, 0}$ were blinded during the analysis. Results in combination with the DES-SN5YR likelihood are very similar (see~Table~\ref{tab:results_LCDM}) and are not shown for conciseness.}
    \label{fig:baseline_corner_plot}
\end{figure*}

In flat $\Lambda$CDM with the Pantheon+ SNe, we obtain $\Hc = \hnotflcdmonea$ \ksmpc using the TDCOSMO-2025 sample alone. When combined with the SLACS and SL2S lenses, the constraint on \Hc improves to $\Hc = 74.3^{+3.1}_{-3.7}$ \ksmpc, at 4.6\% precision. The posterior distribution for the cosmological parameters and deflector population model parameters are shown in \fref{fig:baseline_corner_plot}. The only parameter that differs between the three lens samples is the intrinsic scatter on $\lambda_{\rm int}$ that is found to be consistent with zero for TDCOSMO-2025 but not for SLACS. As discussed in Section~\ref{ssec:previous} and Appendix~\ref{app:SLACS}, the intrinsic scatter somewhat limits the improvement in precision obtained by adding SLACS to TDCOSMO-2025. However, $\alpha_\lambda$, i.e., the parameter describing the slope of $\lambda_{\rm int}$ with $R_{\rm eff}/\theta_{\rm E}$ of the deflector, is consistent across all three lens samples: we find $\alpha_\lambda = 0.00^{+0.05}_{-0.05}$ for TDCOSMO-2025 alone, $\alpha_\lambda = 0.03^{+0.05}_{-0.06}$ for TDCOSMO-2025 + SLACS, and $\alpha_\lambda = -0.01^{+0.05}_{-0.05}$ for TDCOSMO-2025 + SL2S. These values are consistent with zero, providing no evidence that the inferred $\lambda_{\rm int}$ varies as a function of the distance from the center of the lens galaxy.

\subsubsection{DES-SN5YR}
\label{sssec:DESYR5SN}

Similarly to Section~\ref{subsec:Pantheon}, we included the SNe Ia data from the DES \citep{DESSN2024}, without anchors. We adopted the Year-5 data release (i.e., DES-SN5YR\footnote{\url{https://github.com/des-science/DES-SN5YR/tree/main}}). 
In flat $\Lambda$CDM, the $\Omega_{\mathrm{m}}$ prior provided by this dataset is equivalent to $\Omega_{\mathrm{m}} = 0.352 \pm 0.017$. For the TDCOSMO-25 sample, we obtain $\Hc = 71.2^{+3.7}_{-3.6}$ \ksmpc, at 5.1\% precision. In combination with the SLACS and SL2S lenses, we obtain a more precise estimate of $\Hc = 73.9^{+3.4}_{-3.0}$ \ksmpc, at 4.3 \% precision. 

\subsubsection{DESI BAO DR2}

The DESI collaboration recently released their baryon acoustic oscillations (BAO) measurement in \cite{DESIDR2}. We used the 13 distance measurements, including covariances, provided by DESI\footnote{\url{https://github.com/CobayaSampler/bao_data/tree/master/desi_bao_dr2}}, in seven redshift bins covering the range $z=0.295$--2.33. The inferred distances by DESI are relative to the sound horizon $r_{\rm d}$ through the ratios $D_{\rm M}(z)/r_{\rm d}$ and $D_{\rm H}(z)/r_{\rm d}$, where $D_{\rm M}(z)$ is the transverse comoving distance and the Hubble distance $D_{\rm H}(z) = c / H(z)$. To break the degeneracy between $r_{\rm d}$ and \Hc, the DESI collaboration adopted a prior either from Big Bang nucleosynthesis (BBN) or from the \textit{Planck} CMB data to constrain the sound horizon scale.

In our analysis, we used the BAO uncalibrated distance ruler. We adopted a flat uninformative prior on $r_{\rm d} \sim \mathcal{U}, (0, 300)$ Mpc. The TDCOSMO-2025 dataset provided the absolute distance measurement necessary to anchor the BAO distances and break the \Hc– $r_{\rm d}$ degeneracy.

In flat \lcdm, our measurement, combining DESI BAO with the TDCOSMO-2025, SLACS, and SL2S lensing datasets, yields $\Hc = 74.8^{+3.5}_{-3.4}$~\ksmpc at 4.6\% precision. We also obtain a measurement of the sound horizon scale, $135.6^{+6.7}_{-6.1}$ Mpc, independent of assumptions about early-Universe physics. This measurement differs from that of \textit{Planck} by $\sim1.8\sigma$ and is another aspect of the Hubble tension as discussed in, e.g., \citet{Arendse2020}. 

\subsubsection{\textit{Planck}}

As our previous time-delay cosmography analyses (TDCOSMO-1 and TDCOSMO-4) showed significant tension with \textit{Planck}'s determination of \Hc, we refrained from combining our results with \textit{Planck} in flat \lcdm. However, in Section~\ref{sec:alternative_cosmo}, we explored different extensions of the \lcdm model where this tension is less significant, allowing us to combine the TDCOSMO-2025 constraint with the CMB observations from \textit{Planck}. In this case, we adopted the {\tt plikHM\_TTTEEE\_lowl\_lowE} chains from \citet{Planck2020} in various extensions of the \lcdm model \footnote{The {\tt COM\_CosmoParams\_fullGrid\_R3.01} dataset used in this work can be downloaded at \url{https://wiki.cosmos.esa.int/planck-legacy-archive/index.php/Cosmological_Parameters}}.

\renewcommand*\arraystretch{1.5}
\setlength{\tabcolsep}{2pt}
\begin{table*}[hbt!]
\caption{\label{tab:results_LCDM}
Cosmological parameters in flat \lcdm.}
\begin{minipage}{\linewidth}
\centering
\begin{tabular}{lllccc}
\hline \hline
Model ID &
Experiments/datasets & Lensing datasets &
$H_{0}$ [km s$^{-1}$ Mpc$^{-1}$] &
$\Om$ & $\lambda_{\rm int}$ \\
\hline
U$\Lambda$CDM1 & TDCOSMO & TDCOSMO-2025 & $73.7^{+4.7}_{-4.4}$ & $0.232^{+0.155}_{-0.120}$ & $1.00^{+0.07}_{-0.06}$  \\
U$\Lambda$CDM2 & TDCOSMO & TDCOSMO-2025  + SLACS & $76.8^{+3.9}_{-4.6}$ & $0.190^{+0.163}_{-0.110}$ & $1.06^{+0.06}_{-0.06}$  \\
U$\Lambda$CDM3 & TDCOSMO & TDCOSMO-2025  + SL2S & $74.2^{+4.6}_{-4.5}$ & $0.229^{+0.161}_{-0.119}$ & $1.00^{+0.07}_{-0.06}$  \\
U$\Lambda$CDM4 & TDCOSMO & TDCOSMO-2025  + SLACS  + SL2S & $77.8^{+3.7}_{-4.7}$ & $0.172^{+0.175}_{-0.094}$ & $1.06^{+0.06}_{-0.06}$  \\ \hline
$\Lambda$CDM1a & TDCOSMO + Pantheon+ & TDCOSMO-2025 & $71.6^{+3.9}_{-3.3}$ & $0.332^{+0.018}_{-0.018}$ & $0.97^{+0.06}_{-0.05}$  \\
$\Lambda$CDM1b & TDCOSMO + Pantheon+ & TDCOSMO-2025  + SLACS & $74.1^{+3.6}_{-3.3}$ & $0.336^{+0.017}_{-0.019}$ & $1.03^{+0.05}_{-0.06}$  \\
$\Lambda$CDM1c & TDCOSMO + Pantheon+ & TDCOSMO-2025  + SL2S & $72.1^{+3.6}_{-3.4}$ & $0.333^{+0.019}_{-0.018}$ & $0.98^{+0.05}_{-0.05}$  \\
$\Lambda$CDM1d & TDCOSMO + Pantheon+ & TDCOSMO-2025  + SLACS  + SL2S & $74.3^{+3.1}_{-3.7}$ & $0.335^{+0.017}_{-0.020}$ & $1.03^{+0.04}_{-0.04}$  \\
$\Lambda$CDM2a & TDCOSMO + DES-SN5YR & TDCOSMO-2025 & $71.2^{+3.7}_{-3.6}$ & $0.353^{+0.016}_{-0.017}$ & $0.96^{+0.05}_{-0.05}$  \\
$\Lambda$CDM2b & TDCOSMO + DES-SN5YR & TDCOSMO-2025  + SLACS  + SL2S & $73.9^{+3.4}_{-3.0}$ & $0.353^{+0.018}_{-0.019}$ & $1.03^{+0.05}_{-0.04}$  \\
$\Lambda$CDM3a & TDCOSMO + DESI BAO & TDCOSMO-2025 & $72.4^{+3.9}_{-3.6}$ & $0.297^{+0.009}_{-0.008}$ & $0.98^{+0.06}_{-0.06}$  \\
$\Lambda$CDM3b & TDCOSMO + DESI BAO & TDCOSMO-2025  + SLACS  + SL2S & $74.8^{+3.5}_{-3.4}$ & $0.298^{+0.009}_{-0.008}$ & $1.03^{+0.05}_{-0.04}$  \\ \hline

\end{tabular}
\end{minipage}
\tablefoot{Reported values are medians, with errors corresponding to the 16th and 84th percentiles.}
\end{table*}
\setlength{\tabcolsep}{6pt}
\renewcommand*\arraystretch{1.0}

\renewcommand*\arraystretch{1.5}
\setlength{\tabcolsep}{2pt}
\begin{table*}[hbt!]
\caption{Cosmological parameters for extensions to the \lcdm model, from time-delay cosmography alone, or combined with other probes. \label{tab:results_LCDM_extension}}
\begin{minipage}{\linewidth}
\centering
\begin{tabular}{llcccccccc}
\hline \hline
Model &
Experiments/datasets &
$H_{0}$ [km s$^{-1}$ Mpc$^{-1}$] &
$\Om$ &
$\Ok$ &
$w$ or $w_{0}$ &
$w_{a}$
\\
\hline

Uo$\Lambda$CDM & TDCOSMO & $77.1^{+4.9}_{-4.1}$ & $0.229^{+0.151}_{-0.132}$ & $-0.117^{+0.305}_{-0.245}$ & -- & --  \\
o$\Lambda$LCDM & TDCOSMO + \textit{Planck} & $65.2^{+1.8}_{-2.3}$ & $0.338^{+0.030}_{-0.018}$ & $-0.006^{+0.005}_{-0.006}$ & -- & --  \\ \hline
U$w$CDM & TDCOSMO & $78.0^{+6.6}_{-7.3}$ & $0.178^{+0.194}_{-0.091}$ & -- & $-1.12^{+0.44}_{-0.27}$ & --  \\
$w$CDM1 & TDCOSMO + \textit{Planck} & $79.7^{+3.2}_{-4.3}$ & $0.225^{+0.027}_{-0.017}$ & -- & $-1.38^{+0.13}_{-0.08}$ & --  \\
$w$CDM2 & TDCOSMO + DESI BAO & $74.0^{+3.4}_{-3.6}$ & $0.296^{+0.008}_{-0.008}$ & -- & $-0.93^{+0.08}_{-0.09}$ & --  \\
$w$CDM3 & TDCOSMO + Pantheon+ & $73.7^{+3.5}_{-3.1}$ & $0.277^{+0.072}_{-0.083}$ & -- & $-0.86^{+0.14}_{-0.18}$ & --  \\ \hline
U$w_0w_a$CDM & TDCOSMO & $80.9^{+11.2}_{-8.0}$ & $0.203^{+0.171}_{-0.112}$ & -- & $-0.89^{+0.55}_{-0.43}$ & $-5.23^{+4.55}_{-3.31}$  \\
$w_0w_a$CDM1 & TDCOSMO + \textit{Planck} & $65.9^{+2.7}_{-2.4}$ & $0.333^{+0.024}_{-0.027}$ & -- & $-0.64^{+0.28}_{-0.30}$ & $-1.20^{+0.84}_{-0.81}$  \\
$w_0w_a$CDM2 & TDCOSMO + DESI BAO & $70.4^{+5.5}_{-3.7}$ & $0.348^{+0.046}_{-0.066}$ & -- & $-0.54^{+0.41}_{-0.45}$ & $-1.50^{+1.77}_{-1.43}$  \\
$w_0w_a$CDM3 & TDCOSMO + Pantheon+ & $74.7^{+3.3}_{-3.8}$ & $0.359^{+0.068}_{-0.147}$ & -- & $-0.90^{+0.14}_{-0.16}$ & $-0.79^{+1.30}_{-1.85}$  \\
$w_0w_a$CDM4 & TDCOSMO + DESI BAO + Pantheon+ & $73.2^{+4.0}_{-3.0}$ & $0.303^{+0.016}_{-0.021}$ & -- & $-0.89^{+0.06}_{-0.05}$ & $-0.20^{+0.49}_{-0.45}$  \\ \hline
U$w_\phi$CDM & TDCOSMO & $75.9^{+6.8}_{-7.8}$ & $0.196^{+0.162}_{-0.098}$ & -- & $-0.95^{+0.59}_{-0.39}$ & --  \\
$w_\phi$CDM1 & TDCOSMO + DESI BAO & $73.5^{+3.6}_{-3.1}$ & $0.305^{+0.012}_{-0.013}$ & -- & $-0.86^{+0.11}_{-0.13}$ & --  \\
$w_\phi$CDM2 & TDCOSMO + Pantheon+ & $73.8^{+3.3}_{-3.3}$ & $0.313^{+0.041}_{-0.042}$ & -- & $-0.92^{+0.11}_{-0.13}$ & --  \\

\end{tabular}

\tablefoot{Reported values are medians, with errors corresponding to the 16th and 84th percentiles. In this table, we utilized the full strong lensing data set, which includes TDCOSMO-2025, SLACS, and SL2S lenses.}
\\
\end{minipage}
\end{table*}
\setlength{\tabcolsep}{6pt}
\renewcommand*\arraystretch{1.0}

\section{Alternative cosmological models}
\label{sec:alternative_cosmo}

Time-delay cosmography is primarily sensitive to \Hc, and only weakly dependent, given the current sample size, on parameters used in extensions of flat \lcdm, such as \Ok, $w_0$, and $w_a$. The sensitivity to \Hc and the different direction of residual parameter degeneracies make it highly complementary to other cosmological probes \citep{Linder11}. In this section, we report the cosmological constraints provided by TDCOSMO-2025, in combination with the CMB data from \textit{Planck}, BAO measurements from DESI, and SN Ia data from Pantheon+, in open \lcdm, flat $w$CDM, flat $w_0w_a$CDM, and flat $w_\phi$CDM cosmologies. All priors used in these alternative cosmological models are listed in \tref{tab:cosmo_model_LCDM}, and the parameter constraints are given in \tref{tab:results_LCDM_extension}.

\subsection{Open \lcdm}
A simple modification of the flat \lcdm cosmology is the open \lcdm cosmology, which allows for spatial curvature, \Ok $\neq 0$. We adopted a flat prior on \Ok $\sim \mathcal{U}\, (-0.5, 0.5)$ and imposed the additional condition $\Omega_{\Lambda} = 1 - \Om - \Ok >0$. The energy density in photons and neutrinos can be neglected with the current level of precision.

\begin{figure}
    \centering
    \includegraphics[width=\linewidth]{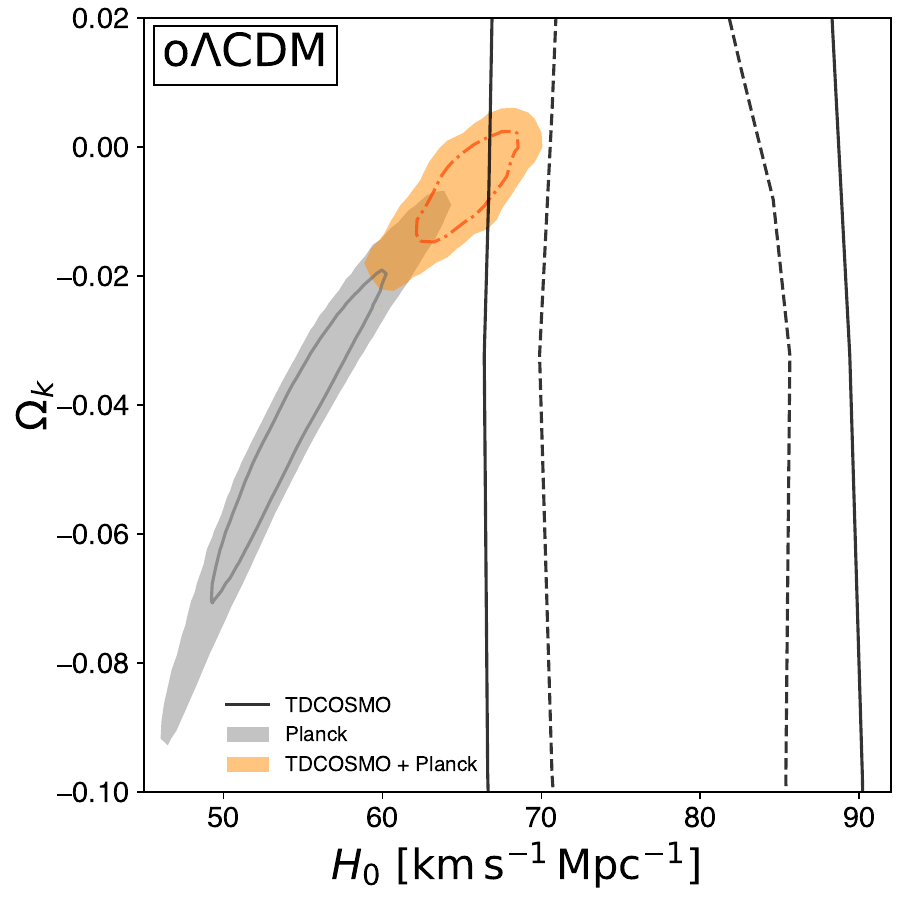}
    \caption{Posterior distribution of \Hc and \Ok in open \lcdm cosmology. The black contours show the constraints from TDCOSMO alone, including the TDCOSMO-2025, SLACS, and SL2S lensing datasets. The gray contours show the constraints from \textit{Planck}. The orange contours correspond to the combination of \textit{Planck} and TDCOSMO. The two datasets exhibit a tension exceeding $3\sigma$, indicating that this joint constraint should be interpreted with caution. The contour levels represent the 1$\sigma$ and 2$\sigma$ constraints.
    }
    \label{fig:oLCDM}
\end{figure}

Using only the TDCOSMO-2025, SLACS, and SL2S lensing datasets, we obtain $\Hc = 77.1^{+4.9}_{-4.1}$~\ksmpc. \Ok is poorly constrained by time-delays alone, with $\Ok = -0.117^{+0.305}_{-0.245}$. However, \Hc remains well constrained, which provides an indirect constraint on the curvature when combining time-delay cosmography with the CMB data. However, our measurement is in $>3\sigma$ tension with that of \textit{Planck}, which might lead to underestimated uncertainties when combining these two datasets. Still, this combination yields $\Hc = 65.2^{+1.8}_{-2.3}$~\ksmpc and $\Ok = -0.006^{+0.005}_{-0.006}$ in open \lcdm cosmology. We show the posterior distribution of \Hc and \Ok in \fref{fig:oLCDM} for TDCOSMO, \textit{Planck}, and their combination.

\subsection{Flat $w$CDM}

We considered the flat $w$CDM cosmological model in which dark energy differs from a cosmological constant and is described by an equation-of-state parameter $w$. When $w$ equals $-1$, this model reduces to the flat \lcdm\ case. For the parameter $w$, we imposed a uniform prior over the interval $[-1.5, 0.5]$, while maintaining the same uniform priors on \Hc\ and $\Omega_{\rm m}$ as used in the flat \lcdm\ model.

\begin{figure}
    \centering
    \includegraphics[width=\linewidth]{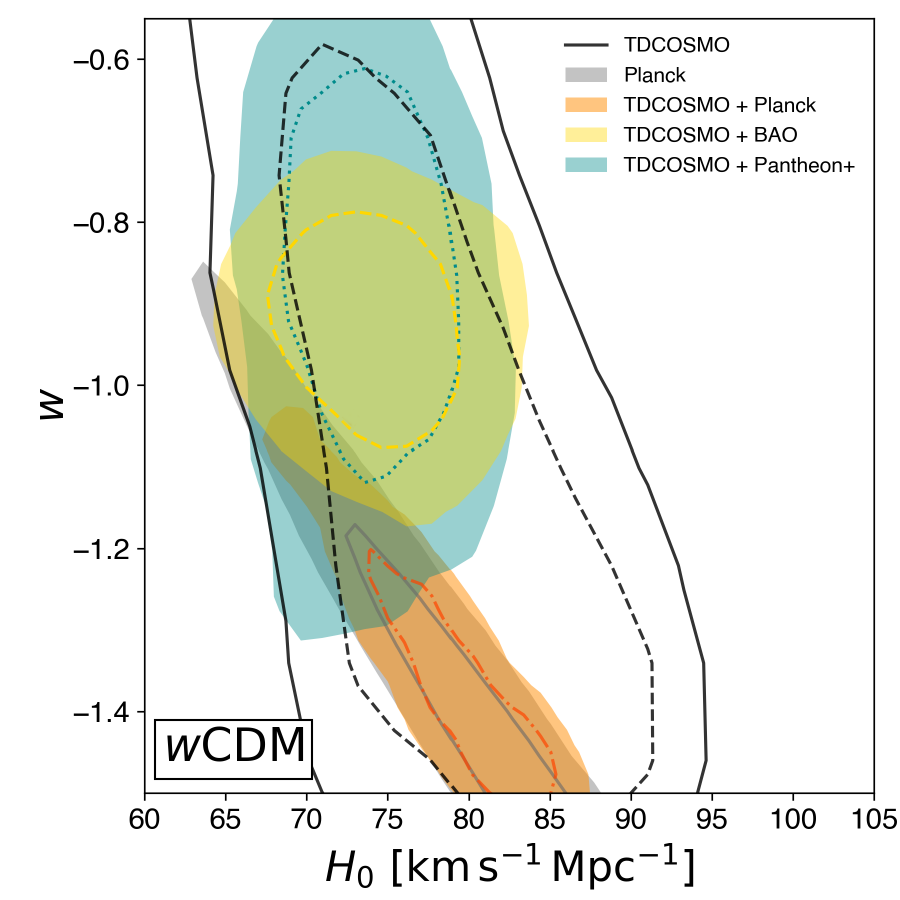}
    \caption{Posterior distribution of \Hc and $w$ in flat $w$CDM cosmology. The black contours show the constraints from TDCOSMO alone, including the TDCOSMO-2025, SLACS, and SL2S lensing datasets. The gray contours show the constraints from Planck. The combination of TDCOSMO with the \textit{Planck}, DESI BAO, and Pantheon+ is shown with colored contours, as described in the figure legend. The contour levels represent the 1$\sigma$ and 2$\sigma$ constraints.}
    \label{fig:wCDM}
\end{figure}

We show the parameters constraints in this cosmology in \fref{fig:wCDM}. The marginalized \Hc posterior for the TDCOSMO-2025+SLACS+SL2S lensing dataset is $\Hc = 78.0^{+6.6}_{-7.3}$~\ksmpc. This is higher and more uncertain than in the flat \lcdm cosmology due to the degeneracy between \Hc and $w$. 

In combination with \textit{Planck}, our constraints on the dark energy equation-of-state is $w = -1.38^{+0.13}_{-0.08}$ and $\Hc = 79.7^{+3.2}_{-4.3}$~\ksmpc. In this combination, the dark energy equation-of-state deviates from a cosmological constant at $\sim4\sigma$. However, this constraint is mainly driven by the tension regarding \Hc between the two datasets, which is already observed in the \lcdm cosmology. 

When combining with the DESI BAO data or Pantheon+, our measurement is in agreement with the \lcdm model, finding no deviation from a cosmological constant. The marginalized posterior distribution on $w$ is $w = -0.93^{+0.08}_{-0.09}$ for TDCOSMO+DESI BAO and $w = -0.86^{+0.14}_{-0.18}$ for TDCOSMO + Pantheon+.

\subsection{Flat $w_0w_a$CDM}

The flat $w_0w_a$CDM cosmology includes a dark energy component characterized by an equation-of-state parameter $w(z)$ that can evolve with time. In this model, the dark energy equation-of-state is parameterized as
\begin{equation}
    w(z) = w_0 + w_a \frac{z}{1+z},
\end{equation}
following \citet{Chevallier2001} and \citet{Linder2003}. In this cosmology we adopt flat priors on $w_0$ $\sim \mathcal{U}\, (-1.5, 0.5)$ and $w_a$ $\sim \mathcal{U}\, (-10, 10)$. 

\begin{figure}
    \centering
    \includegraphics[width=\linewidth]{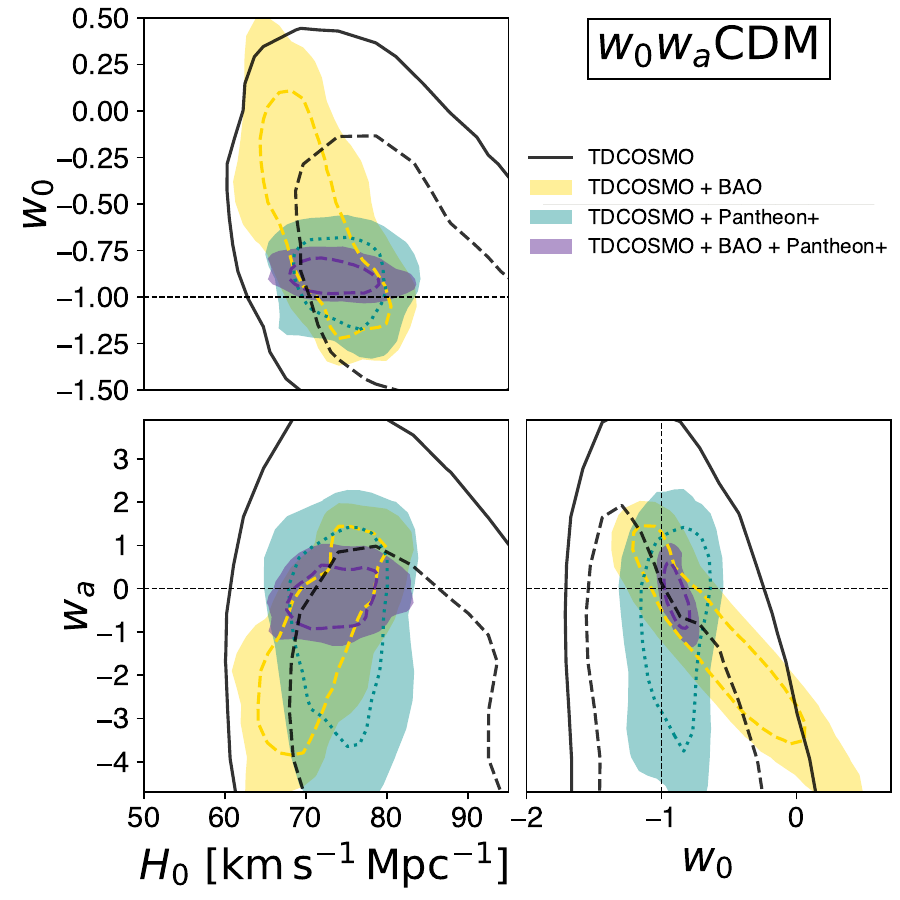}
    \caption{Posterior distribution of \Hc, $w_0$ and $w_a$ in flat $w_0 w_a$CDM cosmology.  The combination of TDCOSMO (including the TDCOSMO-2025, SLACS, and SL2S lensing datasets) with the DESI-DR2 BAO and Pantheon+, and the combination of both external datasets, is shown with colored contours described in the figure legend. The contour levels represent the 1$\sigma$ and 2$\sigma$ constraints.}
    \label{fig:w0waCDM}
\end{figure}

Our lensing-only dataset weakly constrains these two parameters, whose posterior distributions are shown in Fig.~\ref{fig:w0waCDM}. However, \Hc is constrained to be $\Hc = 80.9^{+11.2}_{-8.0}$~\ksmpc. 

The combinations of TDCOSMO individually with \textit{Planck}, DESI-BAO, or Pantheon+ are consistent with a cosmological constant (i.e.,  $w_0=-1$ and $w_a=0$) within 2$\sigma$. This is expected given the weak constraints on the dark energy parameters derived only from the TDCOSMO dataset. The combination of TDCOSMO with both Pantheon+ and DESI-BAO does not improve the precision on the dark energy parameters compared to the results presented in \citet{DESIDR2}, but $\Hc$ is well constrained to be $\Hc = 73.2^{+4.0}_{-3.0}$~\ksmpc in flat $w_0w_a$CDM cosmology. The strong preference for dynamical dark energy becomes evident only when multiple other probes are combined, such as BAO+CMB or BAO+SN+CMB \citep{DESIDR2}. Larger time-delay lens samples in the future will be able to provide precision on the dark energy parameters comparable to other leading probes \citep{Shajib25lsst}.

\subsection{Flat $w_\phi$CDM}

The $w_\phi$CDM model describes the dark energy as originating from a scalar field, such as the pseudo-Nambu-Goldstone bosons \citep{Frieman95, Shajib25c}. This field is allowed to evolve over time, thus giving rise to a dynamic nature in the dark energy. As a result, this is a physics-inspired model of dynamical dark energy, unlike the $w_0w_a$CDM model, which is phenomenological and physics-agnostic, thereby allowing for violation of standard physics, such as the null energy condition, within its parameter space. In the $w_\phi$CDM model, the dynamical behavior of $w(z)$ is well-approximated by the formula \citep{Shajib25c}

\begin{equation}
w(z) = -1 + (1 + w_0) \ \exp (-\alpha z),
\end{equation}

where $w_0$ is the value of $w$ at the current epoch (i.e., $z=0$). The $\alpha$ parameter depends on the form of the potential of the scalar field, and it falls within the narrow range of 1.35--1.55, which we adopted as a uniform prior range. However, all current and near-future datasets would not be sufficiently sensitive to constrain $\alpha$ beyond this prior range \citep{Shajib25c}. As a result, this $w_\phi$CDM model is a pseudo-one-parameter extension to the $\Lambda$CDM.

\begin{figure}
    \centering
    \includegraphics[width=\linewidth]{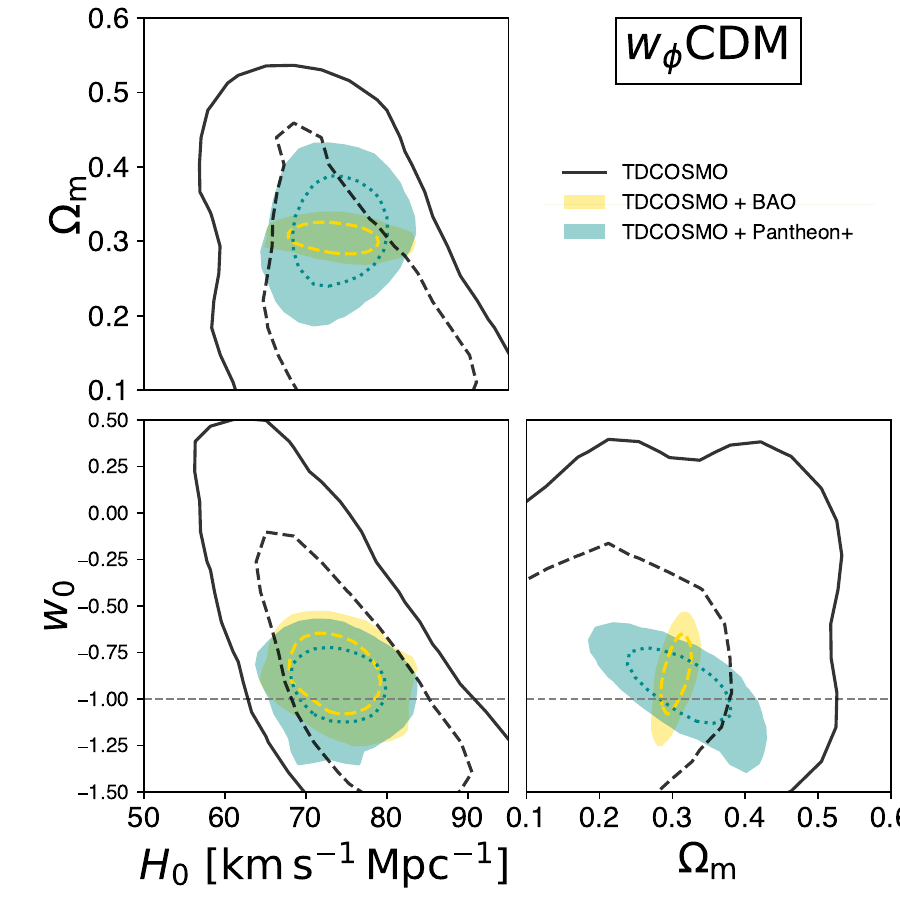}
    \caption{Posterior distribution of \Hc, $\Omega_{\rm m}$ and $w_0$ in flat $w_\phi$CDM cosmology.  The combinations of TDCOSMO (including the TDCOSMO-2025, SLACS, and SL2S lensing datasets) with the DESI-DR2 BAO and Pantheon+ datasets are shown with colored contours, as described in the figure legend. The contour levels represent the 1$\sigma$ and 2$\sigma$ constraints.}
    \label{fig:wphiCDM}
\end{figure}

In this cosmological model, we obtain $\Hc = 75.9^{+6.8}_{-7.8}$~\ksmpc and $w_0 = -0.95^{+0.59}_{-0.39}$, using only the TDCOSMO-2025+SLACS+SL2S lensing dataset (Fig.~\ref{fig:wphiCDM}). In combination with DESI BAO or Pantheon+, we obtain $w_0 = -0.86^{+0.11}_{-0.13}$ and $w_0 = -0.92^{+0.11}_{-0.13}$, respectively. The combination of TDCOSMO-2025 with DESI BAO or Pantheon+ alone does not exclude the cosmological constant (i.e., the case with $w_0 = -1$), similar to what is found also for the $w_0w_a$CDM model above.

\section{Comparison with the literature}
\label{sec:literature}

We now briefly discuss our measurement in comparison with previous work by our collaboration (Section ~\ref{ssec:previous}), by other teams using time-delay cosmography(Section~\ref{ssec:others}), and within the broader context of the Hubble tension (Section~\ref{ssec:tension}).

\subsection{Comparison with our previous work}
\label{ssec:previous}

\begin{figure*}
    \centering
    \includegraphics[width=0.9\textwidth]{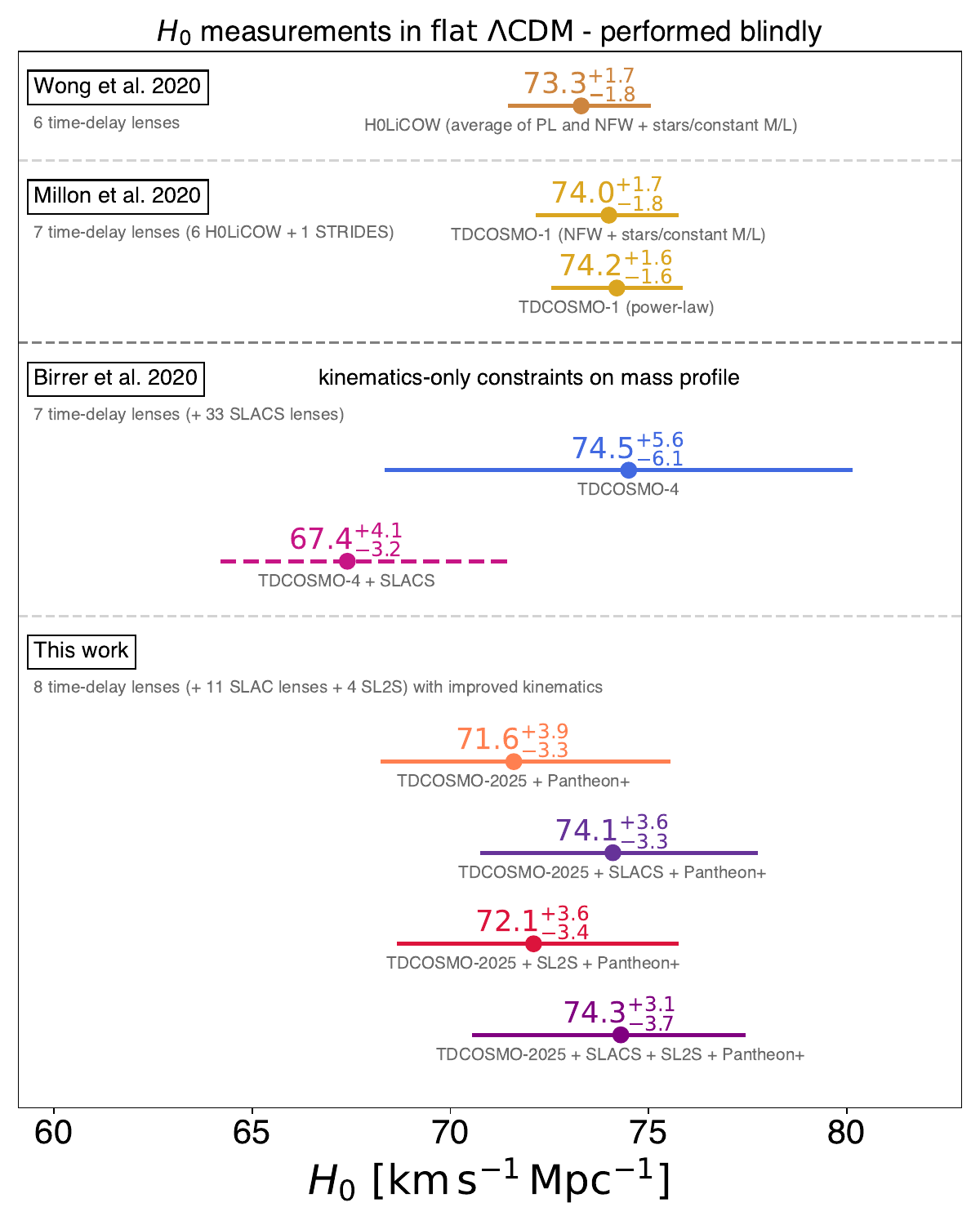}
    \caption{Comparison of the $H_0$ measurements by the TDCOSMO collaboration (and its predecessor H0LiCOW) in chronological order. The measurements at the top by \citet{Wong2019} and \citet{Millon:2020} are "assertive" in the definition of \citet{Treu_review}. They are based on power-law and composite models to describe the mass density profile of the deflector galaxies, thus implicitly breaking the MSD and obtaining $\sim$2\% precision. The TDCOSMO-4 results from \citet{TDCOSMOIV} are "conservative" \citep{Treu_review}. They are based on the same data but obtain larger uncertainties by introducing the internal MSD parameter $\lambda_{\rm int}$, and constraining it with unresolved stellar kinematics. The \citet{TDCOSMOIV} measurement in combination with SLACS is shown as a dashed line for historical purposes but should not be used anymore, as it was later discovered that the stellar velocity dispersions based on low SNR ($\sim$9 \AA$^{-1}$) spectra from SDSS suffer from systematic errors and covariance \citep{Knabel24}. The new measurements presented in this work are shown in the bottom panel. They are "conservative" in terms of uncertainties and constrain the MSD using new stellar kinematics based on high SNR JWST-NIRSpec, VLT-MUSE, and Keck-KCWI spectra, as well as an improved methodology \citep{Knabel25}. 
    \label{fig:H0_results}}
\end{figure*}

In this section, we compare our results with those from our previous work, as summarized by TDCOSMO-1 and TDCOSMO-4 (see Fig.~\ref{fig:H0_results}). In addition to performing the comparison, we aim to provide some intuition for the factors that contributed to the change in the overall precision and those that changed the median value of the TDCOSMO+SLACS analysis. The TDCOSMO-1 and TDCOSMO-4 papers were based on seven out of the eight time-delay lenses studied in this work, with the same identical time delays and lens models. The new additional lens (\WGDtwenty) has considerably larger uncertainties on the time delays \citep{Wong24}, so we expected it not to have a substantial effect on precision and point estimate. Its central estimate of $H_0$ was lower than the average for the other seven. Therefore, we expected it to lower $H_0$ by a modest amount.

\subsubsection{TDCOSMO-only}

Our inferred Hubble constant ($H_0=\hnotflcdmonea$ \ksmpc) is in excellent agreement with the results of both TDCOSMO-1 ($H_0=74.2\pm1.6$ \ksmpc) and TDCOSMO-4 ($H_0=74.5_{-6.1}^{+5.6}$\ksmpc) in the flat $\Lambda$CDM cosmology. The exact comparison depends somewhat on the priors on $\Om$, owing to the mild degeneracy with $H_0$. TDCOSMO-4 imposed $\Om=0.298\pm0.022$ based on the Pantheon SNe, while TDCOSMO-1 used the uniform prior $\Om\in[0.05,0.5]$ that is the same one used by \citet{Wong2019}. Using the same prior as in TDCOSMO-1 instead of Pantheon+, we obtain $\Hc = 73.7^{+4.7}_{-4.4}$ \ksmpc. 

The TDCOSMO-1 analysis was based on stricter assumptions on the mass profile than those assumed here and by TDCOSMO-4. Effectively, the TDCOSMO-1 power-law result is equivalent to imposing $\lambda_{\rm int}=1$ in the analysis of TDCOSMO-4, since the data were almost exactly the same. It is thus not surprising that the precision of TDCOSMO-1 was better than that of TDCOSMO-4, and that in the present analysis, which is based on much improved data. TDCOSMO-4 found $\lambda_{\rm int}=1.02^{+0.08}_{-0.09}$, resulting in the effectively the same central value of the TDCOSMO-only estimate, but with larger uncertainties ($H_0\propto\lambda_{\rm int}$, all other things being equal).  

In this work, we made several changes that affect the parameter $\lambda_{\rm int}$, the anisotropy parameter, the power-law slope $\gamma_{\rm pl}$, and ultimately $H_0$. From a data standpoint, we replaced all stellar velocity dispersions with new measurements based on significantly improved data and methods. Qualitatively, we expected this would change the inference on $H_0$ in the following manner.

First, the new stellar velocity dispersions for the time-delay lenses have an average total uncertainty of 4.0\%, compared with 7.7\% for the values used by TDCOSMO-4. The average total uncertainty is defined as $<\delta \sigma_i/\sigma_i>_i$, averaged over the seven lenses in common with TDCOSMO-4, with $\delta \sigma_i$ being the sum in quadrature of average statistical and systematic errors. All other things being equal, this should have considerably improved the precision on $H_0$, provided that no intrinsic scatter is detected. TDCOSMO-4 did not find any intrinsic scatter in the time-delay lenses. Remarkably, the intrinsic scatter remains consistent with zero even with our improved kinematic measurements. We note that the stellar velocity dispersion is not the only source of uncertainty; the time delays, lens model, and LoS convergence also contribute to the final uncertainty on $H_0$ \citep{Wong2019}. 

Additional information was obtained from the spatially resolved kinematics of \RXJ, which constrain the mass density slope and the internal MSD. As shown by \citet{Shajib23}, the use of the KCWI data alone makes this one system almost as precise ($H_0=77.1_{-7.1}^{+7.3}$\ksmpc) as the seven lenses combined in TDCOSMO-4 ($H_0=74.5_{-6.1}^{+5.6}$\ksmpc). The addition of the JWST data was expected to further improve the precision for this system \citep[further gains can be achieved using the full 2D kinematics;][]{yildirim23, Wang2025}. Therefore, based on these factors alone, we were expecting an improved precision for the TDCOSMO-only measurement, again in the absence of internal scatter on $\lambda_{\rm int}$, depending exactly on the overlap between the individual datasets. As shown in Appendix~\ref{app:rxj}, the KCWI and JWST data are in mutual statistical agreement, improving the overall precision stemming from \RXJ. However, for this one system, the uncertainty on $\kappa_{\rm ext}$ provides an effective noise floor.

In addition, the average ratio of the newly measured stellar velocity dispersions with respect to previously measured values provides us with a posteriori insight into the agreement between the measurement presented here and that presented in TDCOSMO-4. While we did not blind the estimates of the velocity dispersions, in order to minimize confirmation bias, three different team members\footnote{A.J.S.: resolved JWST-NIRSpec and Keck-KCWI measurements for \RXJ, S.K.: the five unresolved JWST-NIRSpec ones, and M.M.: the two VLT-MUSE ones} measured the velocity dispersions independently and without comparing the results. In some cases, the new results differ significantly from the old ones; four are within 1$\sigma$ and three are between 1--2$\sigma$ of the old uncertainties (see Fig.~\ref{fig:oldnew}). Moreover, several other factors besides the velocity dispersions contribute to determining $H_0$, such as spectroscopic apertures and seeing (see the example in Fig.~\ref{fig:KCWIspectra}). Furthermore, $H_0$ depends not only on the measured stellar velocity dispersion, but also on the inferred slope, anisotropy parameter, and external convergence (see example in Appendix~\ref{app:rxj}) in non-trivial ways. The stellar velocity dispersion measurements were frozen before unblinding and before performing the aggregate comparison described in this section.  In a few cases, TDCOSMO-4 combined multiple velocity dispersion measurements for the same system. For the simple comparison we make here, we took an inverse variance weighted average of the four dispersion measurements for \DESzerofour used by TDCOSMO-4, adding the covariance in quadrature. For \PGeleven, we took the central aperture, which is the closest to the JWST one adopted here. For \RXJ, we adopted the stellar velocity dispersion within half of the half-light radius as our point of comparison. In sum, for the seven lenses in common with TDCOSMO-4, the average ratio is $\sigma_{\rm new} / \sigma_{\rm old}=0.986\pm0.049$. In reality, of course, the weight of each lens is different, and those with the best measured time delays and lens models will carry more weight. This simple check provides some insight as to why the overall mean of the $\lambda_{\rm int}$ posterior remained the same within the errors, although only the complete analysis can account for all the changes. In fact, $H_0$ did not scale exactly as $\lambda_{\rm int}$ between TDCOSMO-4 and this work. This is due to two effects. First, the addition of \WGDtwenty that reduces $H_0$ overall, although this is a small contribution as described above, owing to its large uncertainty. Second, and most importantly, the spatially resolved kinematics of \RXJ prefers a higher slope $\gamma_{\rm pl}$ than the lens model, and prefers lower $\lambda_{\rm int}$ than the TDCOSMO-2025 sample average. 

We have also improved the anisotropy model, now accounting for projection effects that affect the model-predicted stellar velocity dispersion when compared with the observed one. This correction shifts the inferred H$_0$ by at most 1-2\% down \citep{Huang:2025}.

In conclusion, the improvement in precision with respect to TDCOSMO-4 can be attributed to new measurements and an enhanced methodology. The good agreement between this blind analysis and TDCOSMO-4 and TDCOSMO-1 provides a powerful validation of the method. Further improvements are possible with improved analysis of 2D kinematics, time delays, LoS characterization, new lens models, and expanded datasets.

\subsubsection{TDCOSMO + external lens samples}

Importantly, $H_0$ from the TDCOSMO-2025+SLACS is in excellent agreement with that from the TDCOSMO-2025 alone, given the uncertainties. Compared to TDCOSMO-4, the TDCOSMO+SLACS inference of $H_0$ moved from $H_0=67.4_{-3.2}^{+4.1}$\ksmpc to $76.8^{+3.9}_{-4.6}$\ksmpc in our U$\Lambda$CDM cosmology (\Hc=$74.8^{+3.5}_{-3.4}$\ksmpc in combination with DESI-DR2 that yields the more similar $\Omega_{\rm m}$ to the one adopted in TDCOSMO-4), to be compared with the TDCOSMO-2025-only result $H_0=73.7^{+4.7}_{-4.4}$\ksmpc in the same cosmology.

The improved precision and change in the median value can be understood as follows. We went from using 33 SLACS lenses with SDSS spectra in TDCOSMO-4 to  \nslacsifusel\ lenses with Keck-KCWI spectra here.  We now know that the SDSS-SLACS velocity dispersion measurements is affected by systematic uncertainty at the level of 3.3\% and covariance at the level of 2.3\% \citep{Knabel24}, making them not suited for cosmography and therefore the old TDCOSMO+SLACS result from TDCOSMO-4 suffered from significant underestimation of uncertainty (random and systematic) and should not be considered as valid anymore. However -- setting aside systematic errors in the SDSS stellar velocity dispersion -- the SNR of the Keck-KCWI spectra is $\sim$18 times higher per lens than that of SDSS. The covariance between the SLACS lenses' stellar velocity dispersion profiles, due to stellar template choices, sets a floor of about 1.5-2\% to the precision on $H_0$ attainable with this sample. Overall, we expected the new KCWI-SLACS dataset to reduce the errors on $H_0$ significantly more than in TDCOSMO-4, even with just \nslacsifusel\ lenses, if no intrinsic scatter in $\lambda_{\rm int}$ was detected. However, our analysis showed that the SLACS lenses require significant scatter in $\lambda_{\rm int}$, as shown in Appendix~\ref{app:SLACS}. The improvement in precision provided by the SLACS sample with respect to TDCOSMO-4 is thus limited not by the quality of the data, but by the intrinsic scatter of the sample. We note that the intrinsic scatter for the TDCOSMO-2025 sample is consistent with zero. However, the posterior distribution of $\sigma(\lambda_{\rm int})$ exhibits a long tail, making it compatible at 95\% confidence level, with the intrinsic scatter detected in the SLACS-KCWI sample. It remains to be investigated whether the intrinsic scatter of the SLACS-KCWI can be reduced with more sophisticated models (e.g., 2D dynamical models and more advanced lens models) or whether it is an irreducible property of the sample.

The SL2S dataset is new and consists of only \nsltwossel\ lenses with integrated stellar velocity dispersion. Thus, we did not expect it to significantly improve the precision of the measurement. Nevertheless, we found it to be in good agreement with the other two samples, which provides an important check of systematics. Further follow-up of the SL2S sample would increase the sample size and potentially contribute more to improving the precision in combination with TDCOSMO, providing a more stringent validation test.

\subsection{Comparison with other work}
\label{ssec:others}

We provide a brief comparison here with the broader literature on time-delay cosmography, referring the reader to recent review articles for more details \citep[e.g.,][]{Treu_review,TreuShajib_review,Birrer:2024, Suyu2024}.

Measurements based on lensed quasars have been around for decades and have improved considerably since the early days, owing to improvements in data quality and modeling methods. The work presented here is the culmination of years of efforts to perfect the so-called "simply-parametrized" approach, in which the mass distribution or gravitational potential of the main deflector is described by a physically motivated functional form depending on a relatively small number of free parameters ($\sim$10--20 typically), constrained by high resolution images and stellar kinematics \citep[see, e.g.,][for an early result]{Oguri2007}. 

The alternative approach is the so-called "free-form" method, in which the mass distribution or gravitational potential of the main deflector is described by pixels or a basis set \citep{Coles2008,Paraficz2010}. In this approach, the number of free parameters vastly exceeds the number of constraints, requiring regularization to avoid overfitting. These methods have also improved considerably since the early days, with more realistic priors and regularization, and have benefited from better data. With reasonable priors and regularization, \citet{Denzel21} finds 
$H_0=71.8^{+3.9}_{-3.3}$ \ksmpc, although the result and precision are heavily dependent on those choices.

A recent and exciting development has been the discovery of multiply imaged SNe \citep{Kelly15}, the astronomical source originally proposed by \cite{refsdal1964} for time-delay cosmography. The first measurements of $H_0$ based on lensed SNe have been published. \citet{Kelly2023b} find $H_0=66.6_{-3.3}^{+4.1}$ \ksmpc based on SN "Refsdal" \citep[see also][]{Grillo24} and \citet{Pascale25} find $H_0 = 71.8^{+ 9.2}_{‑8.1}$ \ksmpc based on SN "H0pe". Although the uncertainties based on just these two systems are too large to settle the Hubble tension, large numbers of lensed SNe are expected to be discovered in the near future \citep[e.g.,][]{Arendse24} and will further boost the precision of time-delay cosmography \citep{Suyu2020, BDS22, Shajib25lsst}.

The results presented here are overall in excellent agreement with recent work on time-delay cosmography. SN "Refsdal" is the most apparently discrepant result, and yet it is consistent with our measurement within 2$\sigma$. Considering the widely different sources, deflectors, and methods involved in these independent studies, the overall agreement can be viewed as a validation of time-delay cosmography.

\subsection{Comparison with non-lensing measurements}
\label{ssec:tension}

\begin{figure*}
    \centering
    \includegraphics[width=\textwidth]{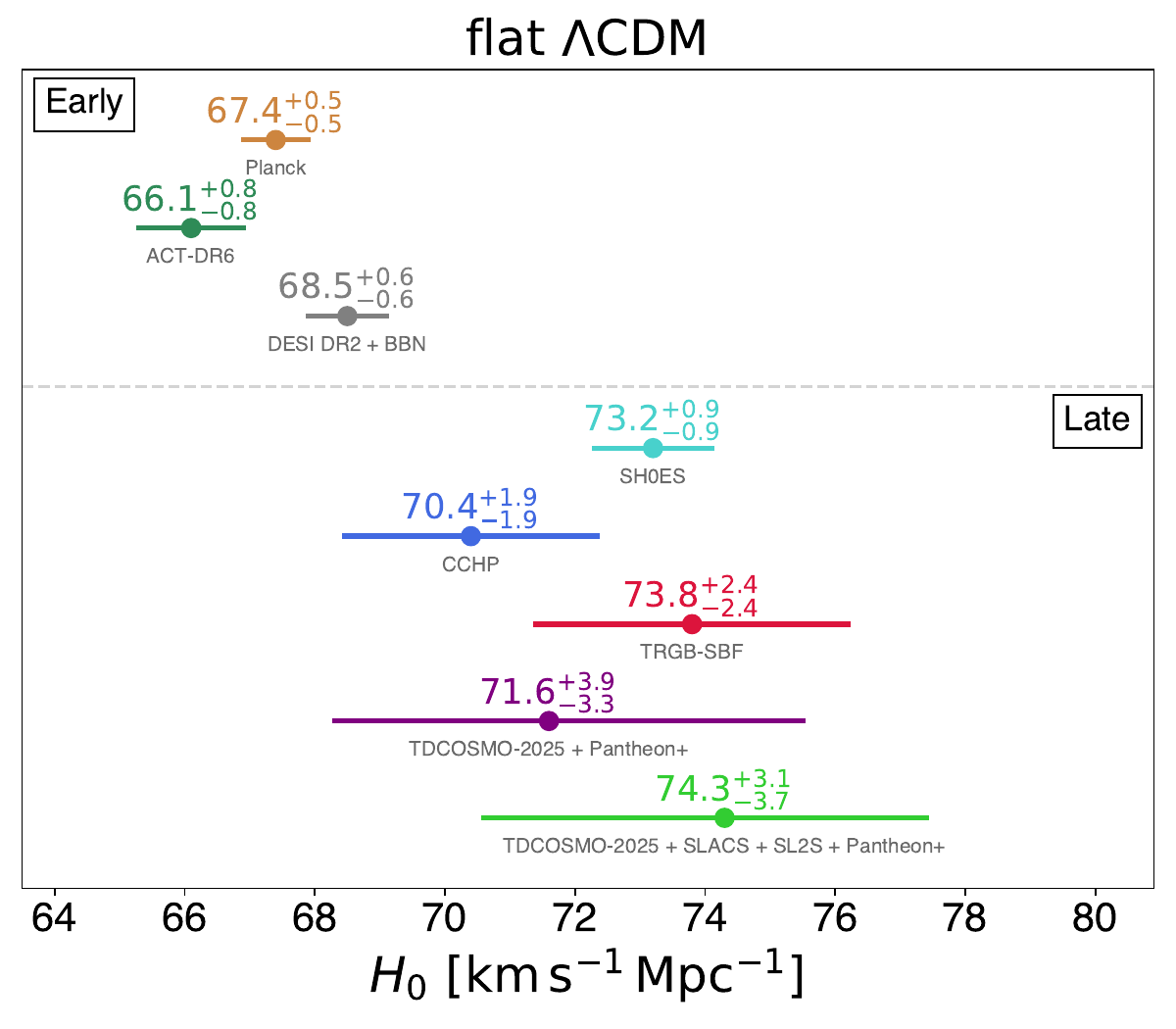}
    \caption{Illustration of the current state of the "Hubble tension" in flat $\Lambda$CDM. Only a selection of recent measurements is shown for the early- and late-Universe probes, in order to avoid overcrowding the plot \citep[for an exhaustive list, see][]{cosmoverse25}. Among the early-Universe probes, we show measurements from the CMB \citep[{\it Planck} and ACT-DR6;][]{Planck2020,ACTDR6}, and baryonic acoustic oscillations (BAO) plus Big Bang nucleosynthesis (BBN) \citep{DESIDR2}. Among the late-Universe probes, we show the Cepheid+SN measurement from the SH0ES team \citep{Breuval2024}, the tip of the red giant branch (TRGB) plus SN measurement from the CCHP team \citep{Freedman24}, the TRGB plus surface brightness fluctuation (SBF) measurement by \citet{Jensen2025},  along with the TDCOSMO-2025 and TDCOSMO-2025+SLACS+SL2S time-delay cosmography measurements in combination with the Pantheon+ likelihood \citep{Scolnic2022, Brout:2022}.}
    \label{fig:H0_tension}
\end{figure*}

In Fig.~\ref{fig:H0_tension}, we show a selection of recent measurements of the Hubble constant in flat $\Lambda$CDM \citep[for a complete list, see the recent review by][]{cosmoverse25}. While the selection is by no means unique, we aimed to represent a range of independent methods. As discussed in the Introduction, the early-Universe probes clearly prefer \Hc\ around 66--68 \ksmpc, while late-Universe probes prefer values above 70 \ksmpc. Our measurements are in excellent agreement with other late-Universe probes. Importantly, this is true also for our measurement \Hc$=77.8^{+3.7}_{-4.7}$ \ksmpc from TDCOSMO-2025+SLACS+SL2S without utilizing any SN information, which is thus completely independent of all other methods. However, given the uncertainties, the results from the early-Universe probes are also consistent within 3$\sigma$ (see discussion in Section~\ref{sec:alternative_cosmo} for a more general comparison with the CMB results). Further improvements in the precision of the TDCOSMO measurements are planned and will sharpen the contribution of time-delay cosmography to the "Hubble tension" debate.

\section{Systematics}
\label{sec:syst}

In this section, we discuss residual systematic errors and our efforts over the past few years to quantify and mitigate them (Section~\ref{ssec:modelass}). We also discuss and quantify potential selection effects (Section~\ref{ssec:sel}).

\subsection{Model assumptions}
\label{ssec:modelass}

In addition to those discussed in this paper, several other sources of systematic uncertainties should be considered when interpreting our results. A brief summary is provided here, referring the reader to the original references for more details. The important point is that all of these uncertainties are much smaller than those associated with our measurement of $H_0$, and are random and uncorrelated. Therefore, they will not change our result by more than a fraction of our error bar. Below, we list the sources of uncertainty that are not yet included in our baseline analysis, unless explicitly stated otherwise:

\begin{itemize}
    \item \textit{Parametrization of the mass model:} TDCOSMO-1 showed that using a so-called "composite" model, i.e., a constant stellar mass-to-light ratio plus an NFW dark matter halo, changes $H_0$ by 0.2\% with respect to adopting a power-law model. This is important because the composite model has the flexibility to represent mass density profiles that differ from a power law. In practice, however, the data reveal that the two components combine to form an almost perfect power-law profile in the radial range around the Einstein radius, a phenomenon known as the "bulge-halo conspiracy" \citep{TreuKoopmans2004,DuttonTreu2014}.
    \item \textit{Uncertainties on the lens models:} Time-delay cosmography requires accurate mass models to infer the difference in the Fermat potential, under the assumption of an elliptical power-law model. Random errors are at the 1--5\% level \citep{Wong2019} for data of our quality. We performed two tests of potential systematics. \citet{Shajib2022} presented a blind analysis of \WGDtwenty\ based on the two independent softwares used by our collaboration \citep[{\sc glee} and {\sc lenstronomy}, respectively;][]{SuyuHalkola2010, Suyu2012, Birrer2018}, independent modeling choices, and PSF reconstruction. They found the results to agree within the errors. \citet{TDCOSMO20} modeled new JWST-NIRcam images of \WFItwenty and found the results to be consistent within the error to those obtained by \cite{Rusu2020} using HST images.
    \item \textit{Dark matter substructure:} In our mass models, we assumed the mass distribution of the deflector could be described by a smooth function. In reality, cold dark matter is clumpy on subgalactic scales. However, this approximation introduces only $\sim 1$\% random scatter in the observed time-delay distance \citep{TDCOSMO3}, which is negligible given the other sources of uncertainty. 
    \item \textit{Peculiar velocities with respect to the CMB reference frame} affect the time delay distances by a small amount. Accounting for both the motion of the Milky Way with respect to the CMB and peculiar velocities, \citet{Dalang2023} estimated the effect to be $\sim 0.25$ \% for the TDCOSMO-1 sample. 
    \item \textit{A central supermassive black hole} could affect the interpretation of the stellar velocity dispersion at the JWST-NIRSpec angular resolution \citep{Wang2025}. This is not accounted for in this work and will be included in future works, when we will analyze 2D JWST kinematic maps in detail.
    \item \textit{Departures for spherical symmetry in dynamical models:} As discussed by \citet{Huang:2025}, after we apply our correction for projection effects, residual random errors are at the level of 1--2\% in $\lambda_{\rm int}$, which we account for in this work. Impacts on the deprojection from a potential triaxiality are below the 1\% level. 
    \item \textit{Our time-delay measurement method} has been tested in a blind data challenge \citep{Liao2015}. The COSMOGRAIL group's technique, namely \textsc{PyCS3} \citep{Tewes2013, Millon2020c}, has consistently produced accurate and precise time delays, with residual systematics and biases of less than 1\% \citep{Bonvin2016}. Systematic errors due to the microlensing time-delay effect \citep{Tie2017} are on the order of a few hours and have only a small impact on our measurements, given that the typical time delays in our sample of lensed quasars span weeks to months \citep{Chen2018}.
    \item \textit{Testing hierarchical inference:} \cite{TDCOSMO8} tested the ability of the hierarchical framework to infer an unbiased $H_0$ by using mock lenses with a composite (dark matter + baryons) mass distribution that they modeled with a power-law mass profile. The framework successfully recovered $H_0$ within 1.5$\sigma$ of the fiducial value when using time-delay lenses alone. The precision was improved to $\sim$ 2\% and accuracy to 0.7\% median offset when combining time-delay lenses with an external sample of non-time-delay systems at lower $z$, similar to the SLACS sample. This confirmed that non-time-delay lenses enhance constraints on $H_0$ when assuming a shared population, despite systematic differences in radii probed by the lensed images (i.e. difference in $R_{\rm{eff}} / \theta_{\rm {E}}$). 
    \item \textit{Departure from elliptical symmetry in the lens models:} \citet{TDCOSMO7} showed that for realistic values of azimuthal structure higher than order 2 (boxyness and diskyness), the effect on $H_0$ is less than 1\%.
    \item \textit{Nearby perturbers in the SLACS and SL2S lens models:} Whereas for the TDCOSMO-2025 lenses we modeled explicitly all the nearby perturbers that contribute more than $5\times10^{-4}$ in the flexion shift \citep{McCuly2017}, this was not done for the SLACS and SL2S lenses. Visual inspection of the selected SLACS and SL2S indicates no major perturber, indicating that the effect is likely to be small. However, this step of the analysis will likely need to be implemented in the external lens datasets as precision improves, to further homogenize the treatment of the two samples.
\end{itemize}

\subsection{Selection effects}
\label{ssec:sel}

Selection effects could potentially bias our results if they are large enough and not properly accounted for \citep[see, e.g.][for a comprehensive discussion of strong lensing selection functions and selection effects]{Sonnenfeld23}. There are two kinds of selection effects that could play a role: those affecting the time-delay lenses in an absolute sense and differential ones affecting the combination of time-delay and non-time-delay lenses. 

\subsubsection{Selection effects in time-delay lenses}

The TDCOSMO-2025 sample is composed almost exclusively of quadruply imaged quasars (seven out of eight). This is for good reasons: quads have significantly more information content per system than doubles. Therefore, they have been prioritized for follow-up and analysis.
However, quads could, in principle, be a biased subset of galaxies; for example, they preferentially select systems with more ellipticity or external shear that subtend a larger area within the four-image caustic. This would have implications only insofar as priors or other information from non-quads is used. The bias, excluding a caustic area prior in the individual lens models, is estimated to be sub-percent, and even lower for well-constrained lens models based on our high-resolution imaging data \citep{BaldwinSchechter2024}. In our analysis, this is a small effect, and it is mostly present in the correction for projection along the LoS in the modeling of stellar velocity dispersion \citep{Huang:2025}. Since this is a percent-level correction, we expect the effect to be sub-percent on $H_0$. To quantify the effect more precisely, one needs to know the parent population and the selection function, both of which are difficult to determine. Based on a simple model for the deflector and source population, \citet{CC16} estimate the potential bias to be at the sub-percent level. \citet{Sonnenfeld23} reaches a similar conclusion that the residual bias is probably sub-percent for current samples. \citet{Li25} claim a larger effect, but their result is not relevant because their model fails to reproduce the actual observations. Their model predicts $\lambda_{\rm int}=0.87^{+0.07}_{-0.06}$ for the TDCOSMO-4 quads, in stark contrast to the measured value there, and that obtained in this paper. Their inference of a lower $H_0$ than that presented in TDCOSMO-4 is almost entirely due to the underestimated $\lambda_{\rm int}$, and insufficiently describes the selection effects that may affect real lenses. 

In conclusion, based on the work done so far, it seems that any residual bias due to unmodeled selection effects is smaller than our uncertainties. However, it is clear that more work remains to be done on this front, as the precision improves. On the one hand, even more realistic descriptions of the parent population and selection function can be implemented in theoretical models. On the other hand, samples of doubles with similar precision to those of quads would provide an empirical test of residual biases.

\subsubsection{Differential selection effects between time-delay and non-time-delay lenses}

A potential source of bias when combining with external lens samples is due to any differences in the selection function of the time-delay and non-time-delay lenses. In this work, we have assumed that the TDCOSMO-2025, SL2S, and SLACS samples have been drawn from the same population, after applying cuts to match observable properties as closely as possible, such as projected ellipticity and lensing velocity dispersion. Potential differences between the populations are described by allowing $\lambda_{\rm int}$ to depend on the ratio of effective and Einstein radius (Eq.~\ref{eqn:lambda_scaling}), which is different for the three samples.

However, the selection functions are inherently different for the three samples. For example, \citet{Sonnenfeld2025} shows that the SLACS follow-up procedure -- prioritizing systems with higher stellar velocity dispersion from SDSS -- introduces a bias in some of the properties of the SLACS lenses compared to the parent population. The most important for our purposes is that the SLACS lenses' stellar velocity dispersion 
could be biased high by up to 5\% when matching the mass profile. Part of that bias stems from noisy SDSS spectra used for the pre-selection, in the sense that velocity dispersions that were affected by upward noise fluctuations were preferentially selected. That component is eliminated by our analysis as we replaced in this work the SDSS spectra with much higher SNR KCWI ones. A residual bias, due to the intrinsic scatter in the velocity dispersion of elliptical galaxies, could still be present, though a no-bias scenario is also consistent with the data \citep[see Fig.~4 of][]{Sonnenfeld2025}. It is for this reason that we present separately the results based on the TDCOSMO-2025 lenses only and those in combination with external lens samples.  

Our mutually blind analysis shows excellent agreement between the TDCOSMO-2025, SLACS, and SL2S inference of $\lambda_{\rm int}$ within the uncertainties.  However, as the precision of the measurement improves, more work will be needed to quantify, mitigate, and correct absolute and differential biases due to the lenses' selection function in order to ensure that they remain below the statistical errors.

\section{Summary and conclusions}
\label{sec:summary}

We presented cosmological constraints based on the analysis of eight multiply imaged quasars with measured time delays and two external samples of non-time delay lenses. This is the culmination of 5 years of work by the TDCOSMO collaboration, including new data, new methods, and many significant improvements with respect to our earlier works based on 7 time delay lenses \citep{Millon:2020,TDCOSMOIV}. Following \citet{TDCOSMOIV}, we parametrized the mass density profile of the main deflector in terms of the mass-sheet-degeneracy parameter $\lambda_{\rm int}$. Using lensing-only observables, this parametrization is completely degenerate with the Hubble constant and thus maximizes the uncertainty, providing a conservative estimate of the error budget. 

Within this framework, the degeneracy is broken by stellar velocity dispersion.  Therefore, a major effort went into obtaining higher-quality data and improving the methodology to measure stellar velocity dispersion with respect to our previous work \cite{Knabel25}. We obtained new JWST-NIRspec spectra for six out of eight time-delay lenses and VLT-MUSE spectra for the other two. For the time delay lens \RXJ, spatially resolved velocity dispersion measurements from both JWST-NIRSpec \citep{Shajib25b} and Keck-KCWI spectroscopy \citep{Shajib23} were used. New spectra and spectral measurements were obtained for the external lens samples. For the SLACS lenses, SDSS spectroscopy was replaced by much higher SNR ($\sim160$ \AA$^{-1}$) spatially resolved kinematics obtained with KCWI at Keck \citep{Knabel24}. For the SL2S lenses, a re-analysis of Keck and VLT data was performed \citep{Mozumdar2025}. New lens models were obtained for both SLACS and SL2S datasets, with improved methods and data \citep{Tan24,Sheu25}. Stringent quality cuts were imposed to guarantee the high quality of all measurements, and to match the observable properties of the TDCOSMO-2025 SLACS and SL2S samples. We also accounted for projection effects in the interpretation of the stellar velocity dispersions \citep{Huang:2025} and improved our description of the velocity dispersion anisotropy and surface brightness profiles. 

The main results of this study can be summarized as follows:

\begin{itemize}
    \item In a flat \lcdm model, using the TDCOSMO-2025 sample in combination with constraints on $\Om$ from the Pantheon+ likelihood, we find $H_0=\hnotflcdmonea$ \ksmpc. A consistent result \Hc$=71.2^{+3.7}_{-3.6}$ \ksmpc is found by combining with the DES-SN5YR sample instead of Pantheon+.
    \item The TDCOSMO-2025, SLACS, and SL2S samples yield consistent constraints and can thus be combined to improve the precision. In a flat \lcdm model, using the TDCOSMO-2025+SLACS+SL2S sample in combination with constraints on $\Om$ from the Pantheon+ sample, we find $H_0=\hnotflcdmoned$ \ksmpc.
    \item The constraints on $H_0$ from time-delay cosmography alone are robust to the choice of cosmological model. For the TDCOSMO-2025+SLACS+SL2S dataset we find \Hc=$77.8^{+3.7}_{-4.7}$ \ksmpc in flat \lcdm, $H_0=77.1^{+4.9}_{-4.1}$ \ksmpc in open $\Lambda$CDM, $78.0^{+6.6}_{-7.3}$ \ksmpc in $w$CDM, $80.9^{+11.2}_{-8.0}$ \ksmpc in $w_0w_a$CDM, and $75.9^{+6.8}_{-7.8}$ in $w_\phi$CDM. 
    \item The TDCOSMO data are consistent with the DESI-DR2 BAO data, the Pantheon+ SN, and DES-SN5YR data in all the cosmologies considered here.
    \item In combination with the {\it Planck} likelihood, the TDCOSMO data push the CMB posterior towards flatness in open \lcdm $\Omega_k=-0.006^{+0.005}_{-0.006}$, and towards $w=-1$ in $w$CDM $w=-1.38^{+0.13}_{-0.08}$.
\end{itemize}

These results illustrate the power of time-delay lenses to constrain \Hc independently of the local distance ladder, and of other cosmological parameters, thereby contributing to the current scientific debate on the "Hubble tension". The overall precision has almost doubled with respect to TDCOSMO-4, owing mainly to the use of better spectroscopic data and methods for stellar kinematics. Importantly, replacing the SDSS-based spectra with much higher SNR spectra from the Keck-KCWI brought the SLACS sample into excellent agreement with the TDCOSMO-2025 dataset.

Looking ahead, the TDCOSMO collaboration is working on multiple fronts to further improve our precision accuracy towards our ultimate goal of 1\%. First, we are expanding the number of lensed quasars with measured time delays \citep{Dux:2025}, cosmography grade models, and ancillary data. Second, using recent improvements in modeling speed \citep{Wang2025}, we plan to take advantage of spatially resolved kinematics from space and from the ground to implement axisymmetric Jeans modeling \citep[see][for a first illustration]{Shajib23}. Third, we are expanding the sample of non-time delay lenses with data of sufficient quality to be used for cosmography. These improvements will be the subject of future papers. 
    
\bibliographystyle{aa}
\bibliography{biblio}

\newpage

\begin{appendix}

\section{Main properties of the lenses}
\label{app:lenses}

In Table~\ref{tab:quality_lens_property} in this appendix, we summarize the main properties of the TDCOSMO time delay lenses and of the selected SLACS and SL2S lenses.

\begin{table*}[h!]
\caption{\label{tab:quality_lens_property}
Physical properties for the lens samples included in the joint lensing+kinematics cosmology inference.
}
\centering
\renewcommand{\arraystretch}{1.5}
\begin{tabular}{lccccccccc}
\hline \hline
Lens & $z_\mathrm{lens}$ & $z_\mathrm{source}$ & $\sigma^\mathrm{ap}_{\rm los}$ [km s$^{-1}$] & $R_\mathrm{eff}$ [\arcsec] & $\theta_{\rm E}$ [\arcsec] & $\gamma_{\rm pl}$ & $q_{\rm light}$ & $\mathcal{I}$ \\
\hline

\DESzerofour & 0.597 & 2.375 & 242.3$\pm$12.2 & 1.940 & 1.92 & 1.90$\pm$0.03 & 0.80 & --  \\ 

\HEzero & 0.455 & 1.693 & 226.6$\pm$5.8 & 1.800 & 1.22 & 1.93$\pm$0.02 & 0.93 & --  \\ 

\PGeleven & 0.311 & 1.722 & 235.7$\pm$6.6 & 0.450 & 1.08 & 2.17$\pm$0.05 & 0.95 & -- \\ 

\RXJ & 0.295 & 0.654 & 303.0$\pm$8.3 & 1.910 & 1.63 & 1.95$\pm$0.05 & 0.94 & -- \\ 

\Jtwelve & 0.745 & 1.789 & 290.5$\pm$9.5 & 0.290 & 1.25 & 1.95$\pm$0.05 & 0.85 & -- \\ 

\Bsixteen & 0.630 & 1.394 & 305.3$\pm$11.0 & 0.590 & 0.81 & 2.08$\pm$0.03 & 1.00 & -- \\ 

\WFItwenty & 0.657 & 1.662 & 210.7$\pm$10.5 & 1.970 & 0.94 & 1.95$\pm$0.02 & 0.83 & -- \\ 

\WGDtwenty & 0.228 & 0.777 & 254.7$\pm$16.3 & 2.223 & 1.38 & 2.30$\pm$0.02 & 0.85 & -- \\ 

\hline

SDSS J0029$-$0055 & 0.227 & 0.931 & 208.2$\pm$2.8 & 2.386 & 0.94 & 2.69$\pm$0.25 & 0.94 & 128.3 \\ 

SDSS J0037$-$0942 & 0.195 & 0.632 & 278.8$\pm$2.9 & 3.334 & 1.48 & 2.38$\pm$0.11 & 0.87 & 240.1 \\ 

SDSS J1112$+$0826 & 0.273 & 0.629 & 276.4$\pm$3.4 & 1.623 & 1.50 & 1.92$\pm$0.18 & 0.76 & 254.9 \\ 

SDSS J1204$+$0358 & 0.164 & 0.631 & 260.2$\pm$2.9 & 1.544 & 1.29 & 2.02$\pm$0.05 & 0.99 & 204.9 \\ 

SDSS J1250$+$0523 & 0.232 & 0.795 & 243.4$\pm$2.6 & 1.748 & 1.12 & 2.18$\pm$0.09 & 0.92 & 225.1 \\ 

SDSS J1306$+$0600 & 0.173 & 0.472 & 229.7$\pm$2.8 & 2.369 & 1.31 & 1.97$\pm$0.07 & 0.88 & 121.1 \\ 

SDSS J1402$+$6321 & 0.205 & 0.481 & 282.6$\pm$2.9 & 2.720 & 1.37 & 1.57$\pm$0.30 & 0.76 & 56.5 \\ 

SDSS J1531$-$0105 & 0.160 & 0.744 & 275.8$\pm$3.6 & 3.416 & 1.71 & 2.07$\pm$0.26 & 0.85 & 118.7 \\ 

SDSS J1621$+$3931 & 0.245 & 0.602 & 262.7$\pm$3.5 & 2.402 & 1.27 & 2.02$\pm$0.07 & 0.87 & 265.2 \\ 

SDSS J1627$-$0053 & 0.208 & 0.524 & 263.5$\pm$3.3 & 2.981 & 1.22 & 1.93$\pm$0.14 & 0.92 & 185.6 \\ 

SDSS J1630$+$4520 & 0.248 & 0.793 & 279.4$\pm$3.1 & 2.012 & 1.79 & 1.92$\pm$0.09 & 0.83 & 161.2 \\  

\hline

SL2SJ0226$-$0420 & 0.494 & 1.230 & 310.0$\pm$14.0 & 0.389 & 1.15 & 2.05$\pm$0.15 & 0.73 & 224.1 \\ 

SL2SJ0855$-$0147 & 0.365 & 3.390 & 196.0$\pm$11.0 & 0.470 & 0.96 & 2.25$\pm$0.11 & 0.91 & 177.5 \\ 

SL2SJ0904$-$0059 & 0.611 & 2.360 & 232.0$\pm$9.0 & 0.713 & 1.41 & 1.98$\pm$0.15 & 0.81 & 319.7 \\ 

SL2SJ2221$+$0115 & 0.325 & 2.350 & 277.0$\pm$13.0 & 0.532 & 1.27 & 2.14$\pm$0.13 & 0.77 & 195.3 \\

\hline
\end{tabular}

\tablefoot{From top to bottom, the time-delay TDCOSMO-2025 lenses, the SLACS sample, and the SL2S sample are separated by horizontal dividers. From left to right, the listed quantities are: deflector redshift $z_\mathrm{lens}$, background source redshift $z_\mathrm{source}$, stellar velocity dispersion measured within an aperture $\sigma^\mathrm{ap}_\mathrm{los}$, half-light radius $R_\mathrm{eff}$, Einstein radius \thetaE, lensing power-law slope $\gamma_\mathrm{pl}$, observed axis ratio of light profile of the deflector $q_\mathrm{light}$, and lensing information $\mathcal{I}$ calculated by \citet{Sheu25}. The uncertainties in the stellar velocity dispersions $\sigma^\mathrm{ap}_{\rm los}$ include both statistical and systematic uncertainties, summed in quadrature. For the lenses with IFU kinematic data (\RXJ and the quality SLACS sample), we list the deflector's stellar velocity dispersion $\sigma^\mathrm{ap}_{\rm los}$ measured within apertures with radii equal to half the effective radii \citep[Table 2 in][]{Knabel24} for comparison with lenses with aperture stellar velocity dispersions.
The stellar velocity dispersion $\sigma^\mathrm{ap}_{\rm los}$ for the SL2S sample is taken from \citet{Mozumdar2025}. For the half-light radius $R_\mathrm{eff}$, we estimated the uncertainty to be 5\% in the inference. For the Einstein radius \thetaE, we estimated the uncertainty to be 2\%. For lenses with aperture kinematic data, the uncertainty for the lensing power-law slope $\gamma_\mathrm{pl}$ was marginalized over in the inference. For the SLACS lenses with velocity dispersion profiles from spatially resolved data, the power-law slope inference is driven by the kinematic data.}
\end{table*}

\begin{figure}
    \centering
    \includegraphics[width=\linewidth]{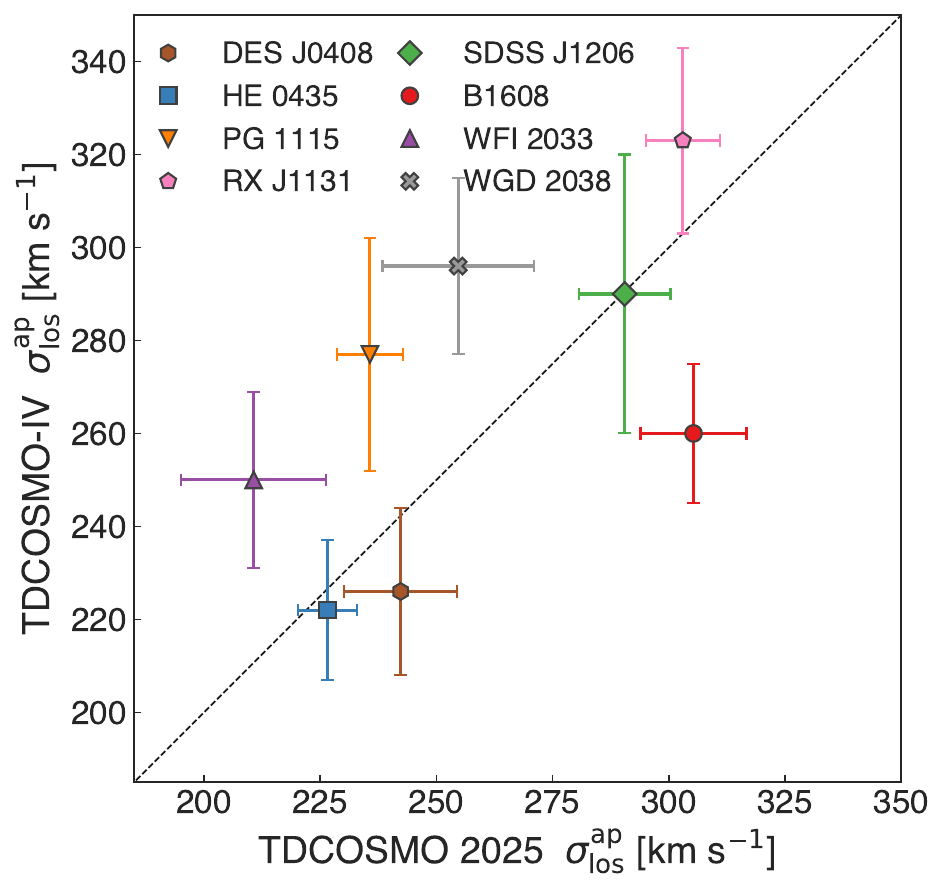}
    \caption{Comparison of the stellar velocity dispersions measurement used in this paper and those in TDCOSMO-4 and \citet[][for \WGDtwenty]{Wong24}. Note that the measurement apertures and PSF resolutions differ between those presented in this paper and those in previous ones. Thus, the measurements are not expected to match exactly. For \RXJ, we illustrate here the value reported in Table~\ref{tab:quality_lens_property} that corresponds to the stellar velocity dispersion measured within half the effective radius.} 
    \label{fig:oldnew}
\end{figure}

In Fig.~\ref{fig:oldnew}, produced after unblinding, we compare the newly measured stellar velocity dispersion to those used by TDCOSMO-4 and \citet{Wong24}.

\section{Modeling of \RXJ}
\label{app:rxj}

In this Appendix, we provide a more detailed description of the analysis of the time delay lens \RXJ, the one for which spatially resolved kinematics are available from JWST and KCWI spectroscopy. One important factor is that the time-delay distance \Ddt based on the lens model by \citet{Suyu2013} depends on the inferred power-law slope $\gamma_{\rm pl}$. This correlation was taken into account when combining the lens model likelihood with the likelihoods from the dynamical analysis, that constrain $\gamma_{\rm pl}$ but not \Ddt.

\begin{figure*}
    \centering
    \includegraphics[width=\linewidth]{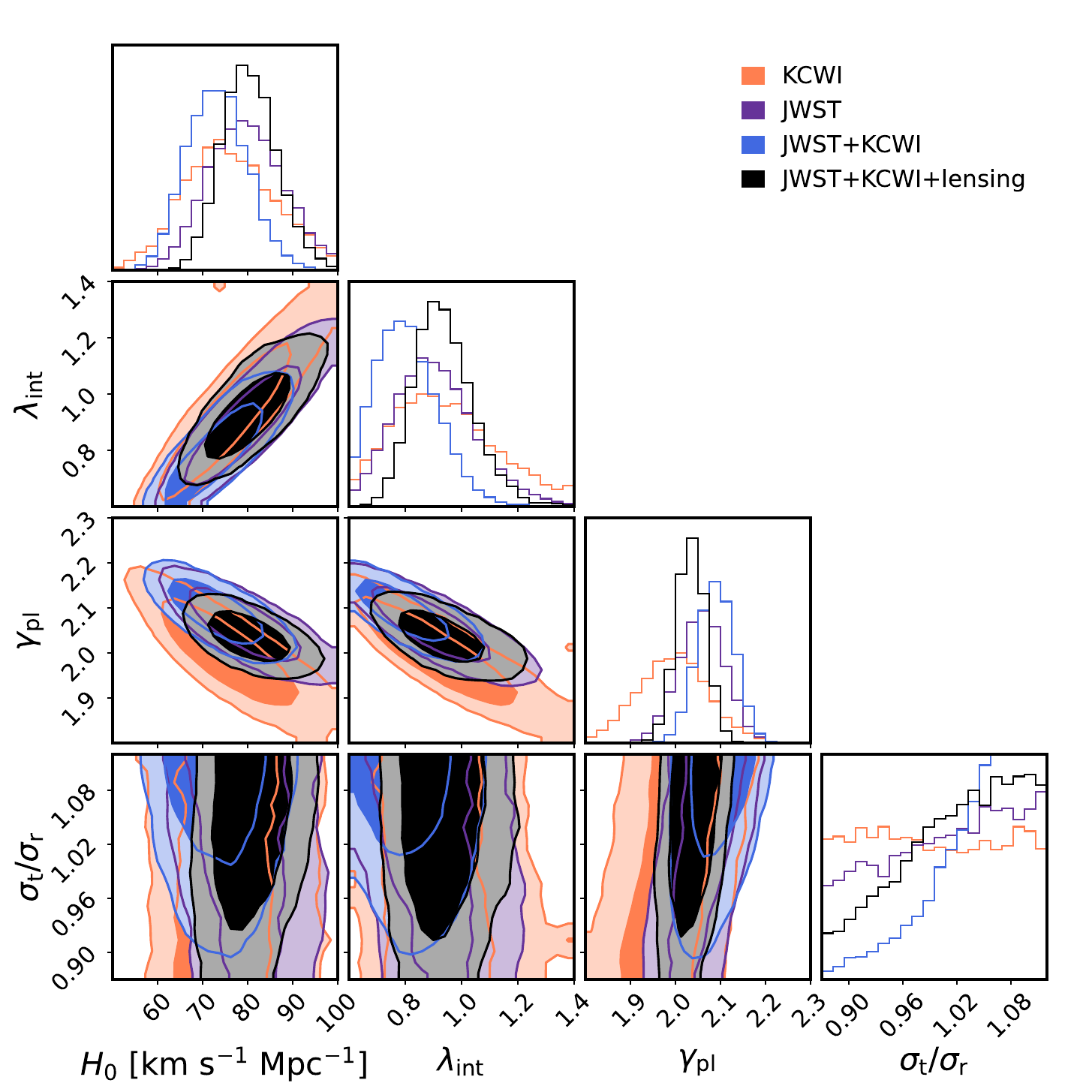}
    \caption{Posterior distribution function of key parameters describing the time-delay lens \RXJ, based on velocity dispersion profiles presented in this paper from the Keck-KCWI and JWST-NIRSpec data, and in combination with the lens model inferred power-law density slope by \citet{Suyu2013}. The individual analysis of this system is shown to demonstrate that the three datasets are mutually consistent and can therefore be combined, and to illustrate the key degeneracies between the parameters. The inferred $H_0$ from this system is consistent with that obtained by \citet{Shajib23} based on the analysis of 2D kinematic maps from the KCWI. In the hierarchical analysis presented in the main body of the paper, $H_0$ and $\lambda_{\rm int}$ are population-level parameters, jointly constrained by all the lenses in the sample.}
    \label{fig:RXJanalysis}
\end{figure*}

In Fig.~\ref{fig:RXJanalysis}, we show the results of the fit to the \RXJ data alone, without combining with the other lenses in our hierarchical analysis. This analysis is similar to that presented by \citet{Shajib23}, except that one-dimensional velocity dispersion profiles are considered instead of the 2D maps.
The figure shows that the KCWI and JWST data yield consistent results, which can be described by a single mass model and combined. The likelihoods can also be combined with the inference of $\gamma_{\rm pl}$ from the lens model, yielding the tighter constraints in black. We note that the contours in black are consistent with an isothermal power law ($\lambda_{\rm int}=1$, $\gamma_{\rm pl}$=2). The constraints on $H_0$ from this system alone are consistent with those obtained by \citet{Shajib23}.

\section{Kinematic fits to the KCWI stellar velocity dispersion profiles of the SLACS lenses }
\label{app:SLACS}

\begin{figure*}
    \centering
    \includegraphics[width=\linewidth]{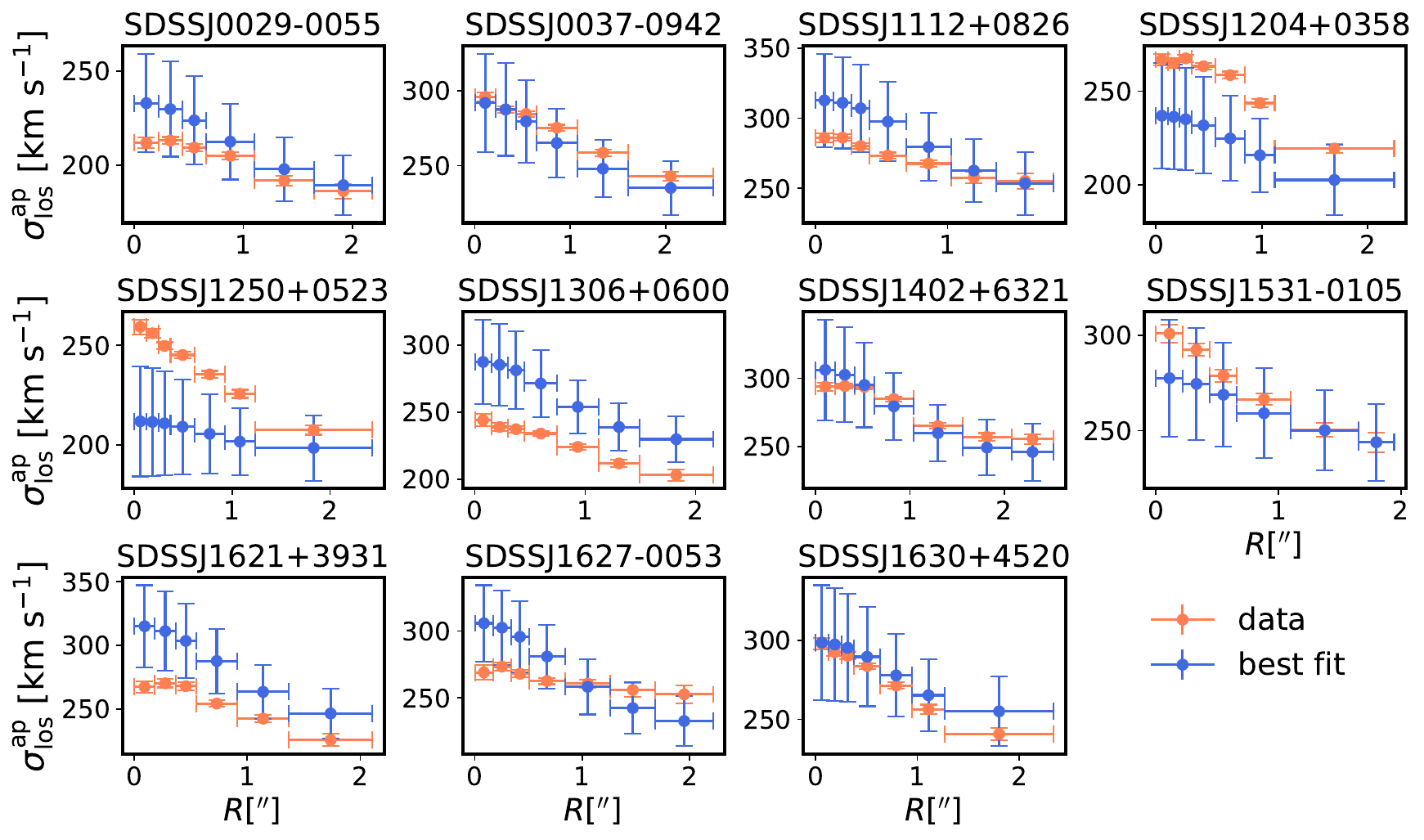}
    \caption{For each SLACS lens used in our analysis, we show the observed stellar velocity dispersion profile and the best fit found in the hierarchical analysis. The data for individual lenses is described very well by our model when applied to a single lens. However, in the hierarchical analysis, they were all required to share a single population-level value of $\lambda_{\rm int}$ (Section~\ref{sec:hierarch}) with the time delay lenses and among each other. The offsets between the data and the best-fit model are due to mismatches between the population-level value of $\lambda_{\rm int}$ and the individual $\lambda_{\rm int}$ that would be required to fit each lens separately. The offsets are reflected in the non-zero scatter in $\lambda_{\rm int}$ found for the SLACS population.}
    \label{fig:KCWIfits}
\end{figure*}

In Fig.~\ref{fig:KCWIfits} we show the velocity dispersion profiles measured for the SLACS lenses based on KCWI data and compare them with the results of the hierarchical fit. 

\section{Author contributions and acknowledgments}
\begin{authorcontributions}
The authors of this paper are listed in alphabetical order. Here, we specify each author's contribution to the paper.

S.B.~developed the hierarchical framework, led the development of \textsc{hierArc}, contributed to the analysis and writing, and oversaw the Stony Brook group. E.J.B.-G.~performed the LoS analysis for \DESzerofour and \WGDtwenty. M.C.~contributed to the stellar kinematic measurements, dynamical modeling, and paper writing and served as the internal reviewer of the paper. F.C.~established and led the time-delay campaigns, supervised the EPFL group. F.D.~measured the time delay for \WGDtwenty. C.D.F.~provided suggestions for data analysis, contributed to the development of environmental analysis code, and supervised the development of one of the velocity dispersion codes as the PhD supervisor of P.M.~and P.W., as well as the acquisition of imaging data used for environmental analysis of several lenses. J.F.~served as the internal reviewer of the paper and led the DES and U Chicago groups. A.G.~served as the internal code reviewer, participated in the development of \textsc{squirrel}, and provided comments on the analysis and the paper. D.G.~quantified uncertainties in the time delays from dark-matter substructure. X.-Y.H.~managed the hierarchical analysis pipeline, led the axisymmetric correction in kinematic modeling, and participated in writing and code review. S.K.~acquired, reduced, and measured KCWI spatially-resolved kinematics of the SLACS sample, reduced and measured aperture-integrated kinematics of JWST-NIRSpec data, contributed to developing the stellar kinematics analysis method used in this work, dynamical modeling, and paper writing. D.L.~performed preliminary velocity dispersion measurements for \DESzerofour and \WGDtwenty from the MUSE data, and provided general indirect contributions to kinematic measurements. H.L.~performed data acquisition and analysis for \WGDtwenty and \DESzerofour. M.M.~performed the MUSE kinematic measurements, implemented the probe combination analysis, contributed to the time-delay measurement campaigns, and contributed to the analysis and writing. T.M.~planned the observations of JWST-GTO-1198 and did initial reduction of the NIRCam imaging and NIRSpec IFU data. V.M.~worked on time-delay campaigns and data acquisition. P.M.~measured the stellar kinematics of the SL2S sample and contributed to the development of the stellar kinematics analysis method used. E.P.~worked on time delay campaigns, data acquisition, and reduction. A.J.S.~performed resolved kinematic measurements of \RXJ, wrote part of the paper and made a subset of the figures, developed the \textsc{squirrel} pipeline for kinematics extraction that was used for the TDCOSMO-2025 sample, developed the \textsc{RegalJumper} pipeline for JWST data reduction, performed lens modeling of \DESzerofour and \WGDtwenty, oversaw Project Dinos that provided external lens models for the SLACS and SL2S systems, and acquired a subset of the JWST data used as co-PI of JWST-GO-2974. W.S.~contributed the model parameters for the systems in the SL2S sample. D.S.~performed data reduction and extraction of the MUSE data, designed and contributed to some systematic tests, and conducted the internal review of this paper. A.S.~contributed to the velocity dispersion measurements of the SL2S lenses. C.S.~contributed to the development of the stellar kinematic measurement method used in this paper. MS carried out the observation and initial reduction of the IFU data for three of the targets and participated in discussions on how to improve the pipeline products for IFU data. S.H.S.~led JWST data acquisition of \RXJ, contributed to lens mass modeling, developed mass modeling software \textsc{glee}, served as internal reviewer of the paper, and oversaw the TUM/MPA group. C.Y.T.~provided lens model parameters for the SLACS and SL2S lens systems. T.T.~led the JWST, HST, and Keck data acquisition, contributed to the analysis and stellar kinematic measurements and methods, contributed to the hierarchical analysis and quality control, contributed to the modeling of the TDCOSMO-2025, SLACS, and SL2S lenses, contributed to the writing of the paper, served as final reader of the paper, and oversaw the UCLA group. L.V.~worked on lensing degeneracies and extraction of the MUSE data of \DESzerofour. H.W.~worked on testing the preliminary NIRSpec kinematic map of \RXJ and the degeneracy between anisotropy and black hole mass. P.W.~performed the LoS analysis of the SL2S Sample. D.W.~verified modeling systematics by modeling \WFItwenty's JWST NIRCam data, and contributed to the determination of external convergence. K.C.W.~contributed to lens mass modeling and provided parameter chains for \WGDtwenty. 
\end{authorcontributions}

\begin{acknowledgements}
We are grateful to the many friends and colleagues who contributed to the TDCOSMO collaboration over the years, including A.~Agnello, T.~Anguita, M.W.~Auger, R.D.~Blandford, V.~Bonvin, C.-F.~Chen, L.~Christensen, T.~Collett, S.~Ertl, M.~Gomer, S.~Hilbert,  L.V.E.~Koopmans, C.~Lemon. P.J.~Marshall, S.~Mukherjee, C.E.~Rusu, T.~Schmidt, O.~Tihhonova, and A.~Yıldırım. 

This work is based on observations made with the NASA/ESA/CSA James Webb Space Telescope. The data were obtained from the Mikulski Archive for Space Telescopes at the Space Telescope Science Institute, which is operated by the Association of Universities for Research in Astronomy, Inc., under NASA contract NAS 5-03127 for JWST. These observations are associated with programs \#1198, 1794, and 2974. The specific observations analyzed can be accessed via \url{https://dx.doi.org/10.17909/m1m8-rk48}, \url{https://dx.doi.org/10.17909/wjva-0750}, and \url{https://dx.doi.org/10.17909/gj7b-p622}.
Some of the data presented herein were obtained at Keck Observatory, which is a private 501(c)3 non-profit organization operated as a scientific partnership among the California Institute of Technology, the University of California, and the National Aeronautics and Space Administration. The Observatory was made possible by the generous financial support of the W.~M.~Keck Foundation. 
The authors wish to recognize and acknowledge the very significant cultural role and reverence that the summit of Maunakea has always had within the Native Hawaiian community. We are most fortunate to have the opportunity to conduct observations from this mountain.
Based in part on observations collected at the European Southern Observatory data obtained from the ESO Science Archive Facility.

This research made use of \textsc{Astropy}, a community-developed core \textsc{python} package for Astronomy \citep{Astropy2013, Astropy2018}, the 2D graphics environment \textsc{matplotlib} \citep{Hunter2007}, \textsc{emcee}, a \textsc{python} implementation of an affine invariant MCMC ensemble sampler \citep{Foreman2013}, \textsc{hierArc} version 1.2.0, a package to hierarchically sample cosmological and lens population parameters \citep{TDCOSMOIV}, \textsc{lenstronomy} version 1.13.1 \citep{Birrer2018, Birrer2021lenstronomy}, and the \textsc{squirrel} pipeline \citep{Shajib25b} to streamline kinematic measurements using \textsc{pPXF} \citep{Cappellari04}. 

%
S.B.~and X.-Y.H.~acknowledge support by the Department of Physics \& Astronomy, Stony Brook University.
F.C.~is supported in part by the Swiss National Science Foundation. 
F.C., D.S., and L.V.~acknowledge the support of the European Research Council (ERC) under the European Union’s Horizon 2020 research and innovation programme (COSMICLENS: grant agreement No 787886). 
C.D.F.~acknowledges support for this work from the National Science Foundation under Grant No.~AST-2407278.
A.G.~acknowledges support by the SNSF (Swiss National Science Foundation) through mobility grant P500PT\_211034.
D.L.~was supported by research grants (VIL16599, VIL54489) from VILLUM FONDEN. 
M.M.~acknowledges support by the SNSF (Swiss National Science Foundation) through mobility grant P500PT\_203114 and return CH grant P5R5PT\_225598. 
T.M.~received support from NASA through the STScI grant JWST-GO-3990. 
E.P.~was supported by JSPS KAKENHI Grant Number JP24H00221. 
V.M.~acknowledges support from ANID FONDECYT Regular grant number 1231418, Millennium Science Initiative, AIM23-0001, and Centro de Astrof\'{\i}sica de Valpara\'{\i}so CIDI N21.
Support for this work was provided by NASA through the NASA Hubble Fellowship grant HST-HF2-51492 awarded to A.J.S.~by the STScI, which is operated by the Association of Universities for Research in Astronomy, Inc., for NASA, under contract NAS5-26555. A.J.S.~also received support from NASA through STScI grants HST-GO-16773 and JWST-GO-2974. 
D.S.~acknowledges the support of the Fonds de la Recherche Scientifique-FNRS, Belgium, under grant No.~4.4503.1.
M.S.~acknowledges partial support from NASA grant 80NSSC22K1294. 
S.H.S.~thanks the Max Planck Society for support through the Max Planck Fellowship. This work is supported in part by the Deutsche Forschungsgemeinschaft (DFG, German Research Foundation) under Germany's Excellence Strategy -- EXC-2094 -- 390783311.  
C.Y.T.~was supported by NASA through the Space Telescope Science Institute (STScI) grant HST-AR-16149.  
T.T. acknowledges support by NSF through grants NSF-AST-1906976, NSF-AST-1836016, and NSF-AST-2407277, and from the Moore Foundation through grant 8548. 
K.C.W.~is supported by JSPS KAKENHI Grant Numbers JP20K14511, JP24K07089, JP24H00221. 

\end{acknowledgements}

\end{appendix}

\end{document}